\documentclass[%
 prd,
 twocolumn,
 superscriptaddress,
 numerical,
 showpacs,
 amsmath,amssymb,
 aps,
nofootinbib,
longbibliography,
 floatfix
]{revtex4-2}

\usepackage{bm}

\usepackage{bbm}
\usepackage{slashed}
\usepackage{verbatim}
\usepackage{graphicx}
\usepackage{combelow}
\usepackage{mathtools}
\usepackage[usenames,dvipsnames,svgnames,table]{xcolor}
\usepackage[colorlinks=True,linkcolor=blue,citecolor=blue,urlcolor=blue]{hyperref}
\usepackage{braket}
\let\oldciteauthor=\citeauthor
\def\citeauthor#1{\hypersetup{citecolor=black}\oldciteauthor{#1}}

\let\oldciten=\onlinecite
\def\onlinecite#1{\hypersetup{citecolor=blue}\oldciten{#1}}

\let\oldcite=\cite
\def\cite#1{\hypersetup{citecolor=blue}\oldcite{#1}}

\newcommand{\om}[1]{{\overline{\mathcal{#1}}}}

\newcommand{\bk}{{\mathbf{k}}}

\allowdisplaybreaks
\begin{document}

\title{Dirac fermions under imaginary rotation}

\author{Tudor P\u{a}tuleanu}
\affiliation{Department of Physics, West University of Timi\cb{s}oara,  Bd.~Vasile P\^arvan 4, Timi\cb{s}oara 300223, Romania}

\author{Amalia Dariana Fodor}
\affiliation{Department of Physics, West University of Timi\cb{s}oara,  Bd.~Vasile P\^arvan 4, Timi\cb{s}oara 300223, Romania}

\author{Victor E. Ambru\cb{s}}
\thanks{Corresponding author: victor.ambrus@e-uvt.ro.}
\affiliation{Department of Physics, West University of Timi\cb{s}oara,  Bd.~Vasile P\^arvan 4, Timi\cb{s}oara 300223, Romania}

\author{Cosmin Crucean}
\affiliation{Department of Physics, West University of Timi\cb{s}oara,  Bd.~Vasile P\^arvan 4, Timi\cb{s}oara 300223, Romania}

\begin{abstract}
In the present study, we investigate the properties of an ensemble of free Dirac fermions, at finite inverse temperature $\beta$ and finite chemical potential $\mu$, undergoing rigid rotation with an imaginary angular velocity $\Omega=i\Omega_I$. Our purpose is to establish the analytical structure of such states, as well as the prospects (and dangers) of extrapolating results obtained under imaginary rotation to the case of real rotation. We show that in the thermodynamic limit, the state of the system is akin to a stationary system with modified inverse
temperature $\beta_q = q\beta$ and the same chemical potential, where $q$ is the denominator of the irreducible fraction $\nu = \beta \Omega_I / 2\pi = p/q$. The temperature of the system becomes a fractal function of the rotation parameter, as in the case of the scalar field. The chemical potential breaks the fractalization of fermions. We also compute the thermodynamic potential $\Phi$ and associated thermodynamic functions, showing that they also exhibit fractal behavior. Finally, we evaluate the axial and helical fluxes through the transverse plane, generated through the vortical effects, and show that they diverge in the thermodynamic limit,
in the case
when $\nu = 1/q$ and $q \to \infty$.
\end{abstract}

\maketitle

\section{Introduction}

Due to the spin-orbit coupling included in the Dirac equation, a fermionic plasma under rotation supports dissipationless transport phenomena known as the chiral \cite{Banerjee:2008th,Erdmenger:2008rm,Torabian:2009qk,Son:2009tf,Kharzeev:2015znc} and helical \cite{Ambrus:2019khr,Ambrus:2019ayb} vortical effects. Due to the close connection to the axial anomaly appearing in perturbation theory at the level of triangle diagrams, such effects are called anomalous transport \cite{Landsteiner:2011cp,Landsteiner:2012kd}. Concretely, the vortical effects refer to the emergence of a fermionic current (be it the vector, axial, or chiral current) in the presence of rotation, oriented parallel to the vorticity vector:
\begin{equation}
 \mathbf{J}_\ell = \sigma^\omega_\ell \boldsymbol{\omega},
\end{equation}
where $\ell \in \{V, A, H\}$ labels the corresponding vector ($V$), axial ($A$) or helical ($H$) charge, while the vortical conductivities $\sigma^\omega_\ell$ have the following high-temperature, low-mass expressions:
\begin{gather}
 \sigma^\omega_V = \frac{2\mu_H T}{\pi^2} \ln 2 +
 \frac{\mu_V \mu_A}{\pi^2}, \quad \sigma^\omega_H = \frac{2\mu_V T}{\pi^2} \ln 2 +
 \frac{\mu_H \mu_A}{\pi^2}, \nonumber\\
 \sigma^\omega_A = \frac{T^2}{6} +
 \frac{\mu_V^2 + \mu_A^2 + \mu_H^2}{2\pi^2},
\end{gather}
where $\mu_{V/A/H}$ represent chemical potentials associated with the vector, axial and helical charges, respectively. For the remainder of this paper, we consider only the case of finite vector potential, $\mu_V \equiv \mu$, and $\mu_A = \mu_H = 0$.

Originally, the emergence of a parity-violating current (akin to the axial current) due to spin-orbit coupling was discussed by Vilenkin in the late '70s \cite{Vilenkin:1978is,Vilenkin:1979ui,Vilenkin:1980zv}, with applications in the context of the rotating Kerr black holes.

Vortical effects have seen a signifcant rise in interest after the STAR collaboration confirmed the persistent polarization of the $\Lambda$ hyperons produced in ultrarelativistic heavy-ion collisions \cite{STAR:2017ckg,STAR:2018gyt}. At present, there are multiple measurements for both the global polarization \cite{STAR:2023nvo}, directed along the total angular momentum of the system, and the local polarization \cite{STAR:2019erd}, calculated as a function of momentum, and achieving quantitative agreement between theoretical calculations and experimental measurements represents an intense current avenue of research \cite{Karpenko:2016jyx,Becattini:2020ngo,Becattini:2021iol,Fu:2021pok}.

On the other hand, the local vorticity may have an effect on the thermodynamics of the plasma. Within thermal field theory, one may compute the energy-momentum tensor of the free Dirac field \cite{Ambrus:2014uqa}, thereby establishing the role of quantum corrections due to vorticity, acceleration \cite{Becattini:2015nva,Prokhorov:2019cik,Ambrus:2019khr,Becattini:2020qol,Palermo:2021hlf} and even space-time curvature \cite{Flachi:2014jra,Flachi:2017vlp,Ambrus:2021eod}. The obtained results are consistent with the effective field theory of hydrodynamics \cite{Kovtun:2018dvd,Jain:2020hcu}. A natural question that arises is to what extent such quantum corrections affect strongly-interacting matter, and whether they have an effect on the QCD phase diagram.

Answering this question has been the subject of intense efforts. Starting from the seminal paper of Yamamoto and Hirono \cite{Yamamoto:2013zwa}, opening the path to lattice QCD simulations in rotating systems, many other studies employed similar techniques and in all cases, the study must be performed on the Euclidean manifold
\cite{Braguta:2021jgn,Chen:2022smf,Chernodub:2022veq,Braguta:2023yjn,Azuma:2024bbi}.
In order to keep the co-rotating line element and/or the measured observables real, one must replace the rotation angular velocity $\Omega$ with its imaginary counterpart, $\Omega = i \Omega_I$, while keeping $\Omega_I$ real.
Such studies indicate exotic properties of rotating strongly-interacting matter, such as negative moment of inertia, as well as suppression of chiral restoration \cite{Jiang:2016wvv,Chernodub:2017ref,Chernodub:2020qah,Sun:2023kuu,Singha:2024tpo}. On the other hand, studies done within effective models at real values of the rotation parameters fail to capture such properties, unanimously confirming the enhancement of chiral symmetry restoration in the presence of rotation. Notable exceptions are: Ref.~\cite{Chen:2024tkr}, obtaining agreement with lattice calculations in the perturbative regime of QCD; and Ref.~\cite{Jiang:2023zzu}, obtaining a non-monotonic dependence of the transition temperature on rotation when a non-trivial running of the coupling strength is taken into account.
%\cite{Braguta:2021jgn,Chen:2022smf,Chernodub:2022veq}.

The recent work of Chernodub \cite{Chernodub:2022qlz} has shown that a system under imaginary rotation develops fractal features in the thermodynamic (infinite volume) limit. This behavior was confirmed in Ref.~\cite{Ambrus:2023bid} in the context of a free scalar field, through detailed thermal field theory calculations, revealing the local dependence of, e.g., the energy-momentum tensor as a function of the distance $\rho$ to the rotation axis, and the subsequent collapse onto its asymptotic, fractal limit as $\rho \rightarrow \infty$, governed only by the denominator $q$ of the irreducible fraction $p / q$ representing the rotation parameter $\nu = \beta \Omega_I / 2\pi$ ($\beta$ is the inverse temperature).

In the present study, we make a step forward starting from Ref.~\cite{Ambrus:2023bid} and consider the properties of free Dirac fermions at finite chemical potential, undergoing rotation with an imaginary angular velocity. We start the discussion in Sec.~\ref{sec:RKT} with an analysis of the equivalent classical system, treated within relativistic kinetic theory (RKT), where no fractalization appears. In Sec.~\ref{sec:QFT}, we review the formalism of thermal field theory employed in this work for the construction of rigidly-rotating fermionic thermal states, focussing on the case of real rotation. We make the transition to imaginary rotation in Sec.~\ref{sec:OI}, pointing out the emergence of the fractal structure in finite-temperature observables in Sec.~\ref{sec:fractal}. We discuss the volume-averaged thermodynamic functions corresponding to the system under imaginary rotation in Sec.~\ref{sec:thermo} and we discuss the properties of the polarization fluxes through the transverse plane in Sec.~\ref{sec:flux}. Appendices~\ref{app:Li} and \ref{app:psi} present mathematical details related to the polylogarithm and polygamma functions, required for the analysis in the main text. Section~\ref{sec:conc} concludes this paper.

\section{Kinetic theory analysis} \label{sec:RKT}

In this section, we consider a relativistic kinetic theory analysis of rigidly-rotating distributions of free fermions in the $3+1$-D Minkowski space. We start with a brief overview of thermal equilibrium in relativistic kinetic theory (RKT) in Subsec.~\ref{sec:RKT:FD}. We discuss the properties of an equilibrium system under rigid rotation with a real angular frequency $\Omega \rightarrow \Omega_R \in \mathbb{R}$ in Subsect.~\ref{sec:RKT:real}. In Subsec.~\ref{sec:RKT:imaginary}, we analyze the properties of the same system under rotation with an imaginary angular frequency $\Omega \rightarrow i \Omega_I$ with $\Omega_I \in \mathbb{R}$. Finally, we discuss the moment of inertia and deformation coefficient derived in the limit of slow rotation in Sec.~\ref{sec:RKT:slow}.

\subsection{Rigidly-rotating Fermi-Dirac distribution} \label{sec:RKT:FD}

In classical (non-quantum) non-equilibrium statistical mechanics, a system of fermions can be described by the one-particle distribution function $f^\sigma_{\bk,\lambda} \equiv f^\sigma_{\bk,\lambda}(x)$, where $\sigma = \pm 1$ distinguishes between particles ($\sigma = 1$) and antiparticles ($\sigma = -1$), $\lambda = \pm 1/2$ represents the particle helicity, while $\bk$ represents the momentum three-vector. The dynamics of $f_\bk^{\sigma,\lambda}$ can be modeled via the  relativistic Boltzmann equation \cite{Groot.1980,Cercignani.2002,Rezzolla.2013}:
\begin{equation}
 k^\mu \partial_\mu f^\sigma_{\bk,\lambda} = C^\sigma_{\bk,\lambda} [f],\label{eq:RKT_boltz}
\end{equation}
where $k^\mu = (k^0, \bk)$ is the four-momentum of an on-shell particle with rest mass $M$, normalized according to $k^2 = k_0^2 - \bk^2 = M^2$, while $C^\sigma_{\bk,\lambda}[f]$ is the collision term.

In this paper, we consider global equilibrium states undergoing rigid rotation. Under thermodynamic equilibrium, the collision term vanishes and $f^\sigma_{\bk,\lambda}$ is given by the Fermi-Dirac distribution,
\begin{align}
 f^{{\rm FD};\sigma}_{\bk,\lambda} &= \left\{ \exp[\beta_\mu (x) k^\mu - \sigma \alpha] + 1\right\}^{-1} \nonumber\\
 &= \left\{\exp[\beta (k^0 - \Omega J^z) - \sigma \alpha] + 1\right\}^{-1},
 \label{eq:RKT_fFD}
\end{align}
where on the second line we replaced the inverse temperature four-vector $\beta^\mu (x) = u^\mu (x) / T(x)$ with the Killing vector corresponding to rigid rotation,
\begin{equation}
 \beta^\mu (x) \partial_\mu = \beta (\partial_t + \Omega \partial_\varphi),
 \label{eq:betamu}
\end{equation}
where $\beta \equiv \beta (0)$ is the temperature on the rotation axis and $J^z = \rho^2 k^\varphi = -k_\varphi$ represents the component of the angular momentum projected onto the axis of rotation.
The local temperature $T_\rho \equiv T(x)$ and four-velocity $u^\mu (x)$ satisfy
\begin{gather}
 T_\rho = \frac{\gamma}{\beta}, \quad
 u^\mu \partial_\mu = \gamma(\partial_t + \Omega \partial_\varphi),\nonumber\\
 \gamma = \frac{1}{\sqrt{1- \rho^2 \Omega^2}},
 \label{eq:TE_gamma}
\end{gather}
where $\gamma \equiv \gamma(\rho)$ represents the Lorentz factor of a rigidly-rotating observable.
Finally, in global equilibrium, $\alpha = \mu_\rho / T_\rho = \beta \mu$ is a constant, where $\mu \equiv \mu_V$ is the (vector) chemical potential describing particle imbalance, while its coefficient $\sigma$ takes the values $+1$ and $-1$ for particles and anti-particles, respectively. In what follows next, the position dependence of local quantities will be omitted.

The global equilibrium state described above can be characterized by the vector charge current $J^\mu_V$ and the energy-momentum tensor $T^{\mu\nu}$, obtained from the one-particle distribution via phase-space integration:
\begin{align}
 J^\mu_V &= \sum_{\sigma,\lambda} \sigma \int dK\, k^\mu f^{\sigma}_{\bk,\lambda}, \nonumber\\
 T^{\mu\nu} &= \sum_{\sigma,\lambda} \int dK\, k^\mu k^\nu f^{\sigma}_{\bk,\lambda},
 \label{eq:RKT_JT}
\end{align}
where $dK = g d^3k / [(2\pi)^3 k^0]$ is the Lorentz-invariant integration measure and $g$ is the degeneracy factor ($g = 1$ in this paper).
Since dissipative effects are absent, $J^\mu_V$ and $T^{\mu\nu}$ assume their perfect fluid form,
\begin{equation}
 J_V^{\mu}= Q u^\mu, \quad
 T^{\mu\nu} = \epsilon u^\mu u^\nu - P \Delta^{\mu\nu},
\end{equation}
where $\Delta^{\mu\nu} = g^{\mu\nu} - u^\mu u^\nu$. The vector charge density $Q \equiv Q_V$, as well as the entropy density $s$, can be derived from the hydrostatic pressure $P$ via
\begin{equation}
 Q = \left(\frac{\partial P}{\partial \mu_\rho}\right)_{T_\rho}, \quad
 s = \left(\frac{\partial P}{\partial T_\rho}\right)_{\mu_\rho},
\end{equation}
while the energy density $\epsilon$ can be obtained from the Euler relation,
\begin{equation}
 \epsilon = s T_\rho + \mu_\rho Q - P.
 \label{eq:Euler}
\end{equation}
The pressure itself can be computed via
\begin{subequations}\label{eq:RKT_macro}
\begin{equation}
 P = -\frac{1}{3} \sum_{\sigma,\lambda} \int dK\, (\Delta_{\alpha\beta} k^\alpha k^\beta) f^{\sigma}_{\bk,\lambda}.
 \label{eq:RKT_P}
\end{equation}
The charge and energy densities can be obtained by contracting $J_V^\mu$ and $T^{\mu\nu}$ with $u^\mu$,
\begin{align}
 \epsilon &= \sum_{\sigma,\lambda} \int dK\, (k \cdot u)^2 f^{\sigma}_{\bk,\lambda},\label{eq:RKT_eps}\\
 Q &= \sum_{\sigma,\lambda} \sigma \int dK\, (k \cdot u) f^{\sigma}_{\bk,\lambda}.
 \label{eq:RKT_Q}
\end{align}
\end{subequations}
In the case of massless fermions, $P$ evaluates to \cite{Ambrus:2019khr}
\begin{equation}
 P = -\frac{T_\rho^4}{\pi^2} \sum_{\sigma,\lambda} {\rm Li}_4(-e^{\sigma \mu_\rho / T_\rho}),
\end{equation}
where ${\rm Li}_s(z) = \sum_{n = 1}^\infty z^n / n^s$ is the polylogarithm.
Using Eq.~\eqref{eq:Li4_id_aux}, the pressure evaluates to
\begin{equation}
 P = \frac{7 \pi^2 T_\rho^4}{180} + \frac{T_\rho^2 \mu_\rho^2}{6} + \frac{\mu_\rho^4}{12\pi^2},
 \label{eq:RKT_m0_P}
\end{equation}
such that $Q = \partial P / \partial \mu_\rho$ and $s = \partial P / \partial T_\rho$ become
\begin{equation}
 Q = \frac{\mu_\rho T_\rho^2}{3} + \frac{\mu_\rho^3}{3\pi^2}, \quad
 s = \frac{7\pi^2 T_\rho^3}{45} + \frac{T_\rho \mu_\rho^2}{3}.
 \label{eq:RKT_m0_sQV}
\end{equation}
Taking into account that $\epsilon = 3P$ for massless fermions, it can be seen that the above quantities satisfy the Euler relation \eqref{eq:Euler}.
Due to their dependence on $T_\rho$ and $\mu_\rho$, it is clear that $P$, $Q$ and $s$ depend on the distance $\rho$ as
\begin{equation}
 P = P_0 \gamma^4, \qquad
 s = s_0 \gamma^3, \qquad
 Q = Q_0 \gamma^3,
 \label{eq:RKT_obs_re}
\end{equation}
where $P_0$, $s_0$ and $Q_{0}$ represent the pressure, entropy density and charge density on the rotation axis.
Moreover, since $\gamma(\rho)$ diverges as $\rho \rightarrow \Omega^{-1}$ (near the light cylinder), also the above quantities become singular in this limit.
%We are thus led to conclude that the rigidly-rotating state cannot achieve thermal equilibrium in unbounded Minkowksi space.
Anticipating the relation ${\rm FC} = T^\mu{}_\mu / M$, we also report the trace of the energy-momentum tensor in the small mass limit:
\begin{equation}
 \frac{\epsilon - 3P}{M^2} \to \frac{T_\rho^2}{6} + \frac{\mu_\rho^2}{2\pi^2}.
 \label{eq:RKT_FC}
\end{equation}

\subsection{Thermodynamics of a rotating system} \label{sec:RKT:real}

Let us now consider that the system is confined within a fictitious cylinder of radius $R$ and vertical extent $L_z$. The grand canonical potential of such a system is
\begin{equation}
 \Phi = -\frac{1}{\beta} \int d^3x \sum_{\sigma,\lambda} \int \frac{d^3k}{(2\pi)^3} \ln(1 + e^{-\beta \widetilde{\mathcal{E}}^\sigma_{\bk,\lambda}}),
 \label{eq:RKT_PhiG}
\end{equation}
where $\beta = 1/T(0)$ is the inverse temperature on the rotation axis and
$\widetilde{\mathcal{E}}^\sigma_{\bk,\lambda} = \mathcal{E}^\sigma_{\bk,\lambda} - \Omega J^z$ is the corotating effective energy, with $J^z = -k_\varphi = xk^y - yk^x$ the angular momentum, while
\begin{equation}
 \mathcal{E}^\sigma_{\bk,\lambda} = k^0 - \sigma \mu.
 \label{eq:RKT_effective_E}
\end{equation}
Since $k^{0}-\Omega J^z=k_\mu u^\mu/\gamma$, the integral in (\ref{eq:RKT_PhiG}) can be written as:
\begin{equation}
 \Phi = -\frac{1}{\beta} \int d^3x \sum_{\sigma,\lambda} \int dK\, k^0 \ln(1 + e^{-\frac{k_{\mu}u^{\mu}}{T_\rho} + \sigma \alpha}),
\end{equation}
where $dK = d^3k / [(2\pi)^3 k^0]$ is the Lorentz-invariant integration measure. Performing a Lorentz boost in the $dK$ integral to the local rest frame, where $u^\mu=(1,0,0,0)$, we arrive at
\begin{equation}
 \Phi = -\frac{1}{\beta} \int d^3x \sum_{\sigma,\lambda} \gamma \int \frac{d^3k}{(2\pi)^3} \ln(1 + e^{-\frac{k^0}{T_\rho} + \sigma \alpha})
\end{equation}
The momentum integral can be performed using spherical coordinates,
$d^3k= k^2 dk d\Omega_k$, leading to
\begin{equation}
 \Phi =- \int \frac{\gamma d^3x}{2\pi^2 \beta} \sum_{\sigma,\lambda} \int^{\infty}_{0} dk
 \, k^2 \ln(1 + e^{-\frac{k^0}{T_\rho} + \sigma \alpha}).
\end{equation}
Using the series expansion of the logarithm, $\ln(1 + x) = -\sum_{n = 1}^\infty (-x)^n / n$, and changing the integration variable to $z=k^0/M$, we arrive at
\begin{multline}
 \Phi = \frac{M^3}{2\pi^2\beta} \sum_{\sigma,\lambda} \sum_{n = 1}^\infty \frac{(-e^{\sigma \alpha})^n}{n} \int d^3x \, \gamma \\\times
\int^{\infty}_{1} dz\,z\, \sqrt{z^2-1} e^{-n M z / T_\rho}.
\end{multline}
It is tempting to solve the $z$ integral in terms of the modified Bessel function of the second kind $K_2(x)$,
\begin{equation}
 \int^{\infty}_{1} e^{-\frac{nMz}{T_\rho}} z \sqrt{z^2-1} \ dz = \frac{T_\rho}{nM} K_{2}\left(\frac{nM}{T_\rho}\right).
\end{equation}
However, it will prove convenient to first perform the spatial integration. Writing $d^3x = \rho d\rho d\varphi dz$, the $\varphi$ and $z$ integrals give a prefactor $2\pi L_z$. The $\rho$ integration can be performed by changing the integration variable to $\zeta = 1/ \gamma$, with $d\zeta = -\rho \Omega^2 \gamma d\rho$, leading to
\begin{multline}
  \Phi = \frac{M^2 V}{\pi^2 R^2 \beta^2 \Omega^2} \sum_{\sigma,\lambda}\sum_{n=1}^{\infty} \frac{(-e^{\sigma \alpha})^n}{n^2} \\\times
  \int_{1}^{\infty} dz \sqrt{z^2 - 1} \left(e^{-nMz / T_R} - e^{-n M \beta z}\right),
\end{multline}
where $V = \pi R^2 L_z$ is the system volume. Using the integral representation
\begin{equation}
 K_1(\alpha) = \alpha \int_1^\infty dz\, \sqrt{z^2 - 1} e^{-\alpha z},
\end{equation}
we arrive at
\begin{multline}
  \Phi = \frac{M V}{\pi^2 R^2 \beta^3 \Omega^2} \sum_{\sigma,\lambda}\sum_{n=1}^{\infty} \frac{(-e^{\sigma \alpha})^n}{n^3} \\\times
  \left[\gamma(R) K_1\left(\frac{n M \beta}{\gamma(R)}\right) - K_1(n M \beta)\right].
  \label{eq:RKT_Phi}
\end{multline}

Using the thermodynamic relation
\begin{equation}
 d\Phi = \beta^{-2} \mathcal{S} d\beta - \mathcal{P} dV - Q d\mu - \mathcal{M} d\Omega,
\end{equation}
the average entropy $\om{S} = \mathcal{S} / V$, charge $\om{Q}$, and angular momentum $\om{M}$ can be obtained from the average thermodynamic potential $\overline{\Phi} = \Phi / V$ as
\begin{align}\label{sqpm}
 \om{S} &= \beta^2 \frac{\partial \overline{\Phi}}{\partial \beta}, &
 \om{Q} &= -\frac{\partial \overline{\Phi}}{\partial \mu}, &
 \om{M} &= -\frac{\partial \overline{\Phi}}{\partial \Omega}.
\end{align}
Since in this setup, the transverse and vertical directions are not equivalent, the thermodynamic pressure $\mathcal{P}$ must be replaced by two different pressures via
\begin{equation}
 \mathcal{P} dV \rightarrow \mathcal{P}_\perp V \frac{2dR}{R} + \mathcal{P}_z V \frac{dL_z}{L_z},
\end{equation}
where we took into account that $V = \pi R^2 L_z$ for a cylinder of radius $R$ and height $L_z$. Then, we have
\begin{equation}
 \mathcal{P}_\perp = -\frac{R}{2V} \frac{\partial \Phi}{\partial R}, \quad
 \mathcal{P}_z = -\frac{L_z}{V} \frac{\partial \Phi}{\partial L_z}.
\end{equation}
The Euler relation \eqref{eq:Euler} becomes
\begin{equation}
 \om{E} = \overline{\Phi} + \beta^{-1} \om{S} + \Omega \om{M} + \mu \om{Q}.
 \label{eq:Euler_Phi}
\end{equation}
The derivatives with respect to $\beta$ and $R$ act also on the modified Bessel functions, in which case the following formula proves useful:
\begin{align}
 K'_n(z) &= -K_{n-1}(z) - \frac{n}{z} K_n(z)\nonumber\\
 &= -K_{n+1}(z) + \frac{n}{z} K_n(z),
\end{align}
however the exact relations remain uninsightful.

Focussing instead on the massless limit, where $\overline{\Phi}$ evaluates to
\begin{subequations}\label{eq:RKT_thermo}
\begin{align}
 \overline{\Phi} &= \frac{2 \gamma^2(R)}{\pi^2 \beta^4} \sum_{\sigma = \pm 1} {\rm Li}_4(-e^{\sigma \alpha})
 = -\gamma^2(R) P_0,
 \label{eq:RKT_thermo_Phi}
\end{align}
one can easily derive
\begin{align}
 \om{S} &= \gamma^2(R) s_0, &
 \om{Q} &= \gamma^2(R) Q_{0}, \nonumber\\
 \om{M} &= 2 \Omega R^2 \gamma^4(R) P_0, &
 \mathcal{P}_\perp &= \gamma^4(R) P_0,
\end{align}
while $\mathcal{P}_z = -\overline{\Phi} =
 \gamma^2(R) P_0$.
Finally, the Euler relation \eqref{eq:Euler_Phi} gives
\begin{equation}
 \om{E} = \gamma^2(R) [2\gamma^2(R) + 1] P_0.
 \label{eq:RKT_thermo_E}
\end{equation}
\end{subequations}
\subsection{Imaginary rotation} \label{sec:RKT:imaginary}

We now turn to the case of imaginary rotation. Setting $\Omega = i \Omega_I$ with real $\Omega_I$ in $f^{\sigma,\lambda}_{\rm FD;\bk}$ leads to a complex-valued distribution function. In order for such a distribution to describe a physical system, we must instead employ an average over clockwise and counterclockwise rotations,
\begin{align}
 f^{{\rm im};\sigma}_{\bk,\lambda} & = \frac{1}{2} \left[f^{{\rm FD};\sigma}_{\bk,\lambda}(i \Omega_I) + f^{{\rm FD};\sigma}_{\bk,\lambda}(-i \Omega_I)\right] \nonumber\\
 &= \frac{e^{\beta \mathcal{E}^\sigma_{\bk,\lambda}} \cos(\beta \Omega_I k_\varphi) - 1}
 {e^{2\beta \mathcal{E}^\sigma_{\bk,\lambda}} - 2 e^{\beta \mathcal{E}^\sigma_{\bk,\lambda}} \cos(\beta \Omega_I k_\varphi) + 1}.
 \label{eq:RKT_fim}
\end{align}
As discussed in Ref.~\cite{Ambrus:2023bid}, the state described by $f^{{\rm im};\sigma}_{\bk,\lambda}$ is not an equilibrium state, since the collision term $C^\sigma_{\bk,\lambda}[f^{{\rm im};\sigma}_{\bk,\lambda}] \neq 0$.

We now consider the consequences of imaginary rotation and note that when the distribution $f^{\rm FD;\sigma}_{\mathbf{k}, \lambda}$ is employed with $\Omega = i \Omega_I$, the charge current and energy momentum tensor retain their forms in Eq.~\eqref{eq:RKT_JT}, with $u^\mu \partial_\mu = \gamma_I (\partial_t + i \Omega_I \partial_\varphi)$ and $\gamma_I(\rho)$ being the imaginary-rotation equivalent of the Lorentz factor,
\begin{equation}
 \label{eq:RKT_im_gamma}
 \gamma_I = \frac{1}{\sqrt{1 + \rho^2 \Omega_I^2}}.
\end{equation}
The macroscopic quantities in Eq.~\eqref{eq:RKT_obs_re} become
\begin{equation}
 P = P_0 \gamma_I^4, \qquad
 s = s_0 \gamma_I^3, \qquad
 Q = Q_0 \gamma_I^3,
 \label{eq:RKT_obs_im}
\end{equation}
now decreasing to $0$ as the distance to the rotation axis is increased.

As remarked in Ref.~\cite{Ambrus:2023bid}, due to the continuous nature of the classical angular momentum, $k_\varphi$, the Euclidean version $\gamma_I$ of the Lorentz factor is incompatible with the quantum Tolman-Ehrenfest effect \cite{Tolman:1930ona,Tolman:1930zza}, by which \cite{Chernodub:2022veq}:
\begin{align}
 \gamma^{\mathrm{TE}}_I = \frac{1}{\sqrt{1 + \rho^2 \beta^{-2} [\beta \Omega_I]^2_{2\pi}}},
 \label{eq_TE_im_gamma}
\end{align}
with $[x]_{2\pi} = x + 2 \pi k \in (-\pi,\pi]$ for $k \in {\mathbb Z}$, as required under the periodicity of the imaginary rotation.

We now consider the large-volume limit of our system. For simplicity, we focus on the case of massless particles. The average energy inside a cylinder of radius $R$ is simply
\begin{align}
 \om{E} &= \frac{2}{R^2} \int_0^R d\rho\, \rho\, T^{tt}
 = \gamma_I^2(R) [2 \gamma_I^2(R) + 1] P_0,
\end{align}
which agrees with the expression in Eq.~\eqref{eq:RKT_thermo_E} under the substitution $\gamma(R) \rightarrow \gamma_I(R)$. The grand potential can be evaluated as shown in Eq.~\eqref{eq:RKT_PhiG} by using the statistics in Eq.~\eqref{eq:RKT_fim}, or analogously, by integrating the following relation:
\begin{equation}
 \om{E} = \left[\frac{\partial(\beta \overline{\Phi})}{\partial \beta}\right]_{\beta \mu, \beta \Omega_I}.
\end{equation}
Applying the same thermodynamic relations as described for the real rotation case in the previous Subsec.~gives expressions for quantities analogous to the system pressure, entropy and angular momentum $\om{M}_I = - \partial \overline{\Phi} / \partial \Omega_I$ as
\begin{align}
 \om{S} &= \gamma_I^2(R) s_0, &
 \om{Q} &= \gamma_I^2(R) Q_{0}, \nonumber\\
 \om{M}_I &= -2 \Omega_I R^2 \gamma_I^4(R) P_0, &
 \mathcal{P}_\perp &= \gamma_I^4(R) P_0,
 \label{eq:RKT_im_thermo}
\end{align}
while $\mathcal{P}_z = \gamma_I^2(R) P_0$.
The above quantities are compatible with an Euler-like relation,
\begin{equation}
 \om{E} = \overline{\Phi} + \beta^{-1} \om{S} + \Omega_I \om{M}_I + \mu \om{Q},
 \label{eq:RKT_im_Euler}
\end{equation}
formulated now for a system under imaginary rotation.

\subsection{Shape coefficients for slow rotation} \label{sec:RKT:slow}

Considering now the limit of slow rotation, we expand the grand potential in a series with respect to the velocity $v_R = \Omega R$ of a corotating particle on the cylinder surface \cite{Braguta:2023yjn}:
\begin{equation}
 \Phi(\Omega) = \Phi(0) \sum_{n = 0}^\infty \frac{v_R^{2n}}{(2n)!} \mathcal{K}_{2n},
 \label{eq:RKT_PhiG_exp}
\end{equation}
where $\mathcal{K}_{2n}$ are $\Omega$-independent dimensionless coefficients and we took into account that $\Phi(\Omega)$ is an even function of $\Omega$. By construction, we demand that $\mathcal{K}_0 = 1$. Comparing Eqs.~\eqref{eq:RKT_PhiG_exp} and \eqref{eq:RKT_Phi}, it can be seen that
\begin{align}
 \overline{\Phi}(0) &= \frac{M^2}{2\pi^2 \beta^2} \sum_{\sigma,\lambda} \sum_{n = 1}^\infty \frac{(-e^{\sigma \alpha})^n}{n^2} K_2(n M \beta),\nonumber\\
 \om{K}_2 &= \frac{M^3}{4\pi^2 \beta} \sum_{\sigma,\lambda} \sum_{n = 1}^\infty \frac{(-e^{\sigma \alpha})^n}{n} K_3(n M \beta),\nonumber\\
 \om{K}_4 &= \frac{M^4}{2\pi^2} \sum_{\sigma,\lambda} \sum_{n = 1}^\infty (-e^{\sigma \alpha})^n K_4(n M \beta),
 \label{eq:RKT_shape_m}
\end{align}
where $\overline{\Phi} = \Phi / V$ and $\om{K}_n = \overline{\Phi} \mathcal{K}_n$.
In the massless limit, when $\Phi$ is given by Eq.~\eqref{eq:RKT_thermo_Phi}, one may employ the expansion
\begin{equation}
 \gamma^2(R) = \sum_{n = 0}^\infty v_R^{2n},
\end{equation}
to show that
\begin{equation}
 \overline{\Phi}(0) = -P_0, \quad
 \mathcal{K}_{2n} = (2n)!.\label{eq:RKT_K2n_def}
\end{equation}

\section{Rotating fermions in unbounded space-time} \label{sec:QFT}

In this section, we review previous results concerning the rigidly-rotating thermal distribution of free Dirac particles in the unbounded Minkowski spacetime. We first review the Dirac equation and its mode solutions in cylindrical coordinates, in Subsec.~\ref{sec:QFT:modes}. In Subsec.~\ref{sec:Dirac:tevs}, we briefly outline the thermal field theory formalism necessary to construct thermal states. Finally, in Subsec.~\ref{sec:QFT:tevs} we summarize the formal expressions for the thermal expectation values of our observables, under the assumption of a real rotation parameter $\Omega$.

\subsection{Mode solutions}
\label{sec:QFT:modes}

We consider free Dirac fermions described by the Lagrangian density
\begin{equation}
 \mathcal{L} = \frac{i}{2} \bar{\psi} \overleftrightarrow{\slashed{\partial}} \psi - M \bar{\psi} \psi,
\end{equation}
where $f \overleftrightarrow{\partial_\mu} g = f \partial_\mu g - (\partial_\mu f) g$ is the bilateral derivative and the Feynman slash $\slashed{\partial} = \gamma^\mu \partial_\mu$ denotes contraction with the Dirac matrices $\gamma^\mu$. In this paper, we take the gamma matrices in the Dirac representation,
\begin{equation}
 \gamma^0 = \begin{pmatrix}
      1 & 0 \\ 0 & -1
 \end{pmatrix}, \quad
 \boldsymbol{\gamma} = \begin{pmatrix}
  0 & \boldsymbol{\sigma} \\
  -\boldsymbol{\sigma} & 0
 \end{pmatrix},
\end{equation}
with the Pauli matrices $\boldsymbol{\sigma} = \{\sigma^1, \sigma^2, \sigma^3\}$ given by
\begin{equation}
 \sigma^1 = \begin{pmatrix}
    0 & 1 \\ 1 & 0
 \end{pmatrix}, \quad
 \sigma^2 = \begin{pmatrix}
     0 & -i \\ i & 0
 \end{pmatrix}, \quad
 \sigma^3 = \begin{pmatrix}
     1 &0 \\ 0 & -1
 \end{pmatrix},
\end{equation}
and adopt the metric signature $(+, -, -, -)$.

The Dirac equation can be obtained by demanding the extremization of the Dirac action $S = \int d^4x\, \mathcal{L}$, via the Euler-Lagrange formalism
\begin{equation}
 (i\slashed{\partial} - M) \psi = 0.
\end{equation}

To facilitate the representation of rigidly-rotating states, we consider the rotation axis to be aligned with the $z$ axis in cylindrical coordinates $x^\mu = (t, \rho, \varphi, z)$, with respect to which the Minkowski metric takes the form $g_{\mu\nu} = \operatorname{diag} (1, -1, -\rho^2, -1)$. Thus, we construct mode solutions of the Dirac equation that are eigenfunctions of the Hamiltonian $H = i \partial_t = -i \gamma^0 \boldsymbol{\gamma} \cdot \boldsymbol{\nabla} + M \gamma^0$, vertical momentum $P^z = -i \partial_z$, angular momentum along the $z$ axis $J^z = -i \partial_\varphi + S^z$, and helicity operator $h = \mathbf{S} \cdot \mathbf{P} / P$, where the spin matrix $\boldsymbol{S}$ is given by
\begin{equation}
 \boldsymbol{S} = \frac{1}{2} \gamma^5 \gamma^0 \boldsymbol{\gamma} = \frac{1}{2}
 \begin{pmatrix}
  \boldsymbol{\sigma} & 0 \\
  0 & \boldsymbol{\sigma}
 \end{pmatrix}.
\end{equation}

Taking particle-like solutions $U_j$ with positive energy eigenvalue $E_j > 0$, we impose the following eigenvalue equations
\begin{align}
 H U_j &= E_j U_j, & P^z U_j &= k_j U_j, \nonumber\\
 J^z U_j &= m_j U_j, &
 h U_j &= \lambda_j U_j,
 \label{eq:U_eigen}
\end{align}
where the label $j = \{E_j, k_j, m_j, \lambda_j\}$ collectively denotes all eigenvalues corresponding to the mode $U_j$.

The mode solutions of the Dirac equation satisfying all of the above eigenvalue equations were derived in Ref.~\cite{Ambrus:2014uqa} (see also Ref.~\cite{Ambrus:2019cvr})
\begin{align}
 U_j &= \frac{e^{-i E_j t + i k_j z}}{2\pi} u_j, &
 u_j &= \frac{1}{\sqrt{2}} \begin{pmatrix}
  \mathfrak{E}_j^+ \phi_j \\
  2\lambda_j \mathfrak{E}_j^- \phi_j
 \end{pmatrix},\label{eq:U}
\end{align}
where $\mathfrak{E}_j^\pm = \sqrt{1 \pm M / E_j}$, while the Pauli two-spinor $\phi_j$ is given by
\begin{equation}
 \phi_j = \frac{1}{\sqrt{2}}
 \begin{pmatrix}
  \mathfrak{p}^+_j e^{i(m_j - \frac{1}{2}) \varphi} J_{m_j - \frac{1}{2}}(q_j \rho) \\
  2i \lambda_j \mathfrak{p}^-_j e^{i(m_j + \frac{1}{2}) \varphi} J_{m_j + \frac{1}{2}}(q_j \rho)
 \end{pmatrix}.
\end{equation}
In the above, $\mathfrak{p}^\pm_j = \sqrt{1 \pm 2\lambda_j k_j / p_j}$, where $p_j = \sqrt{E_j^2 - M^2}$ and $q_j = \sqrt{p_j^2 - k_j^2} = \sqrt{E_j^2 - k_j^2 - M^2}$ represent the magnitudes of the particle's total and transverse momenta. The modes $U_j$ satisfy the orthogonality relation
\begin{equation}
 \langle U_j, U_{j'} \rangle = \delta(j,j') \equiv \delta_{\lambda_j, \lambda_{j'}} \delta_{m_j, m_{j'}} \delta(k_j - k_{j'}) \frac{\delta(E_j - E_{j'})}{E_j},
 \label{eq:U_ortho}
\end{equation}
defined with respect to the Dirac inner product $\langle \psi, \chi \rangle = \int d^3x\, \psi^\dagger \chi$.

The anti-particle modes $V_j$ can be obtained via charge conjugation, $V_j = i \gamma^2 U_j^*$, and satisfy
\begin{align}
 H V_j &= -E_j V_j, & P^z V_j &= -k_j V_j, \nonumber\\
 J^z V_j &= -m_j V_j, &
 h V_j &= \lambda_j V_j.
 \label{eq:V_eigen}
\end{align}
Hence, $V_j$ with $j = (E_j, k_j, m_j, \lambda_j)$ can be related to the particle modes $U_{\bar{\jmath}}$ with $\bar{\jmath} = \{-E_j, -k_j, -m_j, \lambda_j\}$ via
\begin{equation}
 %V_j = i (-1)^{m_j - \frac{1}{2}} U_{\bar{\jmath}}.
 V_j = e^{i \pi m_j} U_{\bar{\jmath}},
\end{equation}
where $e^{i \pi m_j} = i (-1)^{m_j - \frac{1}{2}}$. Explicitly,
\begin{align}
 V_j &= \frac{e^{i E_j t - i k_j z}}{2\pi} v_j, &
 v_j &= \frac{1}{\sqrt{2}} \begin{pmatrix}
  2\lambda_j \mathfrak{E}_j^- \phi^c_j \\
  -\mathfrak{E}_j^- \phi^c_j
 \end{pmatrix},
 \label{eq:V}
\end{align}
where $\phi^c_j = i \sigma_2 \phi^*_j$ is related to the two-spinors $\phi_j$ via
\begin{equation}
 %\phi^c_j = (-1)^{m_j - \frac{1}{2}} 2i \lambda_j \phi_{\bar{\jmath}}.
 \phi^c_j = 2\lambda_j e^{i \pi m_j} \phi_{\bar{\jmath}}.
\end{equation}
The normalization \eqref{eq:U_ortho} is inherited also for the anti-particle modes,
\begin{equation}
 \langle V_j, V_{j'} \rangle = \delta(j, j'). \label{eq:V_ortho}
\end{equation}

\subsection{Thermal states}\label{sec:Dirac:tevs}

In this paper, we will consider three independent charge currents: the vector, axial and helical charge currents $J^\mu_{V/A/H}$, defined as:
\begin{gather}
 J^\mu_V = \bar{\psi} \gamma^\mu \psi, \quad
 J^\mu_A = \bar{\psi} \gamma^\mu \gamma^5 \psi, \nonumber\\
 J^\mu_H = \bar{\psi} \gamma^\mu h \psi + \overline{h \psi} \gamma^\mu \psi.
 \label{eq:CC_def}
\end{gather}
It can be shown that for free fermions, both $J^\mu_V$ and $J^\mu_H$ are conserved:
\begin{equation}
 \partial_\mu J^\mu_V = \partial_\mu J^\mu_H = 0,
\end{equation}
while the conservation of the axial current is explicitly broken by the fermion mass:
\begin{equation}
 \partial_\mu J^\mu_A = 2i M\bar{\psi} \gamma^5 \psi.
\end{equation}
It should be noted that, once interactions are switched on, both the helicity and the axial currents are no longer conserved: the first due to helicity-violating particle annihilation (HVPA) processes \cite{Ambrus:2019khr}; and the second due to triangle anomalies \cite{Adler:1969gk,Bell:1969ts}, however we do not exploit these features for the remainder of this paper. Nevertheless, the vector current remains conserved and the $U(1)_V$ symmetry is preserved.

We now consider the quantum field theory following after the second quantization of the field operator,
\begin{equation}
 \psi \rightarrow \hat{\psi} = \sum_j [U_j(x) \hat{b}_j + V_j(x) \hat{d}^\dagger_j],
 \label{eq:Fock_psi}
\end{equation}
where the mode sum $\sum_j$ reads
\begin{equation}\label{eq:sumj}
 \sum_j = \sum_{\lambda_j = \pm \frac{1}{2}} \sum_{m_j = -\infty}^\infty \int_M^\infty dE_j\, E_j\, \int_{-p_j}^{p_j} dk_j.
\end{equation}
In Eq.~\eqref{eq:Fock_psi}, we used a hat to denote Fock space operators. The one-particle operators $\hat{b}_j$ and $\hat{d}_j$ satisfy canonical anticommutation relations,
\begin{equation}
  \{\hat{b}_j, \hat{b}^\dagger_{j'}\} =
  \{\hat{d}_j, \hat{d}^\dagger_{j'}\} = \delta(j,j'),
\end{equation}
with all other anticommutators vanishing.

We now consider rigidly-rotating thermal states described by the density operator
\begin{equation}
 \hat{\rho} = \exp\left(-\beta :\widehat{H} - \mu \widehat{Q} - \Omega \widehat{J}^z:\right),
 \label{eq:rho}
\end{equation}
where the colons denote normal (Wick) ordering, i.e. $:\widehat{A}: \equiv \widehat{A} - \braket{0|\widehat{A}|0}$.
In the above, $\Omega$ represents the rotation parameter and $\mu$ denotes the chemical potential associated with the vector charge operator, denoted by $\widehat{Q}$.
The operators appearing under the exponential in Eq.~\eqref{eq:rho} admit the following Fock space representation with respect to the modes $U_j$ and $V_j$ defined in Eqs.~\eqref{eq:U_eigen} and \eqref{eq:V_eigen}:
\begin{align}
 :\widehat{Q}: &= \sum_j (\hat{b}^\dagger_j \hat{b}_j -
 \hat{d}^\dagger_j \hat{d}_j),\nonumber\\
 :\widehat{H}: &= \sum_j E_j (\hat{b}^\dagger_j \hat{b}_j +
 \hat{d}^\dagger_j \hat{d}_j),\nonumber\\
 :\widehat{J}^z: &= \sum_j m_j (\hat{b}^\dagger_j \hat{b}_j +
 \hat{d}^\dagger_j \hat{d}_j).
\end{align}

The operators $\widehat{H}$, $\widehat{J}^z$ and $\widehat{Q}$ satisfy the following commutation relations with the one-particle operators
\begin{align}
 [\widehat{Q}, \hat{b}^\dagger_j] &= \hat{b}^\dagger_j, &
 [\widehat{Q}, \hat{d}^\dagger_j] &= - \hat{d}^\dagger_j, \nonumber\\
 [\widehat{H}, \hat{b}^\dagger_j] &= E_j \hat{b}^\dagger_j, &
 [\widehat{H}, \hat{d}^\dagger_j] &= E_j \hat{d}^\dagger_j, \nonumber\\
 [\widehat{J}^z, \hat{b}^\dagger_j] &= m_j \hat{b}^\dagger_j, &
 [\widehat{J}^z, \hat{d}^\dagger_j] &= m_j \hat{d}^\dagger_j.
\end{align}
It is thus not difficult to establish the relations
\begin{align}
 \hat{\rho} \hat{b}^\dagger_j \hat{\rho}^{-1} &= e^{-\beta \widetilde{\mathcal{E}}_j^+} \hat{b}^\dagger_j, \nonumber\\
 \hat{\rho} \hat{d}^\dagger_j \hat{\rho}^{-1} &= e^{-\beta \widetilde{\mathcal{E}}_j^-} \hat{d}^\dagger_j,
\end{align}
where $\widetilde{\mathcal{E}}_j^\sigma = \mathcal{E}_j^\sigma - \Omega m_j$ and $\mathcal{E}^\sigma_j$ represent the quantum analogues of the effective energies introduced in Eq.~\eqref{eq:RKT_effective_E}, with
\begin{equation}
 \mathcal{E}^\sigma_j = E_j - \sigma \mu.
\end{equation}

\subsection{Thermal expectation values}\label{sec:QFT:tevs}

We are now in a position to evaluate thermal expectation values, defined as
\begin{equation}
 \langle\widehat{A} \rangle = \mathcal{Z}^{-1} {\rm Tr}(\hat{\rho} \widehat{A}),
\end{equation}
where $\mathcal{Z} = {\rm Tr}(\hat{\rho})$ is the partition function and the trace runs over Fock space. The building blocks for the expectation values that form the object of the present study involve the products of two one-particle operators:
\begin{equation}
 \langle \hat{b}^\dagger_j \hat{b}_{j'} \rangle = \frac{\delta(j,j')}{e^{\beta \widetilde{\mathcal{E}}^+_j} + 1}, \quad
 \langle \hat{d}^\dagger_j \hat{d}_{j'} \rangle = \frac{\delta(j,j')}{e^{\beta \widetilde{\mathcal{E}}^-_j} + 1}.
\end{equation}
Since all operators considered in this paper are bilinear in the field operator, their thermal expectation value can be summarized simply as
\begin{align}
 A &\equiv \langle :\widehat{A}: \rangle = \sum_j \left(\frac{\mathcal{A}(U_j, U_j)}{e^{\beta \widetilde{\mathcal{E}}^+_j} + 1} - \frac{\mathcal{A}(V_j, V_j)}{e^{\beta \widetilde{\mathcal{E}}^-_j} + 1}\right) \nonumber\\
 & = \sum_{j, \sigma} \mathcal{C}_{\sigma} (\mathcal{A}) \frac{\mathcal{A}(U_j, U_j)}{e^{\beta \widetilde{\mathcal{E}}^\sigma_j} + 1},
 \label{eq:A_sumj_gen}
\end{align}
where $\mathcal{A}(\psi, \chi)$ is the sesquilinear form associated with the operator $\widehat{A}$. On the second line, a summation over particle and anti-particle species with $\sigma = +1$ and $-1$, respectively, with energies $\mathcal{E}^{\sigma}_j$ has been introduced and the anti-particle sesquilinear forms $\mathcal{A}(V_j, V_j)$ have been related to those for particles through
\begin{equation}
 \mathcal{A}(V_j, V_j) = -\mathcal{A}(U_j, U_j) \mathcal{C}_{-}(\mathcal{A}),
\end{equation}
where we interpret $\mathcal{C}_{-} (\mathcal{A})$ as the parity under charge conjugation of the operator $\widehat{A}$, being either $+1$ or $-1$ for an even or an odd operator, respectively, whereas $\mathcal{C}_+(\mathcal{A}) = 1$ is taken by convention for any sesquilinear form $\mathcal{A}$.

In this work, we compute the thermal expectation values for the fermion condensate,
\begin{equation}
 \mathcal{FC} (\psi, \chi) = \bar{\psi} \chi,
\end{equation}
for the charge currents,
\begin{gather}
 \mathcal{J}^\mu_V(\psi, \chi) = \bar{\psi} \gamma^\mu \chi, \quad
 \mathcal{J}^\mu_A(\psi, \chi) = \bar{\psi} \gamma^\mu \gamma^5 \chi,\nonumber\\
 \mathcal{J}^\mu_H(\psi, \chi) = \bar{\psi} \gamma^\mu h \chi + \overline{h \psi} \gamma^\mu \chi,
\end{gather}
and for the energy-momentum tensor,
\begin{equation}
 \mathcal{T}^{\mu\nu} (\psi, \chi) = \frac{i}{2} \bar{\psi} \gamma^{(\mu} \overleftrightarrow{\partial^{\nu)}} \chi.
\end{equation}

Using the relation $V_j = i \gamma^2 U_j^*$, it can be shown that
\begin{gather}
 \overline{V}_j V_j = -(\overline{U}_j U_j)^*, \quad
 \mathcal{J}^\mu_A(V_j, V_j) = -[\mathcal{J}^\mu_A(U_j, U_j)]^*, \nonumber\\
 \mathcal{T}^{\mu\nu}(V_j, V_j) = -[\mathcal{T}^{\mu\nu}(U_j, U_j)]^*.
\end{gather}
Since the quantities appearing on the right-hand sides of the above relations are real numbers (see below), it is not difficult to see that
\begin{equation}
\label{eq: charge_conjugation_equal_1}
\mathcal{C}_\sigma (\mathcal{F C}) = \mathcal{C}_\sigma (\mathcal{J}^\mu_A) = \mathcal{C}_\sigma(\mathcal{T}^{\mu\nu}) = 1.
\end{equation}
Similarly,
\begin{align}
 \mathcal{J}^\mu_V(V_j, V_j) &= [\mathcal{J}^\mu_V(U_j, U_j)]^*, \nonumber\\
 \mathcal{J}^\mu_H(V_j, V_j) &= [\mathcal{J}^\mu_H(U_j, U_j)]^*
\end{align}
and
\begin{equation}
\label{eq: charge_conjugation_equal_sigma}
\mathcal{C}_\sigma (\mathcal{J}^\mu_V) = \mathcal{C}_\sigma (\mathcal{J}^\mu_H) = \sigma.
\end{equation}

We now summarize the thermal expectation values of the operators mentioned above, as computed in Ref.~\cite{Ambrus:2019ayb}. In what follows, we employ the notation
\begin{subequations}\label{eq:Jpmx}
\begin{align}
 J_{m_j}^\pm(q_j\rho) &= J_{m_j -\frac{1}{2}}^2(q_j \rho) \pm J_{m_j + \frac{1}{2}}^2(q_j \rho), \label{eq:Jpm}\\
 J_{m_j}^\times(q_j \rho) &= 2 J_{m_j-\frac{1}{2}}(q_j \rho) J_{m_j + \frac{1}{2}}(q_j \rho).
 \label{eq:Jx}
\end{align}
\end{subequations}

In the case of the fermion condensate, we have
\begin{equation}
 \mathcal{FC}(U_j, U_j) = \frac{M}{8\pi^2 E_j} \left[J^+_{m_j}(q_j \rho) + \frac{2\lambda_j k_j}{p_j} J^-_{m_j}(q_j \rho)\right].
 \label{eq:FC_sesqi}
\end{equation}
This allows the fermion condensate to be written as
\begin{equation}\label{eq:fc}
 {\rm FC} = \frac{M}{8 \pi^2} \sum_{j, \sigma} \frac{1}{E_j(e^{\beta\widetilde{\mathcal{E}}^\sigma_j} + 1)} J_{m_j}^+(q_j\rho),
\end{equation}
where we used $\mathcal{C}_-(\mathcal{FC}) = 1$. The second term in Eq.~\eqref{eq:FC_sesqi} is proportional to $\lambda_j$ and $k_j$, being odd under both $\lambda_j \to -\lambda_j$ and $k_j \to -k_j$. Therefore, this term vanishes under summation with respect to $\lambda_j$ and/or integration with respect to $k_j$.

The sesquilinear forms for the vector and helical charge currents are linked to each other, since $h U_j = \lambda_j U_j$. This means that $\mathcal{J}^\mu_H(U_j, U_j) = 2\lambda_j \mathcal{J}^\mu_V(U_j, U_j)$, with
\begin{align}
 \mathcal{J}^t_V(U_j, U_j) &= \frac{1}{8\pi^2}\left[J_{m_j}^+(q_j\rho) +
 \frac{2\lambda_j k_j}{p_j} J_{m_j}^-(q_j\rho)\right],\nonumber\\
 \mathcal{J}^z_V(U_j, U_j) &=
 \frac{1}{8\pi^2}\left[\frac{k_j}{E_j} J_{m_j}^+(q_j\rho) +
 \frac{2\lambda_j p_j}{E_j} J_{m_j}^-(q_j\rho)\right],\nonumber\\
 \mathcal{J}^\varphi_V(U_j, U_j) &=
 \frac{q_j}{8\pi^2 \rho E_j} J_{m_j}^\times(q_j \rho),
 \label{eq:sesq_CC}
\end{align}
while $\mathcal{J}^\rho_*(U_j, U_j) = 0$, for $* \in \{V,A,H\}$.  Since the vector and helical currents are $C$-odd, we have
\begin{equation}
 \begin{pmatrix}
  J^t_V \\
  J^\varphi_V \\
  J^z_H
 \end{pmatrix} = \frac{1}{8\pi^2} \sum_{j, \sigma}
 \frac{\sigma / E_j}{e^{\beta \widetilde{\mathcal{E}}^\sigma_j} + 1}
 \begin{pmatrix}
  E_j J_{m_j}^+(q_j \rho) \\
  \rho^{-1} q_j J_{m_j}^\times(q_j\rho) \\
  p_j J_{m_j}^-(q_j \rho)
 \end{pmatrix},
 \label{currents}
\end{equation}
while the other components vanish, $J^z_V = J^t_H = J^\varphi_H = 0$.

Similarly, the sesquilinear forms for the axial current are
\begin{align}
 \mathcal{J}^t_A(U_j, U_j) =& \frac{p_j}{8\pi^2 E_j} \left[2 \lambda_j J_{m_j}^+(q_j \rho) + \frac{k_j}{p_j} J_{m_j}^-(q_j \rho) \right], \nonumber\\
 \mathcal{J}^\varphi_A(U_j, U_j) =& \frac{\lambda_j q_j}{4\pi^2 \rho p_j} J_{m_j}^\times(q_j \rho), \nonumber\\
 \mathcal{J}^z_A(U_j, U_j) =& \frac{1}{8\pi^2} \left[J_{m_j}^-(q_j \rho) + \frac{2\lambda_j k_j}{p_j} J_{m_j}^+(q_j \rho) \right],
 \label{eq:sesq_JA}
\end{align}
with $\mathcal{J}^\mu_A(V_j, V_j) = -\mathcal{J}^\mu_A(U_j, U_j)$, as required for a $C$-even operator. Due to the antisymmetry of $\mathcal{J}^t_A(U_j,U_j)$ and $\mathcal{J}^\varphi_A(U_j, U_j)$ under $\lambda_j \rightarrow -\lambda_j$ and $k_j \rightarrow -k_j$, only the $z$ component is non-vanishing:
\begin{equation}
 J^z_A = \frac{1}{8\pi^2} \sum_{j, \sigma}
 \frac{1}{e^{\beta \widetilde{\mathcal{E}}^\sigma_j} + 1} J_{m_j}^-(q_j \rho).
 \label{eq:JAz}
\end{equation}

The sesqulinear forms for the components of the energy-momentum tensor can be found in Eq.~(9.7) of Ref.~\cite{Ambrus:2019ayb}. Without repeating these lengthy expressions, we give below the relevant expressions for the thermal expectation values
\begin{align}\label{eq:emtensor}
 \begin{pmatrix}
     T^{tt} \\ T^{\varphi\varphi} \\ T^{zz}
 \end{pmatrix} &= \frac{1}{8\pi^2} \sum_{j,\sigma} \frac{E_j^{-1}}{e^{\beta \widetilde{\mathcal{E}}^\sigma_j} + 1}
 \begin{pmatrix}
  E_j^2 J_{m_j}^+(q_j \rho) \\
  \rho^{-3} m_j q_j J_{m_j}^\times(q_j \rho) \\
  k_j^2 J_{m_j}^+(q_j \rho)
 \end{pmatrix}, \nonumber\\
 T^{\rho\rho} &= \frac{1}{8\pi^2} \sum_{j,\sigma} \frac{q_j^2 / E_j}{e^{\beta \widetilde{\mathcal{E}}^\sigma_j} + 1}
 \left[
  J_{m_j}^+(q_j \rho) - \frac{m_j}{q_j \rho} J_{m_j}^\times(q_j \rho) \right],\nonumber\\
 T^{t\varphi} &= \frac{1}{16\pi^2 \rho^2} \sum_{j,\sigma} \frac{1}{e^{\beta\widetilde{\mathcal{E}}^\sigma_j} + 1} \Big[m_j J_{m_j}^+(q_j\rho) \nonumber\\
 & - \frac{1}{2} J^-_{m_j}(q_j \rho) + q_j \rho J^{\times}_{m_j}(q_j \rho)\Big],
\end{align}
which make no reference to the type of rotation yet.
Converting in \eqref{eq:emtensor} the $J_{m_j}^\times$ term into $J_{m_j}^+$ via the identity \cite{Ambrus:2019ayb}
\begin{align}
    \label{eq:conversion_integral}
   \frac{1}{2\rho} \int^{p_j}_0 \,dk\, q\, J_{m_j}^\times(q_j \rho)  &= \frac{1}{2 m_j} \int^{p_j}_{0} dk_j (q_j^2 - k_j^2) J_{m_j}^{+} (q_j \rho) \nonumber\\
   &= \int^{p_j}_0 dk_j\,k_j^2 J_{m_j}^{-}(q_j \rho),
\end{align}
shows that $T^{\rho\rho} = T^{z z}$. Also, $J^\varphi_V$, $T^{\varphi\varphi}$ and $T^{t\varphi}$ can be reexpressed as
\begin{subequations}
\begin{align}
 J^\varphi_V &= \frac{1}{4\pi^2} \sum_{j,\sigma} \frac{\sigma k_j^2 J_{m_j}^-(q_j \rho)}{E_j(e^{\beta \widetilde{\mathcal{E}}^\sigma_j} + 1)}, \\
 T^{\varphi\varphi} &= \frac{1}{4\pi^2 \rho^2} \sum_{j,\sigma} \frac{m_j k_j^2 J_{m_j}^-(q_j \rho)}{E_j(e^{\beta \widetilde{\mathcal{E}}^\sigma_j} + 1)},\\
 T^{t\varphi} &= \frac{1}{16\pi^2 \rho^2} \sum_{j,\sigma} \frac{1}{e^{\beta\widetilde{\mathcal{E}}^\sigma_j} + 1} \Big[m_j J_{m_j}^+(q_j\rho) \nonumber\\
 & - \frac{1 - 4\rho^2 k_j^2}{2} J^-_{m_j}(q_j \rho)\Big].
\end{align}
\end{subequations}

While, in this section, we considered fermions of arbitrary mass, for the remainder of this paper we will focus only on the massless case, when $M = 0$ and $E_j = p_j$. Although in this limit, ${\rm FC} = 0$, the ratio ${\rm FC} / M$ will have a finite value as $M \to 0$ and in what follows, we will consider this ratio instead.

\subsection{Vanishing rotation: shape coefficients}\label{sec:QFT:shape}

In the limit of vanishing rotation, the Fermi-Dirac factor in Eqs.~\eqref{currents}, \eqref{eq:JAz} and \eqref{eq:emtensor} becomes independent of $m_j$. Using the summation formulas \cite{Ambrus:2014uqa,Ambrus:2019ayb}:
\begin{gather}
 \sum_{m = -\infty}^\infty J_m^+(z) = 2, \quad
 \sum_{m = -\infty}^\infty m J^\times_{m}(z) = z, \nonumber\\
 \sum_{m = -\infty}^\infty m J^+_m(z) =
 \sum_{m = -\infty}^\infty J^-_m(z) =
 \sum_{m = -\infty}^\infty J^\times_m(z) = 0,
 \label{eq:sumJ0}
\end{gather}
it is not too difficult to see that
\begin{equation}
 T^{\mu\nu} = {\rm diag}(\epsilon, P, P, P), \quad
 J_V^\mu = Q \delta^\mu_0,
\end{equation}
with $\epsilon$, $P$ and $Q$ given in Eqs.~\eqref{eq:RKT_macro},
while $J^\mu_A = J^\mu_H = 0$.
For clarity, we illustrate the procedure for $T^{tt}$. Replacing the formal summation over $j$ using Eq.~\eqref{eq:sumj}, the sum over $m_j$ can be performed using Eq.~\eqref{eq:sumJ0}, leading to
\begin{align}
 T^{tt}_{\Omega = 0} &= \frac{1}{4\pi^2} \sum_{\sigma,\lambda} \int_M^\infty \frac{dE\, E^2}{e^{\beta(E - \sigma \mu)} + 1} \int_{-p}^p dk \nonumber\\
 &= \sum_{\sigma,\lambda} \int \frac{dP\, E^2}{e^{\beta (E - \sigma \mu)} + 1},
 \label{eq:Ttt0}
\end{align}
where in the last step we used $dE\, p \to 2\pi^2 dP$, with $dP = d^3p / [(2\pi)^3 E]$. It is easy to check that Eq.~\eqref{eq:Ttt0} coincides with the relativistic kinetic theory result in Eq.~\eqref{eq:RKT_eps}.

To evaluate the shape coefficients $\mathcal{K}_n$, we first note that the thermodynamic potential can be related to the energy of the system via \cite{Ambrus:2023bid}
\begin{equation}
 \mathcal{E} = \left[\frac{\partial(\beta \Phi)}{\partial \beta}\right]_{\beta \mu, \beta \Omega}.
\end{equation}
Taking $\mathcal{E} = \int_V d^3x\, T^{tt}$ as the energy in the volume $V$, we can introduce the grand potential density via $\Phi = \int_V d^3x\, \phi(x)$, with
\begin{align}
 \phi(x) &= \frac{1}{\beta} \int d\beta\, (T^{tt})_{\beta\mu,\beta\Omega} \nonumber\\
 &= -\frac{1}{8\pi^2 \beta} \sum_{j,\sigma} \ln(1 + e^{-\beta \widetilde{\mathcal{E}}_j}) J_m^+(q_j\rho).
\end{align}
At vanishing rotation, the sum over $m_j$ can be performed using Eq.~\eqref{eq:sumJ0}, leading to the kinetic theory result in Eq.~\eqref{eq:RKT_PhiG}. Considering now that $\phi(x)$ depends on $\Omega$ only through the Fermi-Dirac factor, we can write
\begin{multline}
 \left.\frac{\partial^n \phi}{\partial\Omega^n}\right\rvert_{\Omega = 0} = -\frac{(-1)^n}{8\pi^2 \beta} \sum_{j,\sigma} \left[\frac{d^n}{dE_j^n} \ln(1+ e^{-\beta \mathcal{E}^\sigma_j})\right] \\\times
 m_j^n J_{m_j}^+(q_j\rho).
\end{multline}
Using the relations \cite{Ambrus:2014uqa,Ambrus:2019ayb}
\begin{align}
 \sum_{m_j = -\infty}^\infty m_j^2 J_{m_j}^+(q_j\rho) &= \frac{1}{2} + q_j^2 \rho^2, \nonumber\\
 \sum_{m_j = -\infty}^\infty m_j^4 J_{m_j}^+(q_j\rho) &= \frac{1}{8} + \frac{5}{2} q_j^2 \rho^2 + \frac{3}{4} q_j^4 \rho^4,
\end{align}
and noting that the sums involving odd powers of $m_j$ vanish, i.e. $\sum_{m_j = -\infty}^\infty m_j^{2n+1} J_{m_j}^+(q_j\rho) = 0$, we arrive at
\begin{align}
 \left.\phi(\rho)\right\rvert_{\Omega = 0} &= -\frac{1}{\pi^2 \beta} \int_M^\infty dE\, E p \sum_{\sigma = \pm 1} \ln(1 + e^{-\beta(E - \sigma \mu)}), \nonumber\\
 \left.\frac{\partial^2 \phi(\rho)}{\partial \Omega^2}\right\rvert_{\Omega = 0} &= -\frac{1}{2\pi^2 \beta} \int_M^\infty dE\, E p\left(\frac{1}{2} + \frac{2}{3} p^2 \rho^2\right)\nonumber\\
 & \times \sum_{\sigma = \pm 1} \frac{d^2}{dE^2} \ln(1 + e^{-\beta(E - \sigma \mu)}), \nonumber\\
 \left.\frac{\partial^4 \phi(\rho)}{\partial \Omega^4}\right\rvert_{\Omega = 0} &= -\frac{1}{4\pi^2 \beta} \int_M^\infty dE\, E p\left(\frac{1}{4} + \frac{10}{3} p^2 \rho^2 + \frac{4}{5} p^4 \rho^4\right)\nonumber\\
 & \times \sum_{\sigma = \pm 1} \frac{d^4}{dE^4} \ln(1 + e^{-\beta(E - \sigma \mu)}).
\end{align}
Averaging the above over a cylinder of height $L_z$ and radius $R$, we obtain
\begin{align}
 \overline{\Phi}(0) &= -\frac{1}{\pi^2 \beta} \int_M^\infty dE\, E p \sum_{\sigma = \pm 1} \ln(1 + e^{-\beta(E - \sigma \mu)}), \nonumber\\
 \om{K}_2 &= -\frac{1}{2\pi^2 \beta} \int_M^\infty dE\, E p\left(\frac{1}{2R^2} + \frac{1}{3} p^2\right)\nonumber\\
 & \times \sum_{\sigma = \pm 1} \frac{d^2}{dE^2} \ln(1 + e^{-\beta(E - \sigma \mu)}), \nonumber\\
 \om{K}_4 &= -\frac{1}{4\pi^2 \beta} \int_M^\infty dE\, E p\left(\frac{1}{4R^4} + \frac{5}{3R^2} p^2 + \frac{4}{15} p^4\right)\nonumber\\
 & \times \sum_{\sigma = \pm 1} \frac{d^4}{dE^4} \ln(1 + e^{-\beta(E - \sigma \mu)}),
\end{align}
with $\overline{\Phi}(0) \equiv \left.\overline{\Phi}\right\rvert_{\Omega = 0}$, as in Sec.~\ref{sec:RKT:slow}.
It can be seen that the shape coefficients $\om{K}_n = \overline{\Phi}(0) \mathcal{K}_n = R^{-n} (d^n \Phi / d\Omega^n)_{\Omega = 0}$ receive corrections in inverse powers of the system radius when compared to the classical, relativistic kinetic theory predictions in Eqs.~\eqref{eq:RKT_shape_m}. In the massless limit, when $M = 0$ and $E = p$, the energy integrals can be performed in closed form:
\begin{align}
 &\frac{1}{\beta} \sum_{\sigma = \pm 1} \int_0^\infty dE\, \frac{d^2}{dE^2} \ln(1 + e^{-\beta(E - \sigma \mu)}) = 1, \nonumber\\
 &\frac{1}{\beta} \sum_{\sigma = \pm 1} \int_0^\infty dE\, \ln(1 + e^{-\beta(E - \sigma \mu)}) = \frac{\pi^2}{6\beta^2} + \frac{\mu^2}{2}, \nonumber\\
 &\frac{1}{\beta} \sum_{\sigma = \pm 1} \int_0^\infty dE\, E^2 \ln(1 + e^{-\beta(E - \sigma \mu)}) \nonumber\\
 & \hspace{.4\linewidth} = \frac{7\pi^4}{180\beta^4} + \frac{\pi^2 \mu^2}{6\beta^2} + \frac{\mu^4}{12},
\end{align}
leading to
\begin{align}
 \overline{\Phi}(0) &= -\frac{7\pi^2}{180\beta^4} - \frac{\mu^2}{6\beta^2} - \frac{\mu^4}{12\pi^2}, \nonumber\\
 \om{K}_2 &= 2! \times \overline{\Phi}(0) - \frac{1}{2R^2} \left(\frac{1}{6\beta^2} + \frac{\mu^2}{2\pi^2}\right),\nonumber\\
 \om{K}_4 &= 4! \times \overline{\Phi}(0) - \frac{10}{R^2}\left(\frac{1}{6\beta^2} + \frac{\mu^2}{2\pi^2}\right) - \frac{1}{8\pi^2 R^4}.
\end{align}
At large distances from the rotation axis, $\mathcal{K}_n = \om{K}_n / \overline{\Phi}(0)$ reproduces the classical result in Eq.~\eqref{eq:RKT_K2n_def}: for any non-negative integer $n$, $\mathcal{K}_{2n} \to (2n)!$ as $R \to \infty$, while $\mathcal{K}_{2n+1} = 0$ for any value of $R$.

\section{Imaginary rotation: general considerations}\label{sec:OI}

We now consider the properties of a fermionic system undergoing imaginary rotation, $\Omega = i\Omega_I$, with $\Omega_I$ a real number. We will often refer to a generic observable $A$. In this section, we give the detailed steps required to evaluate the thermal expectation values introduced in the previous section. The key step is the expansion of the Fermi-Dirac factor, made possible when $\Omega = i \Omega_I$, as shown in Subsec.~\ref{sec:OI:FD}. Here, we identify immediately a contribution depending on the chemical potential and independent of both temperature and angular velocity, which we call degenerate and plays a key role at large distances from the rotation axis. These degenerate contributions, denoted $A_{\rm deg.}$, are evaluated explicitly for the observables of interest in Subsec.~\ref{sec:OI:deg}. For the remaining non-degenerate terms, denoted $\Delta A$, we perform the summation over the angular momentum quantum number $m_j$ in Subsec.~\ref{sec:OI:summ}. The integrations over the longitudinal momentum $k_j$ and momentum magnitude $p_j$ are performed in Subsections~\ref{sec:OI:kint} and \ref{sec:OI:pint}, respectively.

\subsection{Expansion of the Fermi-Dirac factor} \label{sec:OI:FD}

The Fermi-Dirac function $1 / (e^x + 1)$ can be expanded as a series of exponentials. If $|e^x| > 1$, then we use the formula
\begin{equation}
 \frac{1}{e^x + 1} = \frac{e^{-x}}{1 + e^{-x}} = \sum_{v = 1}^\infty (-1)^{v+1} e^{-v x}, \quad {\rm Re}(x) > 0.
\end{equation}
If $|e^x| < 1$, then we use
\begin{equation}
 \frac{1}{e^x + 1} = \sum_{v = 0}^\infty (-1)^v e^{vx}, \quad
 {\rm Re}(x) < 0.
\end{equation}
These two cases can be represented in a single expression using the Heaviside step function,
\begin{multline}
 \frac{1}{e^x + 1} = \theta[{\rm Re}(x)] \sum_{v = 1}^\infty (-1)^{v+1} e^{-v x} \\
 - \theta[-{\rm Re}(x)] \sum_{v = 0}^\infty (-1)^{v+1} e^{vx}.
\end{multline}

Applying the above discussion to the Fermi-Dirac factor $1 / (e^{\beta \widetilde{\mathcal{E}}^\sigma_j + 1})$, with $\widetilde{\mathcal{E}}^{\sigma}_j = \mathcal{E}^{\sigma}_j - i \Omega_I m_j$,
we note that ${\rm Re}(x)$ evaluates in this case to $\beta \mathcal{E}^{\sigma}_j$. Since $\beta > 0$, the step function can take as its argument just the effective energy, as follows:
\begin{multline}
\label{eq:theta_expansion}
 \frac{1}{\exp (\beta \widetilde{\mathcal{E}}^{\sigma}_j)+1}
 = \theta (\mathcal{E}^{\sigma}_j) \sum_{v=1}^{\infty} (-1)^{v+1} e^{- v \beta \mathcal{E}^{\sigma}_j} e^{i v \beta \Omega_I m_j} \\
 - \theta (- \mathcal{E}^{\sigma}_j) \sum_{v=0}^{\infty}(-1)^{v+1} e^{v \beta \mathcal{E}^{\sigma}_j} e^{- i v \beta \Omega_I m_j},
\end{multline}
where we considered that $\mathcal{E}^{\sigma}_j = E_j - \sigma \mu$ can be both positive or negative for energies above or below the Fermi level.
The $v = 0$ component of the second term appearing above is independent of temperature and rotation, being determined only by the chemical potential. Starting from Eq.~\eqref{eq:A_sumj_gen}, it is convenient to isolate this ``degenerate'' term as follows:
\begin{subequations}\label{eq:decomposition}
\begin{equation}\label{eq:decomposition_gen}
A = A_{\rm deg.} + \Delta A,
\end{equation}
where
\begin{align}
 A_{\rm deg.} &= \sum_{j, \sigma} \theta(-\mathcal{E}^\sigma_j) \mathcal{C}_{\sigma}(\mathcal{A}) \mathcal{A}_j, \label{eq:decomposition_deg}\\
 \Delta A &= \sum_{v = 1}^\infty (-1)^{v+1}
 \sum_{j, \sigma} e^{-v \beta |\mathcal{E}^\sigma_j|} \mathcal{C}_{\sigma}(\mathcal{A}) \mathcal{A}_j \nonumber\\
 &\times \left[\theta(\mathcal{E}^{\sigma}_j) e^{i v \beta \Omega_I m_j} - \theta(-\mathcal{E}^{\sigma}_j) e^{-iv \beta \Omega_I m_j}\right].
\label{eq:decomposition_DA}
\end{align}
\end{subequations}
We see that the second term inside the square brackets can be obtained from the first by performing the inversion $m_j \rightarrow -m_j$. The behavior of $\mathcal{A}_j = \mathcal{A}(U_j, U_j)$ under such an inversion can be characterized generically by splitting $\mathcal{A}_j = \mathcal{A}^j_{\rm even} + \mathcal{A}^j_{\rm odd}$, with $\mathcal{A}^j_{{\rm even}/{\rm odd}}\rfloor_{m_j \rightarrow -m_j} = \pm \mathcal{A}^j_{{\rm even}/{\rm odd}}$ representing the even and odd components of $\mathcal{A}_j$. Then, Eq.~\eqref{eq:decomposition_DA} becomes
\begin{multline}
 \Delta A = \sum_{v = 1}^\infty (-1)^{v+1}
 \sum_{j, \sigma} e^{-v \beta |\mathcal{E}^\sigma_j|} \mathcal{C}_{\sigma}(\mathcal{A}) e^{i v \beta \Omega_I m_j} \\
 \times \left[{\rm sgn}(\mathcal{E}^{\sigma}_j) \mathcal{A}^j_{\rm even} + \mathcal{A}^j_{\rm odd}\right].
 \label{eq:DA_master}
\end{multline}

\subsection{Degenerate contributions} \label{sec:OI:deg}

In the degenerate case, the sum over particle and anti-particle states can be computed as follows:
\begin{equation}
 \sum_{\sigma = \pm 1} \theta(-\mathcal{E}^\sigma_j) \mathcal{C}_\sigma(\mathcal{A}) = [\theta(\mu) + \theta(-\mu) \mathcal{C}_-(\mathcal{A})] \theta(|\mu| - E_j),
\end{equation}
such that
\begin{multline}
 A_{\rm deg.} = [\theta(\mu) + \theta(-\mu) \mathcal{C}_-(\mathcal{A})] \int_0^{|\mu|} dE_j E_j \int_{-p_j}^{p_j} dk_j \\
 \times \sum_{m_j = -\infty}^\infty \sum_{\lambda_j = \pm 1/2} \mathcal{A}(U_j, U_j),
\end{multline}
where we have set the lower limit for the energy integration to $M = 0$, as we are considering only the massless case.
The summation with respect to $m_j$ can be peformed using the identities \cite{DLMF,Ambrus:2014uqa,Ambrus:2019ayb}
\begin{subequations}
\begin{align}
 \sum_{m_j=-\infty}^{\infty} J_{m_j}^{+}\left(q_j \rho\right) &= 2, &
 \hspace{-8pt} \sum_{m_j=-\infty}^{\infty} m_j J_{m_j}^{+}\left(q_j \rho\right) &= 0,\\
 \sum_{m_j=-\infty}^{\infty} J_{m_j}^{-}\left(q_j \rho\right) &= 0, &
 \hspace{-8pt} \sum_{m_j = -\infty}^{\infty} m_j J^-_{m_j} (q_j \rho) &= 1.
\label{eq: symmetric_Bessel_sums}
\end{align}
\end{subequations}
The summation over $\lambda_j$ leads to the cancellation of the contributions which are odd with respect to $\lambda_j$, while for the ones that are even, it gives an overall factor of $2$, keeping in mind that $\lambda_j^2 = 1/4$. For example, in the case of the fermion condensate, we have
\begin{equation}
 \frac{1}{M}\sum_{m_j = -\infty}^\infty \sum_{\lambda_j = \pm 1/2} \overline{U}_j U_j = \frac{1}{2\pi^2 E_j}.
\end{equation}
After performing the $m_j$ and $\lambda_j$ summations, the integrations over both $E_j$ and $k_j$ can be performed trivially, as the integrands involve only polynomial expressions in both these variables. Using the sesquilinear forms in Eqs.~\eqref{eq:FC_sesqi}, \eqref{eq:sesq_CC} and \eqref{eq:sesq_JA}, as well as those implied by Eq.~\eqref{eq:emtensor}, we obtain
\begin{gather}
 \frac{\rm FC_{\rm deg.}}{M} \to \frac{\mu^2}{2\pi^2}, \quad J^t_{V;{\rm deg.}} = \frac{\mu^3}{3\pi^2},  \nonumber\\
 \frac{1}{3} T^{tt}_{\rm deg.} =
 T^{\rho\rho}_{\rm deg.} =
 \rho^2 T^{\varphi\varphi}_{\rm deg.} =
 T^{zz}_{\rm deg.} = \frac{\mu^4}{12\pi^2},
 \label{eq:deg}
\end{gather}
with all other expectation values vanishing.

\subsection{Summation over angular momentum \texorpdfstring{$m_j$}{mj}}\label{sec:OI:summ}

The summation over $m_j$ appearing in the expression for $\Delta A$, given in Eq.~\eqref{eq:DA_master}, can be performed using the identities
\begin{subequations}\label{eq:sumJ}
\begin{align}
 \sum^{\infty}_{m_j = -\infty} e^{i v\beta \Omega_I m_j} J^{+}_{m_j}(q_j \rho) &= 2 c_v J_{0}\left(2 q_j \rho s_v\right), \label{eq:sumJ_plus}\\
 \sum^{\infty}_{m_j = - \infty} e^{i v \beta \Omega_I m_j} J^{-}_{m_j}(q_j \rho) &= 2 i s_v J_{0}\left(2 q_j \rho s_v\right), \label{eq:sumJ_minus}
\end{align}
\end{subequations}
where we defined
\begin{equation}
 s_v = \sin\left(\frac{v \beta \Omega_I}{2}\right), \quad c_v = \cos\left(\frac{v \beta \Omega_I}{2}\right).
\end{equation}

The identities in Eqs.~\eqref{eq:sumJ} can be obtained using the summation theorem for Bessel functions
[see Eq. (8.531.1) in Ref.~\cite{gradshteyn2015table}; and Eq. (10.23.7) in
Ref.~\cite{DLMF}]:
\begin{equation}
 \sum_{n = -\infty}^\infty e^{-in x} J_n^2 (x) = J_0\left(2z \sin \frac{x}{2}\right),
 \label{eq:sumJ_identity}
\end{equation}
where $n = 0, \pm 1, \pm 2, \dots$ is an integer number. Using the explicit expressions \eqref{eq:Jpm} for $J_{m_j}^\pm(q_j \rho)$, the sums over the odd half-integer $m_j = \pm \frac{1}{2}, \pm \frac{3}{2}, \dots$ appearing in Eq.~\eqref{eq:sumJ} can be converted into sums over the integer $n$ as follows:
\begin{subequations}
\begin{align}
 \hspace{-15pt} \sum^{\infty}_{m_j = -\infty} \hspace{-6pt} e^{i v\beta \Omega_I m_j} J^{+}_{m_j}(q_j \rho) &= 2c_v \hspace{-4pt} \sum_{n = -\infty}^\infty \hspace{-4pt} e^{i v \beta \Omega_I n} J_n^2(q_j \rho), \\
 \hspace{-15pt} \sum^{\infty}_{m_j = -\infty} \hspace{-6pt} e^{i v\beta \Omega_I m_j} J^{-}_{m_j}(q_j \rho) &= 2is_v \hspace{-4pt} \sum_{n = -\infty}^\infty \hspace{-4pt} e^{i v \beta \Omega_I n} J_n^2(q_j \rho).
\end{align}
\end{subequations}
Applying Eq.~\eqref{eq:sumJ_identity} leads to the results in Eqs.~\eqref{eq:sumJ}.

Furthermore, the computation of $T^{\varphi\varphi}$ and $T^{t\varphi}$ requires the following sums,
\begin{subequations}\label{eq:sumJm}
\begin{align}
 \hspace{-15pt} \sum_{m_j = -\infty}^\infty \hspace{-6pt}  e^{i v\beta \Omega_I m_j} m_j J^{+}_{m_j}(q_j \rho) &= \frac{d}{d\Omega_I} \left[\frac{2c_v}{i v \beta} J_0(2q_j \rho c_v)\right], \label{eq:sumJm_plus}\\
 \hspace{-15pt} \sum^{\infty}_{m_j = - \infty} \hspace{-6pt} e^{i v \beta \Omega_I m_j} m_j J^{-}_{m_j}(q_j \rho) &= \frac{d}{d\Omega_I} \left[\frac{2s_v}{v\beta} J_0(2q_j \rho c_v)\right],\label{eq:sumJm_minus}
\end{align}
\end{subequations}
which follow after differentiating Eqs.~\eqref{eq:sumJ} with respect to $\Omega_I$.

Let us illustrate how the above prescription is applied for the simplest case, that of the fermion condensate. Since $\overline{U}_j U_j$, given in Eq.~\eqref{eq:FC_sesqi}, is even with respect to $m_j \to -m_j$, while $\mathcal{C}_\sigma(\mathcal{FC}) = 1$, Eq.~\eqref{eq:DA_master} becomes
\begin{multline}
 \frac{\Delta FC}{M} = \sum_{v = 1}^\infty (-1)^{v+1} \sum_{j,\sigma} e^{-v \beta |\mathcal{E}^\sigma_j|} {\rm sgn}(\mathcal{E}^\sigma_j) e^{iv\beta \Omega_I m_j} \\\times \frac{J_{m_j}^+(q_j \rho)}{8\pi^2 E_j}.
\end{multline}
The summation over $\lambda$ gives a factor of $2$, while replacing the domain for the $k_j$ integral from $[-p_j,p_j]$ to $[0,p_j]$ gives another factor of $2$. Using Eq.~\eqref{eq:sumJ_plus} to perform the summation over $m_j$ leads, in the massless limit, to
\begin{multline}
 \frac{\Delta FC}{M} \to \sum_{v = 1}^\infty \frac{(-1)^{v+1}}{\pi^2} c_v \sum_{\sigma = \pm 1} \int_0^\infty dE_j e^{-v\beta |\mathcal{E}^\sigma_j|} {\rm sgn}(\mathcal{E}^\sigma_j)
 \\\times \int_0^{p_j} dk_j J_0(2q_j \rho s_v).
\end{multline}
For completeness, we list below the results for all observables using the shorthand notation
\begin{multline}
 \Delta A = \sum_{v = 1}^\infty \frac{(-1)^{v+1}}{\pi^2} \sum_{\sigma = \pm 1} \int_0^\infty dE_j e^{-v\beta |\mathcal{E}^\sigma_j|} \\\times
 \int_0^{p_j} dk_j J_0(2q_j \rho s_v) \widetilde{A}^j_{v,\sigma}.
 \label{eq:DA_vsEk}
\end{multline}
We list first the observables for which the sesquilinear forms are odd with respect to $m_j \to -m_j$:
\begin{gather}
 \widetilde{J}^{\varphi;j}_{V;v,\sigma} = 2i\sigma s_v k_j^2, \quad
 \widetilde{J}^{z;j}_{H;v,\sigma} = i \sigma s_v p_j, \quad
 \widetilde{J}^{z;j}_{A;v,\sigma} = i s_v E_j, \nonumber\\
 \widetilde{T}^{t\varphi;j}_{v,\sigma} = -\frac{i c_v E_j}{2v \beta \rho^2 J_0} \frac{dJ_0}{d\Omega_I} + i s_v k_j^2 E_j,
 \label{eq:DA_vsj_odd}
\end{gather}
where $J_0 \equiv J_0(2q_j \rho s_v)$. The observables FC, $J^t_V$, $J^z_A$, $T^{tt}$, $T^{\rho\rho} = T^{zz}$ and $T^{\varphi\varphi}$ whose sesquilinear forms are even with respect to $m_j \to -m_j$ can be factored as $\widetilde{A}^j_{v,\sigma} = \overline{A}^j_{v,\sigma} {\rm sgn}(\mathcal{E}^\sigma_j)$, with
\begin{gather}
 \frac{\overline{\rm FC}^j_{v,\sigma}}{M} = c_v, \quad
 \overline{J}^{t;j}_{V;v,\sigma} = \sigma c_v E_j, \quad
 \overline{T}^{tt;j}_{v,\sigma} = c_v E_j^2, \nonumber\\
 %\overline{J}^{z;j}_{H;v,\sigma} = i \sigma s_v E_j, \quad
 \overline{T}^{\rho\rho;j}_{v,\sigma} = c_v k_j^2, \quad
 \overline{T}^{\varphi\varphi;j}_{v,\sigma} = \frac{2k_j^2}{v\beta \rho^2 J_0} \frac{d(s_v J_0)}{d\Omega_I}.
 \label{eq:DA_vsj_even}
\end{gather}

\subsection{Longitudinal momentum integration}\label{sec:OI:kint}

To perform the $k_j$ integration, we employ polar coordinates in the form
\begin{align}
    k_j &= p_j \sin \theta , &
    q_j &= p_j \cos \theta,
\end{align}
such that $d k_j = p_j \cos \theta d\theta$. We furthermore introduce the notation
\begin{equation}
 x = v\beta p_j, \quad
 X^\sigma_v = x - v \beta \sigma \mu, \quad
 \alpha_v = \frac{2\rho}{v\beta} s_v,
 \label{eq:alphav}
\end{equation}
with the intention of switching the integration variable from $E_j$ to $x$, via $dE_j= dx / v\beta$.

The $k_j$ integration appearing in Eq.~\eqref{eq:DA_vsEk} can be solved in terms of the integrals:
\begin{align}
 \mathcal{I}_n &= \frac{1}{p_j^{2n+1}} \int_0^{p_j} dk_j\, q_j^{2n} J_0(2 q_j \rho s_v) \nonumber\\
 &= \int_0^{\pi/2} d\theta  (\cos\theta)^{2n+1} J_0(\alpha_v x \cos\theta).
\end{align}
For our observables, only the cases $n = 0$ and $n = 1$ are required,
given in Eqs.~(175a) and (175b) in Ref.~\cite{Ambrus:2023bid} and reproduced below:
\begin{subequations}
\begin{align}
 \mathcal{I}_0 &= \frac{\sin(\alpha_v x)}{\alpha_v x}, \\
 \mathcal{I}_1 &= \frac{\cos(\alpha_v x)}{\alpha_v^2 x^2} + (\alpha_v^2 x^2 -1) \frac{\sin(\alpha_v x)}{\alpha_v^3 x^3}.
\end{align}
\end{subequations}

We thus have simple rules for performing the $k_j$ (equivalently, $\theta$) integral starting from the terms $\widetilde{A}^j_{v,\sigma}$ shown in Eqs.~\eqref{eq:DA_vsj_odd}--\eqref{eq:DA_vsj_even}. Denoting
\begin{equation}
 \widetilde{A}^x_{v,\sigma} = \frac{1}{v\beta} \int_0^{p_j} dk_j J_0(2q_j \rho s_v) \widetilde{A}^j_{v,\sigma},
\end{equation}
we obtain $\widetilde{A}^x_{v,\sigma}$ from $\widetilde{A}^j_{v,\sigma}$ by noting that there are only two types of terms: when $k_j$ does not appear explicitly and the $k_j$ integral gives $p_j \mathcal{I}_0$; or when the integrand contains the $k_j^2 = p_j^2 - q_j^2$ factor, in which case the $k_j$ integral gives
\begin{multline}
 \int_0^{p_j} dk_j k_j^2 J_0(2q_j \rho s_v) = \frac{x}{v^3 \beta^3 \alpha_v^2} \left[\frac{\sin(\alpha_v x)}{\alpha_v x} - \cos(\alpha_v x)\right] \\
 = -\frac{1}{v^3\beta^3 \alpha_v} \frac{d}{d\alpha_v} \left[\frac{\sin(\alpha_v x)}{\alpha_v}\right].
\end{multline}

Explicitly, for the terms shown in Eq.~\eqref{eq:DA_vsj_odd} we obtain
\begin{align}
 \widetilde{J}^{\varphi;x}_{V;v,\sigma} &= -\frac{2i\sigma s_v}{v^4 \beta^4 \alpha_v} \frac{d}{d\alpha_v} \left[\frac{\sin(\alpha_v x)}{\alpha_v}\right], \nonumber\\
 \widetilde{J}^{z;x}_{H;v,\sigma} &= \sigma \widetilde{J}^{z;x}_{A;v,\sigma} = \frac{i \sigma s_v x}{v^3 \beta^3 \alpha_v} \sin(\alpha_v x), \nonumber\\
 \widetilde{T}^{t\varphi;x}_{v,\sigma} &= -\frac{i s_v x (1 + c_v^2)}{v^5 \beta^5 \alpha_v} \frac{d}{d\alpha_v} \left[\frac{\sin(\alpha_v x)}{\alpha_v}\right],
 \label{eq:DA_vsx_odd}
\end{align}
where in the evaluation of $\widetilde{T}^{t\varphi;x}_{v,\sigma}$ we pulled the derivative with respect to $\Omega_I$, appearing in the first term in $\widetilde{T}^{t\varphi;j}_{v,\sigma}$ given in Eq.~\eqref{eq:DA_vsj_odd}, in front of the integration with respect to $\theta$.
For the observables represented in Eq.~\eqref{eq:DA_vsj_even}, we use the same notation convention by which $\widetilde{A}^x_{v,\sigma} = \overline{A}^x_{v,\sigma} {\rm sgn}(X^\sigma_v)$, obtaining
\begin{align}
 \overline{T}^{tt;x}_{v,\sigma} &= \frac{x^2}{v^4 \beta^4} \frac{c_v}{\alpha_v} \sin (\alpha_v x),\nonumber\\
 \overline{J}^{t;x}_{V;v,\sigma} &= \frac{\sigma x}{v^3 \beta^3} \frac{c_v}{\alpha_v} \sin (\alpha_v x),\nonumber\\
 \frac{\overline{FC}^x_{v,\sigma}}{M} &\to \frac{c_v}{v^2\beta^2 \alpha_v} \sin(\alpha_v x), \nonumber\\
 \overline{T}^{\rho\rho;x}_{v,\sigma} &= -\frac{c_v}{v^4\beta^4 \alpha_v} \frac{d}{d\alpha_v} \left[\frac{\sin(\alpha_v x)}{\alpha_v}\right],\nonumber\\
 \overline{T}^{\varphi\varphi;x}_{v,\sigma} &= -\frac{c_v}{\rho^2 v^4 \beta^4} \frac{d^2}{d\alpha_v^2}\left[\frac{\sin(\alpha_v x)}{\alpha_v}\right].
 \label{eq:DA_vsx_even}
\end{align}

With the above notation, we can express $\Delta A$ as
\begin{equation}
 \Delta A = \sum_{v = 1}^\infty \frac{(-1)^{v+1}}{\pi^2} \sum_{\sigma = \pm 1} \int_0^\infty dx e^{-|X^\sigma_v|}
 \widetilde{A}^x_{v,\sigma}.
 \label{eq:DA_vx}
\end{equation}
In particular, for the fermion condensate, we obtain
\begin{multline}
 \frac{\Delta {\rm FC}}{M} = \sum_{v =1}^\infty \frac{(-1)^{v+1} c_v}{(\pi v\beta)^2 \alpha_v} \sum_{\sigma = \pm 1} \int_0^\infty dx\, e^{-|X^\sigma_v|} {\rm sgn}(X^\sigma_v) \\\times \sin(\alpha_v x).
 \label{eq:FC_vx}
\end{multline}

\subsection{Momentum magnitude integration}\label{sec:OI:pint}

Moving now to the momentum magnitude integration, we write Eq.~\eqref{eq:DA_vx} as
\begin{equation}
 \Delta A = \sum_{v = 1}^\infty \frac{(-1)^{v+1}}{\pi^2} A_v, \
 A_v = \sum_{\sigma = \pm 1} \int_0^\infty dx\, e^{-|X^\sigma_v|} \widetilde{A}^x_{v,\sigma}.
 \label{eq:DA_v}
\end{equation}
Inspection of Eqs.~\eqref{eq:DA_vsj_odd}--\eqref{eq:DA_vsj_even} shows that the integrands $\widetilde{A}^x_{v,\sigma}$ involve only terms of the form $x^n \sin(\alpha_v x)$, with $n \in \{0, 1, 2\}$, and $x \cos(\alpha_v x)$. These terms can be multiplied by the signum function, ${\rm sgn}(X^\sigma_v)$, or by the particle/anti-particle factor $\sigma = \pm 1$. All these terms can be generated from the following master functions,
\begin{align}
 \begin{pmatrix}
  I_+(\alpha_v) \\ J_+(\alpha_v)
 \end{pmatrix} &= \frac{1}{2} \sum_{\sigma = \pm 1}
 \begin{pmatrix}
  1 \\ \sigma
 \end{pmatrix}
 \int_0^\infty dx\, e^{-|X^\sigma_v|} {\rm sgn}(X^\sigma_v) e^{i \alpha_v x}, \nonumber\\
 \begin{pmatrix}
  I_-(\alpha_v) \\ J_-(\alpha_v)
 \end{pmatrix} &= \frac{1}{2} \sum_{\sigma = \pm 1}
 \begin{pmatrix}
  1 \\ \sigma
 \end{pmatrix}
 \int_0^\infty dx\, e^{-|X^\sigma_v|} e^{i \alpha_v x}.
 \label{eq:IJpm_def}
\end{align}
The terms in Eq.~\eqref{eq:DA_vsx_odd} lead to
\begin{subequations}\label{eq:DA_v_odd_aux}
\begin{align}
 J^\varphi_{V;v} &= -\frac{2s_v}{v^4 \beta^4 \alpha_v} \frac{d}{d\alpha_v}\left[\frac{1}{\alpha_v}[J_-(\alpha_v) - J_-(-\alpha_v)]\right], \\
 J^z_{H;v} &= -\frac{i}{2\rho v^2 \beta^2} \frac{d}{d\alpha_v}\left[J_-(\alpha_v) + J_-(-\alpha_v)\right], \\
 J^z_{A;v} &= -\frac{i}{2\rho v^2 \beta^2} \frac{d}{d\alpha_v}\left[I_-(\alpha_v) + I_-(-\alpha_v)\right],\\
 T^{t\varphi}_{v} &= \frac{i(1 + c_v^2)}{2 \rho v^4 \beta^4} \frac{d}{d\alpha_v} \frac{1}{\alpha_v} \frac{d}{d\alpha_v}[I_-(\alpha_v) + I_-(-\alpha_v)],
\end{align}
\end{subequations}
Similarly, the terms in Eq.~\eqref{eq:DA_vsx_even} lead to
\begin{subequations}\label{eq:DA_v_even_aux}
\begin{align}
 \frac{{\rm FC}_v}{M} &\to -\frac{ic_v}{v^2 \beta^2 \alpha_v}[I_+(\alpha_v) - I_+(-\alpha_v)],\\
 J^t_{V;v} &= -\frac{c_v}{v^3 \beta^3 \alpha_v} \frac{d}{d\alpha_v} [J_+(\alpha_v) + J_+(-\alpha_v)],\\
 T^{tt}_v &= \frac{ic_v}{v^4 \beta^4 \alpha_v} \frac{d^2}{d\alpha_v^2}[I_+(\alpha_v) - I_+(-\alpha_v)],\label{eq:DA_Tttv}\\
 T^{\rho\rho}_v &= \frac{i c_v}{v^4 \beta^4 \alpha_v} \frac{d}{d\alpha_v} \left[\frac{1}{\alpha_v} [I_+(\alpha_v) - I_+(-\alpha_v)]\right], \\
 T^{\varphi\varphi}_v &= \frac{i c_v}{\rho^2 v^4 \beta^4} \frac{d^2}{d\alpha_v^2} \left[\frac{1}{\alpha_v}[I_+(\alpha_v) - I_+(-\alpha_v)]\right].
\end{align}
\end{subequations}
To obtain the above results, we employed the relations
\begin{align}
 x^n \sin(\alpha_v x) &= \frac{1}{2i^{n+1}} \frac{d^n}{d\alpha_v^n} [e^{i \alpha_v x} - (-1)^n e^{-i \alpha_v x}],\nonumber\\
 x^n \cos(\alpha_v x) &= \frac{1}{2i^n} \frac{d^n}{d\alpha_v^n} [e^{i \alpha_v x} + (-1)^n e^{-i \alpha_v x}].
\end{align}

To evaluate the integrals $I_\pm$ and $J_\pm$ defined in Eqs.~\eqref{eq:IJpm_def}, we note that the chemical potential is multiplied by the particle/anti-particle index $\sigma = \pm 1$. Hence, we write $\mu = \sigma_\mu | \mu |$, where $\sigma_\mu = {\rm sgn}(\mu)$ and denote $\sigma^\prime = \sigma \sigma_\mu$. The case $\sigma' = -1$ will lead to $X^{\sigma'=-1}_v = x + v \beta |\mu| > 0$. Therefore, only the term with $\sigma' = 1$ will change sign under the absolute value, since $|X^{\sigma'=1}_v| = -X^{\sigma'=1}_v$ and ${\rm sgn}(X^{\sigma'=1}_v) = -1$ when $x< v \beta |\mu|$. Putting the above into practice, we arrive at
\begin{subequations}\label{eq:IJpm_aux}
\begin{align}
 I_\pm(\alpha_v) &= \frac{1}{2} \int_0^{v\beta|\mu|} dx(\mp e^{x-v\beta|\mu|} + e^{-x - v \beta|\mu|}) e^{i\alpha_v x} \nonumber\\
 & + \cosh(v\beta\mu) \int_{v\beta|\mu|}^\infty dx\, e^{-x + i \alpha_v x}, \label{eq:Ipm_aux} \\
 J_\pm(\alpha_v) &= \frac{\sigma_\mu}{2} \int_0^{v\beta|\mu|} dx(\mp e^{x-v\beta|\mu|} - e^{-x - v \beta|\mu|}) e^{i\alpha_v x}\nonumber\\
 & + \sinh(v\beta\mu) \int_{v\beta|\mu|}^\infty dx\, e^{-x + i \alpha_v x} . \label{eq:Jpm_aux}
\end{align}
\end{subequations}
A straightforward computation leads to
\begin{subequations}\label{eq:IJpm}
\begin{align}
 I_+(\alpha_v) &= \frac{1}{1 + \alpha_v^2} \left(i \alpha_v e^{i v \beta |\mu| \alpha_v} + e^{-v \beta|\mu|}\right), \label{eq:Ip}\\
 I_-(\alpha_v) &= \frac{1}{1 + \alpha_v^2} \left(e^{i v \beta |\mu| \alpha_v} + i \alpha_v e^{-v \beta|\mu|}\right), \label{eq:Im}\\
 J_+(\alpha_v) &= \frac{i \sigma_\mu \alpha_v}{1 + \alpha_v^2} \left(e^{i v \beta |\mu| \alpha_v} - e^{-v \beta|\mu|}\right), \label{eq:Jp}\\
 J_-(\alpha_v) &= \frac{\sigma_\mu}{1 + \alpha_v^2} \left(e^{i v \beta |\mu| \alpha_v} - e^{-v \beta|\mu|}\right). \label{eq:Jm}
\end{align}
\end{subequations}

Applying the above steps for the fermion condensate, given in Eq.~\eqref{eq:FC_vx}, leads to:
\begin{equation}
 \frac{\Delta {\rm FC}}{M} = \sum_{v =1}^\infty \frac{(-1)^{v+1}}{\pi^2} \frac{2c_v c^{v\mu}}{v^2 \beta^2 (1 + \alpha_v^2)}.
 \label{eq:FC_v}
\end{equation}
Above and for future convenience, we employ the following notation:
\begin{equation}
 s^{v\mu} = \sin(v\beta \mu \alpha_v), \quad
 c^{v\mu} = \cos(v \beta \mu \alpha_v).
\end{equation}
The coefficients $A_v$ corresponding to the observables in Eq.~\eqref{eq:DA_v_odd_aux} become
\begin{subequations}\label{eq:DA_v_odd}
\begin{align}
 J^{\varphi}_{V;v} &= -\frac{4 i s_v}{v^4 \beta^4 \alpha_v}
 \frac{d}{d\alpha_v} \left[\frac{s^{v\mu}}{\alpha_v(1 + \alpha_v^2)}\right], \\
 J^{z}_{H;v} &= -\frac{i \sigma_\mu}{\rho v^2 \beta^2} \frac{d}{d\alpha_v} \left(\frac{c^{v\mu} - e^{-v\beta|\mu|}}{1 +\alpha_v^2}\right), \\
 J^{z}_{A;v} &= -\frac{i}{\rho v^2 \beta^2} \frac{d}{d\alpha_v} \left(\frac{c^{v\mu}}{1 +\alpha_v^2}\right), \\
 T^{t\varphi}_v &= \frac{i(1 + c_v^2)}{\rho v^4 \beta^4} \frac{d}{d\alpha_v} \left[\frac{1}{\alpha_v} \frac{d}{d\alpha_v} \left(\frac{c^{v\mu}}{1 + \alpha_v^2}\right)\right].
\end{align}
\end{subequations}
The same coefficients for the observables in Eq.~\eqref{eq:DA_v_even_aux} are
\begin{subequations}\label{eq:DA_v_even}
\begin{align}
 \frac{{\rm FC}_v}{M} &\to \frac{2 c_v c^{v\mu}}{v^2 \beta^2(1 + \alpha_v^2)}, \label{eq:DA_v_FC}\\
 J^t_{V;v} &= \frac{2c_v}{v^3\beta^3 \alpha_v} \frac{d}{d\alpha_v} \left(\frac{\alpha_v s^{v\mu}}{1 + \alpha_v^2}\right), \\
 \label{eq: Ttt_v}
 T^{tt}_v &= -\frac{2 c_v}{v^4\beta^4 \alpha_v} \frac{d^2}{d\alpha_v^2} \left(\frac{\alpha_v c^{v\mu}}{1 + \alpha_v^2}\right), \\
 T^{\rho\rho}_v &= -\frac{2c_v}{v^4 \beta^4 \alpha_v} \frac{d}{d\alpha_v} \left(\frac{c^{v\mu}}{1 + \alpha_v^2}\right), \\
 T^{\varphi\varphi}_v &= -\frac{2 c_v}{\rho^2 v^4 \beta^4} \frac{d^2}{d\alpha_v^2} \left(\frac{c^{v\mu}}{1 + \alpha_v^2}\right),
\end{align}
\end{subequations}
while we remind that $T^{zz}_v = T^{\rho\rho}_v$.

Before performing the summation over $v$, the results in Eqs.~\eqref{eq:DA_v_odd}, \eqref{eq:DA_v_even} indicate that the non-degenerate contributions $\Delta A$ to the thermal expectation value of any observable $A$ decay far from the rotation axis as an inverse power of $\rho$, due to the presence of the function $\alpha_v = (2\rho / v \beta) s_v$. Meanwhile, the degenerate contributions $A_{\rm deg.}$, given in Eq.~\eqref{eq:deg}, are independent of $\rho$ and hence survive far from the rotation axis, in contradiction to the relativistic kinetic theory prediction in Eqs.~\eqref{eq:RKT_obs_im}. We will return to the comparison between the QFT and RKT results in the following subsections.

\subsection{Evaluation on the rotation axis} \label{sec:OI:axis}

The expectation values discussed so far can be evaluated analytically on the rotation axis, where $\rho = 0$. Taking into account Eq.~\eqref{eq:alphav}, defining $\alpha_v = \frac{2\rho}{v\beta} s_v$, Eqs.~\eqref{eq:DA_v_odd}--\eqref{eq:DA_v_even} reduce on the rotation axis to
\begin{gather}
 J^\varphi_{V;v} \to \frac{8i s_v \mu}{v^3\beta^3} + \frac{4i s_v \mu^3}{3v\beta}, \quad
 J^z_{A;v} \to \frac{4 i s_v}{v^3\beta^3} + \frac{2i s_v \mu^2}{v\beta}, \nonumber\\
 J^z_{H;v} \to \frac{4i \sigma_\mu s_v}{v^3\beta^3}(1 - e^{-v\beta|\mu|}) + \frac{2i \sigma_\mu s_v \mu^2}{v\beta}, \nonumber\\
 T^{t\varphi}_v \to 2i s_v(1 + c_v^2) \left(\frac{8}{v^5 \beta^5} + \frac{4\mu^2}{v^3\beta^3} + \frac{\mu^4}{3v \beta}\right), \nonumber\\
 \frac{{\rm FC}_v}{M} \to \frac{2c_v}{v^2\beta^2}, \quad
 J^t_{V;v} \to \frac{4c_v \mu}{v^2 \beta^2}, \nonumber\\
 T^{tt}_v, 3 T^{\rho\rho}_v, 3\rho^2 T^{\varphi\varphi}_v, 3T^{zz}_v \to \frac{12 c_v}{v^4\beta^4} + \frac{6 c_v \mu^2}{v^2 \beta^2}.
 \label{eq:DA_v_axis}
\end{gather}

The summation with respect to $v$ can be performed in terms of the polylogarithm function, ${\rm Li}_s(z) = \sum_{v = 1}^\infty z^v / v^s$, using the summation formulas
\begin{subequations}
\begin{align}
    \sum_{v=1}^{\infty} \frac{(-1)^v}{v^s} s_v
    &= \operatorname{Li}^{\nu, \mathfrak{i}}_s, &
    \sum_{v=1}^{\infty} \frac{(-1)^v}{v^s} c_v
    &= \operatorname{Li}^{\nu, \mathfrak{r}}_s,
\end{align}
\end{subequations}
where $\operatorname{Li}^{\nu, \mathfrak{r}(\mathfrak{i})}_s$ is the real (imaginary) part of $\operatorname{Li}_s (-e^{i \frac{\beta \Omega_I}{2}}) \equiv \operatorname{Li}_s (-e^{i \pi \nu})$, namely:
\begin{align}
 \operatorname{Li}^{\nu, \mathfrak{i}}_s &= \frac{1}{2i} \left[\operatorname{Li}_s (-e^{i \pi \nu}) - \operatorname{Li}_s (-e^{-i \pi \nu})\right], \nonumber\\
 \operatorname{Li}^{\nu, \mathfrak{r}}_s &= \frac{1}{2} \left[\operatorname{Li}_s (-e^{i \pi \nu}) + \operatorname{Li}_s (-e^{-i \pi \nu})\right].
 \label{eq:Li_r_i_def}
\end{align}

Applying the above prescription, our observables in Eqs.~\eqref{eq:DA_v_odd}--\eqref{eq:DA_v_even} evaluate on the rotation axis $\rho = 0$ to
\begin{subequations}\label{eq:DA_axis}
\begin{align}
 \frac{\Delta FC}{M} &\to - \frac{2}{\pi^2 \beta^2} \operatorname{Li}^{\nu, \mathfrak{r}}_2, \quad
 \Delta J_V^t \to -\frac{4 \mu}{\pi^2 \beta^2} \operatorname{Li}^{\nu, \mathfrak{r}}_2, \\
 \Delta J^\varphi_V &\to - \frac{4 i \mu^3}{3 \beta \pi^2} \operatorname{Li}^{\nu, \mathfrak{i}}_1 - \frac{8 i \mu}{\beta^3 \pi^2} \operatorname{Li}^{\nu, \mathfrak{i}}_3, \\
 \Delta J_A^z &\to - \frac{2 i \mu^2}{\pi^2 \beta} \operatorname{Li}^{\nu, \mathfrak{i}}_1 - \frac{4 i}{\pi^2 \beta^3} \operatorname{Li}^{\nu, \mathfrak{i}}_3, \\
 \Delta T^{t t} &\to - \frac{12}{\pi^2 \beta^4} \operatorname{Li}^{\nu, \mathfrak{r}}_4 - \frac{6 \mu^2}{\pi^2 \beta^2} \operatorname{Li}^{\nu, \mathfrak{r}}0_2,
\end{align}
while $\Delta T^{\rho\rho}, \Delta T^{z z}, \rho^2 \Delta T^{\varphi\varphi} \to \Delta T^{tt}/3$. The vertical component of the helicity current, $\Delta J^z_H$, requires special treatment, due to the presence of the exponential function $e^{-v \beta |\mu|}$ appearing in Eq.~\eqref{eq:DA_v_odd}. We find
\begin{multline}
  \Delta J_H^z \to - \frac{2 i \mu^2 \sigma_\mu}{\pi^2 \beta} \operatorname{Li}^{\nu, \mathfrak{i}}_1
  - \frac{4 i \sigma_\mu}{\pi^2 \beta^3} \operatorname{Li}^{\nu, \mathfrak{i}}_3 \\
  + \frac{2 \sigma_\mu}{\pi^2 \beta^3} \left[\operatorname{Li}_3(-e^{i\pi \nu - \beta |\mu|}) - \operatorname{Li}_3(-e^{-i\pi \nu - \beta |\mu|})\right].
  \label{eq:DA_axis_JHz}
\end{multline}
\end{subequations}

In order to simplify the expressions involving polylogarithms, we employ the identities in Eqs.~\eqref{eq:polylogs}.
The values on the rotation axis then become
\begin{subequations}\label{eq:axis}
\begin{align}
 \frac{FC}{M} &\to \frac{1 - 3 \nu^2}{6 \beta^2} + \frac{\mu^2}{2\pi^2},\label{eq:axis_FC}\\
 J_V^t &\to \frac{\mu (1 - 3 \nu^2)}{3\beta^2} + \frac{\mu^3}{3 \pi^2},\\
 J^\varphi_V &\to \frac{2\pi i \nu}{\beta} \left(\frac{\mu(1 - \nu^2)}{3 \beta^2} + \frac{\mu^3}{3 \pi^2}\right), \\
%
 %\Delta\left(J_H^z\right)^{i \Omega_I}_{\rho \to 0}
     % & = i \Omega_I \left(\frac{2 \ln 2 \mu}{\pi^2 \beta} + \frac{\mu^3}{12 \pi^2 \beta} - \frac{\beta \mu}{48 \pi^2} \Omega_I^2\right),\\
%
 J_A^z &\to \frac{2\pi i \nu}{\beta} \left(\frac{1 - \nu^2}{6\beta^2} + \frac{\mu^2}{2\pi^2}\right), \\
 T^{tt} &\to \frac{\pi^2}{\beta^4} \left(\frac{7}{60} - \frac{\nu^2}{2} + \frac{\nu^4}{4}\right) \nonumber\\
 & + \frac{\mu^2(1 - 3\nu^2)}{2\beta^2} + \frac{\mu^4}{4 \pi^2},\label{eq:axis_Ttt}\\
 T^{t\varphi} &\to \frac{2\pi i \nu}{\beta} \left[\frac{\pi^2}{\beta^4} \left(\frac{7}{45} - \frac{8 \nu^2}{9} + \frac{31 \nu^4}{15}\right) \right.\nonumber\\
 &\left. + \frac{2\mu^2(1 - 4\nu^2)}{3\beta^2} + \frac{\mu^4}{3\pi^2} \right],
\end{align}
\end{subequations}
while $T^{\rho\rho}, \rho^2 T^{\varphi\varphi}, T^{zz} \to \frac{1}{3} T^{tt}$.

In the case of the helicity current, Eqs.~\eqref{eq:polylogs} can be applied only for the terms appearing on the first line of Eq.~\eqref{eq:DA_axis_JHz}, leading to
\begin{multline}
 \Delta J^z_H \to \frac{2i \pi \nu \sigma_\mu}{\beta} \left[\frac{1- \nu^2}{6\beta^2} + \frac{\mu^2}{2\pi^2}\right] \\
 + \frac{2 \sigma_\mu}{\pi^2 \beta^3} [\operatorname{Li}_3(-e^{i\pi \nu - \beta |\mu|}) - \operatorname{Li}_3(-e^{-i\pi \nu - \beta |\mu|})].
 \label{eq:axis_JHz}
\end{multline}
We first contrast the above result with that obtained in Eq.~(6.26) of Ref.~\cite{Ambrus:2019ayb}, for the case of real rotation $\Omega = 2\pi \nu_R / \beta$. The vortical conductivities on the rotation axis, given in the aforementioned equation, read
\begin{equation}
 \sigma^\omega_\pm  = \mp \sum_{\varsigma_\pm = \pm 1} \sum_{\varsigma_\Omega = \pm 1} \frac{\varsigma_\pm \varsigma_\Omega}{\pi^2 \beta^3 \Omega}
 \operatorname{Li}_3(-e^{\varsigma_\pm \beta \mu_\pm + \frac{\varsigma_\Omega}{2} \beta \Omega}),
\end{equation}
where $\mu_\pm = \mu_V \pm \mu_H \to \mu_V$ (we consider a vanishing helicity chemical potential) and we replaced $T = 1/\beta$ and we took the massless limit, $M = 0$.
Writing as before $\varsigma_\pm = \sigma_\mu \varsigma_\pm'$, with $\sigma_\mu = {\rm sgn}(\mu)$, we obtain for the vertical component of the helicity current, $J^z_H = \Omega (\sigma^\omega_+ - \sigma^\omega_-) / 2$, the following result:
\begin{multline}
 J^z_H \to \frac{2\pi \nu_R \sigma_\mu}{\beta} \left(\frac{1 + \nu_R^2}{6\beta^2} + \frac{\mu^2}{2\pi^2}\right) \\
 + \frac{2 \sigma_\mu}{\pi^2 \beta^3} [\operatorname{Li}_3(-e^{\pi \nu_R - \beta|\mu|} - \operatorname{Li}_3(-e^{-\pi\nu_R - \beta|\mu|})],
\end{multline}
which coincides with Eq.~\eqref{eq:axis_JHz} upon the replacement $\nu_R \to i \nu$. We therefore do not pursue a discussion of the various limits of Eq.~\eqref{eq:axis_JHz}, and instead give below just the large-T expression, obtained by performing a series expansion in powers of $\beta$, considering $\nu = \beta \Omega_I / 2\pi \sim \beta$:
\begin{equation}
 J^z_H \to \frac{2i \pi \nu \mu}{\beta^2} \left[\frac{2 \ln 2}{\pi^2} - \frac{\nu^2}{12} + \frac{\beta^2 \mu^2}{12\pi^2} + O(\beta^4)\right].
\end{equation}

The relations~\eqref{eq:axis} entail expressions which are polynomial with respect to $\nu$, diverging as $\nu \to \pm \infty$. However, the terms in Eqs.~\eqref{eq:DA_v_axis} are periodic with respect to $\nu \to \nu + 2$, as the dependence on $\nu$ is fully contained in the trigonometric functions $s_v = \sin(\frac{v\beta\Omega_I}{2}) = \sin(v \pi \nu)$ and $c_v = \cos(\frac{v\beta\Omega_I}{2}) = \cos(v \pi \nu)$. Taking as the principal domain $-1 \le \nu < 1$, we define the variable $\tilde{\nu} = \frac{1}{2}(1 + \nu)$, as well as $\{\tilde{\nu}\} = \tilde{\nu} - \lfloor \tilde{\nu} \rfloor \in [0,1)$, representing the fractional part of $\tilde{\nu}$, obtained by subtracting from $\tilde{\nu}$ its integer part $\lfloor \tilde{\nu} \rfloor$. We rewrite the on-axis expectation value of $T^{tt}$ with respect to $\{\tilde{\nu}\}$:
\begin{multline}
 T^{tt} = -\frac{\pi^2}{\beta^4} \left[\frac{2}{15} - 4\{\tilde{\nu}\}^2(1-\{\tilde{\nu}\})^2\right] \\
 - \frac{\mu^2}{\beta^2} \left[1 - 6\{\tilde{\nu}\}(1 - \{\tilde{\nu}\})\right] + \frac{\mu^4}{4\pi^2}.
 \label{eq:axis_Ttt_nup}
\end{multline}

\begin{figure}
\centering
\begin{tabular}{c}
\includegraphics[width=.95\columnwidth]{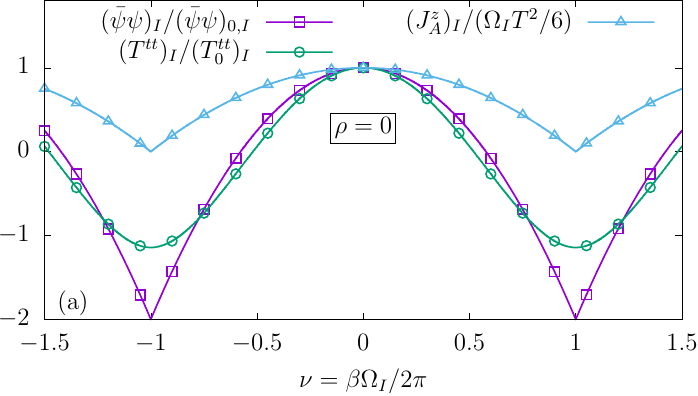} \\
\includegraphics[width=.95\columnwidth]{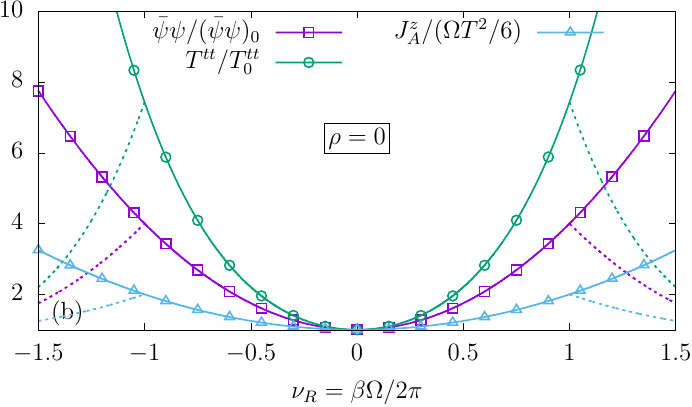}
\end{tabular}
\caption{Thermal expectation values for the observables $\bar{\psi} \psi$ (purple lines with squares), $T^{tt}$ (green lines with circles) and $J_A^z$ (blue lines with triangles), evaluated at vanishing chemical potential ($\mu = 0$) on the rotation axis ($\rho = 0$) using Eqs.~\eqref{eq:axis}, plotted with respect to the (a) imaginary ($\nu$) and (b) real ($\nu_R$) rotation parameter. The normalization is carried out with respect to the $\nu = 0$ value for $\bar{\psi} \psi$ and $T^{tt}$; while $J^z_A$ is normalized to $\Omega T^2 / 6$. The dashed lines in panel (b) represent the expected trends of the expectation values if the periodicity with respect to imaginary rotation $\nu$ would be preserved by the analytic continuation.
}
\label{fig:axis}
\end{figure}

The results obtained in Eqs.~\eqref{eq:axis} can be compared with those obtained under real rotation. For brevity, we only discuss $T^{tt}$ below. Taking the $\rho\to 0$ limit of Eq.~(154a) of Ref.~\cite{Ambrus:2019khr} and using Eqs.~(72a), (72c) and (72e) therein, we find that $T^{tt}(\rho = 0) = 3P(\rho = 0)$ evaluates to
\begin{align}
 T^{tt} &\to \frac{7\pi^2}{60 \beta^4} + \frac{\mu^2}{2\beta^2} + \frac{\mu^4}{4\pi^2} + \frac{3\Omega^2}{4} \left(\frac{1}{6\beta^2} + \frac{\mu^2}{2\pi^2}\right) + \frac{\Omega^4}{64\pi^2} \nonumber\\
 &\hspace{-8pt} = \frac{\pi^2}{\beta^4} \left(\frac{7}{60} + \frac{\nu_R^2}{2} + \frac{\nu_R^4}{4}\right) + \frac{\mu^2(1 + 3\nu_R^2)}{2\beta^2} + \frac{\mu^4}{4\pi^2},
\end{align}
which agrees with Eq.~\eqref{eq:axis_Ttt} upon replacing $\nu_R \to i\nu$, implying that the on-axis results under real rotation can be obtained from those corresponding to imaginary rotation through analytical continuation. This is contrary to the scalar field case considered in Ref.~\cite{Ambrus:2023bid}, where the analysis in Sec.~VD therein reveals odd terms of the form $\nu^3$ in the imaginary rotation case, which are analytically continued to $0$ to obtain the real rotation result. Nevertheless, the periodic structure implied by Eq.~\eqref{eq:axis_Ttt_nup} is absent in the result obtained using real rotation and thus analytical continuation can be trusted only over the principal domain, $-1 \le \nu < 1$, or equivalently, $-2\pi \le \beta \Omega < 2\pi$. To see this more clearly, a comparison between the real and imaginary results (given in (\ref{eq:axis})) on the rotation axis at zero chemical potential can be seen for the fermion condensate ${\rm FC}$, the energy-momentum tensor component $T^{tt}$ and the axial current $J_A^z$ in Fig.~\ref{fig:axis}, which highlights the periodicity with respect to the principal domain for the imaginary case in panel (a) and the analytic continuation to the real case marked by the dashed lines (outside the principal domain) in panel (b).

Let us consider a comparison to the RKT results in Eqs.~\eqref{eq:RKT_obs_im}, evaluated on the rotation axis, where $\gamma_I = 1$. The QFT results in Eqs.~\eqref{eq:axis} display corrections proportional to $\nu^2$ and $\nu^4$ that significantly change the expectation values of all observables, as $\nu$ ranges over its domain of periodicity, as shown in Fig.~\ref{fig:axis}.

\subsection{Large temperature expansion}\label{sec:OI:largeT}

The expectation values $A = A_{\rm deg.} + \Delta A$, with $\Delta A$ given in Eq.~\eqref{eq:DA_v} in terms of the terms $A_v$, can be obtained in the limit of large temperature by expanding $A_v$ from Eqs.~\eqref{eq:DA_v_odd}--\eqref{eq:DA_v_even} with respect to $\beta$. We illustrate this for the (massless limit of the) fermion condensate:
\begin{equation}
 \frac{{\rm FC}_v}{M} \to \frac{2 \gamma_I^2}{v^2 \beta^2} + (\gamma_I^2 - 1) \mu^2 - \frac{\Omega_I^2 \gamma_I^2}{12}(1 + 2 \gamma_I^2) + O(\beta^2),
 \label{eq:largeT_FCv_aux}
\end{equation}
where the Lorentz factor $\gamma_I = (1 + \rho^2 \Omega_I^2)^{-1/2}$ for an observer undergoing rigid imaginary rotation $\Omega = i \Omega_I$ was introduced in Eq.~\eqref{eq:RKT_im_gamma}. Recognizing the squared vorticity and acceleration,
\begin{align}
 \boldsymbol{\omega}^2 &= \Omega^2 \gamma^4 & &\to & \boldsymbol{\omega}_I^2 &= -\Omega_I^2 \gamma_I^4, \nonumber\\
 \mathbf{a}^2 &= \rho^2 \Omega^4 \gamma^4 & &\to & \mathbf{a}_I^2 &= -\Omega_I^2 \gamma_I^2 (\gamma_I^2 - 1),
\end{align}
we can rewrite Eq.~\eqref{eq:largeT_FCv_aux} in the more compelling form
\begin{equation}
 \frac{{\rm FC}_v}{M} \to \frac{2 T_\rho^2}{v^2} + \mu_\rho^2 - \mu^2 + \frac{3\boldsymbol{\omega}_I^2 -\mathbf{a}_I^2}{12} + O(\beta^2),
 \label{eq:largeT_FCv}
\end{equation}
where we denoted the local temperature and chemical potential
\begin{equation}
 T_\rho = \frac{\gamma_I}{\beta}, \quad
 \mu_\rho = \gamma_I \mu,
\end{equation}
as described in Eq.~\eqref{eq:TE_gamma}.
Introducing the above into Eq.~\eqref{eq:DA_v}, the sum over $v$ can be performed in terms of the alternating zeta function $\eta(n) = \sum_{v=1}^\infty (-1)^{v+1} / v^n$, for which we note the following values
\begin{gather}
\eta(0) = \frac{1}{2}, \quad
\eta(1) = \ln 2, \quad
\eta(2) = \frac{\pi^2}{12}, \nonumber\\
\eta(3) = \frac{3}{4} \zeta(3), \quad
\eta(4) = \frac{7\pi^4}{720}.
\end{gather}
For the $n = 0$ series, Abel summation was adopted. In the case of the FC, we obtain
\begin{equation}
 \frac{\rm FC}{M} \to \frac{T_\rho^2}{6} + \frac{\mu_\rho^2}{2\pi^2} + \frac{3 \boldsymbol{\omega}^2_I - \mathbf{a}_I^2}{24\pi^2},
 \label{eq:largeT_FC}
\end{equation}
where we added the degenerate contribution ${\rm FC}_{\rm deg.}$ from Eq.~\eqref{eq:deg}, which cancels exactly the $\mu^2$ term in Eq.~\eqref{eq:largeT_FCv}.

In Eq.~\eqref{eq:largeT_FC}, we omitted the $O(\beta^2)$ tail implied by the large-$T$ expansion in Eq.~\eqref{eq:largeT_FCv}. The reason why this tail is omitted is that the higher-order corrections to ${\rm FC}_v / M$ make vanishing contributions, due to the remarkable property of the (analytical continuation of the) alternating zeta function to negative arguments, satisfying
\begin{equation}
 \eta(-2n) = 0, \qquad \forall n = 1, 2, 3, \dots.
\end{equation}
To apply this property, we note that ${\rm FC}_v / M$ in Eq.~\eqref{eq:DA_v_FC} is an even function of both $v$ and $\beta$. A series expansion with respect to $\beta$ means an expansion in powers of $v \beta \Omega_I$, $v \beta \mu$ and $v\beta/\rho$. Furthermore, we note that $s_v = \sin(v\beta \Omega_I/2)$ and $s^{v\mu} = \sin(v \beta \mu \alpha_v)$ are odd with respect to $\beta v$, while $\alpha_v = \frac{2\rho}{v\beta} s_v$, $c_v = \cos(v\beta \Omega_I/2)$ and $c^{v\mu} = \cos(v\beta \mu \alpha_v)$ are even. Thus, the large-$T$ expansion of ${\rm FC}_v/M$ will contain only even powers of $v$, namely: $v^{-2}$, $v^0$, $v^2$, dots. The first two are evaluated under the summation with respect to $v$ as $\eta(2)$ and $\eta(0)$, while all the remaining terms are of the form $\eta(-2n)$ and therefore vanish.

A quick inspection of Eqs.~\eqref{eq:DA_v_odd}--\eqref{eq:DA_v_even} shows that all terms $A_v$, apart from $J^z_{H;v}$, are similarly even with respect to $\beta v$. Thus, their series expansions at large $T$ contain only even powers of $v$ and, therefore, the series terminates.
We give below these results:
\begin{subequations}\label{eq:largeT}
\begin{align}
 \frac{FC}{M} &= \frac{{\rm FC}_{\rm cl}}{M}  + \frac{3\boldsymbol{\omega}_I^2 - \boldsymbol{a}_I^2}{24 \pi^2},\\
 J^t_V &= \gamma_I \left(Q^{\rm cl}_V + \frac{3\boldsymbol{\omega}_I^2 + \boldsymbol{a}_I^2}{12 \pi ^2}\mu_\rho \right),\\
 J^\varphi_V &= u^\varphi_I \left(Q^{\rm cl}_V + \frac{\boldsymbol{\omega}_I^2 + 3\boldsymbol{a}_I^2}{12 \pi ^2} \mu_\rho \right),\\
 J^z_A &= i \gamma_I^2 \Omega_I \left(\sigma^\omega_{A;{\rm cl}} + \frac{\boldsymbol{\omega}_I^2 + 3\boldsymbol{a}_I^2}{24 \pi ^2}\right),\\
% \left(J^z_H\right)^{i \Omega_I}
% & = i \gamma_I^2 \Omega_I \frac{2 \mu T}{\pi^2} \log 2,\\
 T^{tt} &= (4\gamma_I^2 - 1) P_{\rm cl} \nonumber\\
 &- \frac{3\gamma_I^2 \Omega_I^2}{4} \left(\frac{1}{9} -\frac{16\gamma_I^2}{9} + \frac{8\gamma_I^4}{3}\right) \frac{{\rm FC}_{\rm cl}}{M} \nonumber\\
 & +\frac{\gamma_I^4 \Omega_I^4}{64\pi^2} \left(\frac{17}{45} + \frac{196 \gamma_I^2}{45} - \frac{376 \gamma_I^4}{15} + \frac{64 \gamma_I^6}{3}\right),\\
 T^{\rho\rho} & = P_{\rm cl} + \frac{3\boldsymbol{\omega}_I^2 + \mathbf{a}_I^2}{12} \frac{{\rm FC}_{\rm cl}}{M} \nonumber\\
 &- \frac{\gamma_I^4 \Omega_I^4}{192\pi^2}
 %\left(\boldsymbol{\omega}_I^4 + \frac{122}{15} \boldsymbol{\omega}_I^2 \mathbf{a}_I^2 - \frac{17}{15} \mathbf{a}_I^4\right), \\
 \left(\frac{17}{15} + \frac{88 \gamma_I^2}{15} - 8 \gamma_I^4\right),\\
 T^{\varphi\varphi} &= \frac{4\gamma_I^2 - 3}{\rho^2} P_{\rm cl} - \frac{\gamma_I^2 \Omega_I^2}{4\rho^2}(8\gamma_I^4 - 8\gamma_I^2 + 1) \frac{{\rm FC}_{\rm cl}}{M} \nonumber\\
 &+ \frac{\gamma_I^4 \Omega_{I}^4}{192 \pi^2\rho^2} \left(\frac{17}{5} + \frac{124 \gamma_I^2}{5} - \frac{456}{5} \gamma_I^4 + 64 \gamma_I^6\right),\\
 T^{t\varphi} & = 4 i \Omega_I \gamma_I^2 \left[P_{\rm cl} - \frac{\gamma_I^2 \Omega_I^2}{3} \frac{{\rm FC}_{\rm cl}}{M} \left(\frac{3}{2} \gamma_I^2 -\frac{1}{2}\right) \right.\nonumber\\
 & \left. + \frac{31 \gamma_I^4 \Omega_I^4}{960 \pi^2} \left(\frac{15}{31} - \frac{64\gamma_I^2}{31} + \frac{80 \gamma_I^4}{31}\right)\right],
\end{align}
\end{subequations}
where we used the notations
\begin{gather}
 \frac{{\rm FC}_{\rm cl}}{M} = \sigma^\omega_{A;{\rm cl}} = \frac{T_\rho^2}{6} + \frac{\mu_\rho^2}{2 \pi^2}, \quad
 Q^{\rm cl}_V = \frac{\mu_\rho T_\rho^2}{3} + \frac{\mu_\rho^3}{3 \pi ^2},\nonumber\\
 P_{\rm cl} = \frac{7\pi^2 T_\rho^4}{180} + \frac{\mu_\rho^2 T_\rho^2}{6} + \frac{\mu_\rho^4}{12\pi^2}.
 \label{eq:largeT_notation}
\end{gather}
The results in Eqs.~\eqref{eq:largeT} coincide in the large temperature limit with the relativistic kinetic theory results presented in Sec.~\ref{sec:RKT}. Besides these classical terms, the QFT results shown above exhibit quantum corrections, proportional to $\Omega_I^2$ or $\Omega_I^4$, similar to the on-axis limits discussed in Sec.~\ref{sec:OI:axis}.

The results obtained in Eqs.~\eqref{eq:largeT} match those obtained by performing the calculation under real rotation, as described in Refs.~\cite{Ambrus:2014uqa,Ambrus:2019ayb,Ambrus:2019khr,Ambrus:2021eod,Palermo:2021hlf}. While this agreement seems to further support the possibility of an analytical continuation from imaginary to real rotation, we first remark that the results in Eqs.~\eqref{eq:largeT} do not exhibit the expected periodicity with respect to $\nu \to \nu \pm 2$, or equivalently, $\Omega_I \to \Omega_I \pm 4\pi / \beta$. Therefore, even though the series expansion seemingly provides the full, exact result, it fails to recover this essential, topological behavior. Furthermore, we will show in the next section that the properties of the trigonometric functions appearing in Eqs.~\eqref{eq:DA_v_odd}--\eqref{eq:DA_v_even} lead to the fractalization of the expectation values at large distances from the rotation axis.

Before ending this section, we return to the analysis of the helical current, for which
\begin{equation}
 J^z_{H;v} = \frac{2i \Omega_I \gamma_I^2}{v} T_\rho \mu_\rho + O(\beta v),
\end{equation}
where the series is continued by both even and odd positive powers of $v$. Therefore, the summation with respect to $v$ yields an infinite series, from which we retain just the leading-order contribution:
\begin{equation}
 J^z_H = i \Omega_I \gamma_I^2 \left[\frac{2 \ln 2}{\pi^2} \mu_\rho T_\rho + O(T_{\rho}^{-1})\right].
\end{equation}
Again, the above result agrees with the large temperature expansion of the helicity current derived in Refs.~\cite{Ambrus:2019ayb,Ambrus:2019khr}.

\section{Emergence of fractal structure} \label{sec:fractal}

In sections~\ref{sec:OI:largeT}, we analyzed the expectation values of the fermion condensate, charge currents and energy-momentum tensor in the limit of large temperature. Although our procedure seemed to produce the exact results shown in Eqs.~\eqref{eq:largeT}--\eqref{eq:largeT_notation}, we argued that they were incorrect, as the topologically-required periodicity with respect to $\nu \to \nu \pm 2$ was not preserved. One may artificially impose this periodicity by enforcing $-1 \le \nu < 1$, however this step is insufficient. As we will show in this section, the intrinsic periodicity apparent in Eqs.~\eqref{eq:DA_v_odd}--\eqref{eq:DA_v_even} has far deeper consequences, leading to the emergence of a fractal-like behaviour of all expectation values, visible at large distances from the rotation axis.

\subsection{Summation split due to harmonic periodicity}\label{sec:fractal:split}

The fractal structure appears when $\nu = \beta \Omega_I / 2 \pi$ is considered a rational number, represented by an irreducible fraction, i.e. $\nu = p / q$ for integer $p$ and $q$ that are co-prime (with common greatest divisor $1$). Writing $v = r + qQ$, with $1 \le r \le q$ and $0 \le Q < \infty$, we see that the trigonometric functions $c_v$ and $s_v$ evaluate to
\begin{align}
 s_v &\to s_{r + qQ} = (-1)^{pQ} s_r, &
 s_r = \sin\left(\frac{\pi p r}{q}\right), \nonumber\\
 c_v &\to c_{r + qQ} = (-1)^{pQ} c_r, &
 c_r = \cos\left(\frac{\pi p r}{q}\right),
 \label{eq:fractal_sv_cv}
\end{align}
and thus the dependence on $Q$ is carried only by the $(-1)^{pQ}$ factor. Similarly, $\alpha_v = \frac{2\rho}{v \beta} s_v$ is replaced by
\begin{equation}
 \alpha_v \to \alpha_{r + qQ} = \frac{(-1)^{pQ}}{Q + \frac{r}{q}} x_r, \quad
 x_r = \frac{l s_r}{\pi q},
 \label{eq:shift_properties}
\end{equation}
with $l = 2\pi \rho/ \beta$. The functions $s^{v\mu} = \sin(v \beta \mu \alpha_v)$ and $c^{v\mu} = \cos(v \beta \mu \alpha_v)$ are replaced by
\begin{align}
 s^{v\mu} &\to s^{r+qQ,\mu} = (-1)^{pQ} s^{r\mu}, &
 s^{r\mu} &= \sin(q \beta\mu x_r),\nonumber\\
 c^{v\mu} &\to c^{r+qQ,\mu} = c^{r\mu}, &
 c^{r\mu} &= \cos(q \beta\mu x_r).
 \label{eq:def_srmu}
\end{align}

The series with respect to $v$ in Eq.~\eqref{eq:DA_v} is decomposed as $\sum_{v=1}^{\infty} \equiv \sum_{r + q Q=1}^{\infty} \equiv \sum_{Q=0}^{\infty} \sum_{r=1}^q$. Employing the shift properties \eqref{eq:shift_properties} and interchanging the sum over $r$ with the series over $Q$
leads to\footnote{This is allowed when the series over $v$ is absolutely convergent. We take special care in the case when the series is not absolutely convergent, as described in Appendix~\ref{subsec:P}.}
\begin{align}
 \Delta A &= \sum_{r=1}^q \frac{(-1)^{r + 1}}{\pi^2} \sum_{Q = 0}^\infty (-1)^{kQ} A_{r,Q},\nonumber\\
 A_{r,Q} &= (-1)^{pQ} A_{v = r + qQ},
 \label{eq:fractal_notation}
\end{align}
where we denoted $k = p + q$.
The coefficients $A_{r,Q}$ corresponding to the observables in Eqs.~\eqref{eq:DA_v_odd}--\eqref{eq:DA_v_even} read
\begin{subequations}\label{eq:DA_rQ}
\begin{align}
 J^{\varphi}_{V;r,Q} &= -\frac{4 i s_r}{q^4 \beta^4 x_r}
 \frac{d}{dx_r} \left[\frac{s^{r\mu}}{x_r}\frac{Q + \frac{r}{q}}{(Q + \frac{r}{q})^2 + x_r^2}\right], \\
 J^{z}_{H;r,Q} &= -\frac{i \sigma_\mu}{\rho q^2 \beta^2} \frac{d}{dx_r} \left[\frac{(Q + \frac{r}{q}) (c^{r\mu} - e^{-\beta|\mu|(qQ + r)})}{(Q + \frac{r}{q})^2 + x_r^2}\right], \label{eq:DA_rQ_JHz}\\
 J^{z}_{A;r,Q} &= -\frac{i}{\rho q^2 \beta^2} \frac{d}{dx_r} \left(\frac{c^{r\mu} (Q + \frac{r}{q})}{(Q + \frac{r}{q})^2 + x_r^2}\right), \\
 T^{t\varphi}_{r,Q} &= \frac{i(1 + c_r^2)}{\rho q^4 \beta^4} \frac{d}{dx_r} \frac{1}{x_r} \frac{d}{dx_r} \frac{c^{r\mu}(Q + \frac{r}{q})}{(Q + \frac{r}{q})^2 + x_r^2}, \\
 \frac{{\rm FC}_{r,Q}}{M} &\to \frac{2 c_r c^{r\mu}}{q^2 \beta^2[(Q + \frac{r}{q})^2+ x_r^2]}, \\
 J^t_{V;r,Q} &= \frac{2c_r}{q^3\beta^3 x_r} \frac{d}{dx_r} \frac{x_r s^{r\mu}}{(Q + \frac{r}{q})^2 + x_r^2}, \\
 T^{tt}_{r,Q} &= -\frac{2 c_r}{q^4\beta^4 x_r} \frac{d^2}{dx_r^2} \frac{x_r c^{r\mu}}{(Q + \frac{r}{q})^2 + x_r^2}, \\
 T^{\rho\rho}_{r,Q} &= -\frac{2c_r}{q^4 \beta^4 x_r} \frac{d}{dx_r} \left(\frac{c^{r\mu}}{(Q + \frac{r}{q})^2 + x_r^2}\right), \\
 T^{\varphi\varphi}_{r,Q} &= -\frac{2 c_r}{\rho^2 q^4 \beta^4} \frac{d^2}{dx_r^2} \frac{c^{r\mu}}{(Q + \frac{r}{q})^2 + x_r^2}.
\end{align}
\end{subequations}
From the above terms, the vertical component of the helical current $J^z_H$ deserves special treatment, due to the exponential term $e^{-q \beta |\mu|(Q + \frac{r}{q})}$, and will be addressed at the end of this section.

It is easy to notice that when $r = q$, all coordinate dependence entering Eqs.~\eqref{eq:DA_rQ} through the function $x_r$ disappears, as $x_{r = q} = 0$. It is thus convenient to express the expectation value of the observable $A$ as
\begin{align}
A &= A_q + \sum_{r = 1}^{q-1} \frac{(-1)^{r+1}}{\pi^2}\delta A_r, \nonumber\\
 A_q &= A_{\rm deg.} + \sum_{Q = 0}^\infty \frac{(-1)^{kQ + q + 1}}{\pi^2} A_{q,Q}, \nonumber\\
 \delta A_r &= \sum_{Q = 0}^\infty (-1)^{kQ} A_{r,Q},
 \label{eq:Aq_def}
\end{align}
where $A_q$ contains the coordinate-independent contribution to $A$, due to the degenerate part found in Eq.~\eqref{eq:deg}, as well as due to the fractal part arising when $r = q$. We will perform the summation with respect to $Q$ and discuss the contributions $A_q$ and $\delta A_r$ in Subsections~\ref{sec:fractal:q} and \ref{sec:fractal:dA}, respectively.

We note that the coordinate-independent contribution made by the $r = q$ terms in the sum over $v = r + qQ$ appears only when $\nu = p/q$ is a rational number. When $\nu$ is irrational, the argument $v \beta \Omega_I / 2 = v \pi \nu$ of $s_v$ will never be a multiple of $\pi$ (though it can become arbitrarily close to one). On the other hand, the form of the coefficients $A_v$ in Eqs.~\eqref{eq:DA_v_odd}--\eqref{eq:DA_v_even} indicates that at large $\rho$, and hence large $\alpha_v = (2\rho / v\beta) s_v$, all non-degenerate contributions $\Delta A$ decay to $0$. This is consistent with taking the limit $q \to \infty$ in Eqs.~\eqref{eq:fractal_q}, discussed in the next subsection, i.e. by considering that an irrational number can be represented as a fraction $\mathfrak{p}/\mathfrak{q}$ with an infinite denominator $\mathfrak{q}$. An illustration of the decay of the observables when $\nu$ is irrational is provided in Fig.~14 of Ref.~\cite{Ambrus:2023bid} for the related case of a scalar field under imaginary rotation.

\subsection{Asymptotic terms}\label{sec:fractal:q}

%As the cases when $k = p + q$ is an odd or an even quantity correspond to different solution branches, we will treat them separately in what follows.

As mentioned in the previous subsection, the $l = 2\pi \rho/ \beta$ dependence of the observables considered in this paper disappears in the case when $r = q$ and $x_r = 0$. Concretely, the expressions in Eqs.~\eqref{eq:DA_rQ} reduce when $r = q$ to
\begin{gather}
 J^\varphi_{V;q,Q} = J^z_{H;q,Q} = J^z_{A;q,Q} = T^{t\varphi}_{q,Q} = 0, \nonumber\\
 \frac{{\rm FC}_{q,Q}}{M} \to \frac{2(-1)^p}{q^2 \beta^2 (Q+1)^2}, \quad
 J^t_{V;q,Q} = \frac{4\mu (-1)^p}{q^2 \beta^2 (Q+1)^2}, \nonumber\\
 T^{tt}_{q,Q} = \frac{12(-1)^p}{q^4 \beta^4 (Q+1)^4} + \frac{6\mu^2(-1)^p}{q^2 \beta^2 (Q+1)^2}.
 \label{eq:DA_qQ}
\end{gather}
while $T^{\rho\rho}_{q,Q} = \rho^2 T^{\varphi\varphi}_{q,Q} = T^{zz}_{q,Q} = \frac{1}{3} T^{tt}_{q,Q}$. It is clear that the above terms are independent of coordinates. Plugging the above into $A_q$ introduced in Eq.~\eqref{eq:Aq_def}, the sum over $Q$ can be performed using the relations
\begin{align}
 \sum_{Q = 1}^\infty \frac{1}{Q^2} &= \frac{\pi^2}{6}, &
 \sum_{Q = 1}^\infty \frac{(-1)^{Q+1}}{Q^2} &= \frac{\pi^2}{12}, \nonumber\\
 \sum_{Q = 1}^\infty \frac{1}{Q^4} &= \frac{\pi^4}{90}, &
 \sum_{Q = 1}^\infty \frac{(-1)^{Q+1}}{Q^4} &= \frac{7\pi^4}{720}.
 \label{eq:sumQ_q}
\end{align}
The contributions $A_q$ can be obtained by adding the results in Eq.~\eqref{eq:deg} to those corresponding to Eqs.~\eqref{eq:DA_qQ} and \eqref{eq:sumQ_q}. It can be seen that
\begin{equation}
 J^\varphi_{V;q} = J^z_{H;q} = J^z_{A;q} = T^{t\varphi}_q = 0.
\end{equation}
When $k = p + q$ is odd, we find
\begin{subequations}\label{eq:fractal_q}
\begin{gather}
 \frac{{\rm FC}_q}{M} = \frac{1}{6q^2\beta^2} + \frac{\mu^2}{2\pi^2}, \quad
 J^t_{V;q} = \frac{\mu}{3q^2\beta^2} + \frac{\mu^3}{3\pi^2},\nonumber\\
 T^{tt}_q = \frac{7\pi^2}{60q^4\beta^4} + \frac{\mu^2}{2q^2\beta^2} +
 \frac{\mu^4}{4\pi^2},\label{eq:fractal_q_odd}
\end{gather}
while for even $k = p + q$, we get
\begin{gather}
 \frac{{\rm FC}_q}{M} = -\frac{1}{3q^2\beta^2} + \frac{\mu^2}{2\pi^2}, \quad
 J^t_{V;q} = -\frac{2\mu}{3q^2\beta^2} + \frac{\mu^3}{3\pi^2},\nonumber\\
 T^{tt}_q = -\frac{2\pi^2}{15q^4\beta^4} - \frac{\mu^2}{q^2\beta^2} +
 \frac{\mu^4}{4\pi^2},\label{eq:fractal_q_even}
\end{gather}
\end{subequations}
while $T^{\rho\rho}_q = \rho^2 T^{\varphi\varphi}_q = T^{zz}_q = \frac{1}{3} T^{tt}_q$ in both cases considered above.

The asymptotic terms obtained when $k = p + q$ is odd are consistent with a stationary state at effective inverse temperature $\beta_q = q \beta$ displaying fractal features. The factor $q$ represents the denominator when $\nu = p/q$ is represented as an irreducible fraction and therefore varies wildly for small changes in $\nu$. For example, when $\nu = 0.5 = 1/2$, $q = 2$, while for $\nu = 0.51 = 51/100$, $q = 100$. Contrary to the na\"ive expectation, the chemical potential describing this asymptotic state is the same as that of the original state. In other words, the chemical potential breaks the thermodynamic fractalization.
While the results for even $k$ display similar fractal properties, they do not correspond to a static system, as all thermal contributions come with a negative sign.

\begin{figure*}
\begin{tabular}{c}
\includegraphics[width=0.9\textwidth]{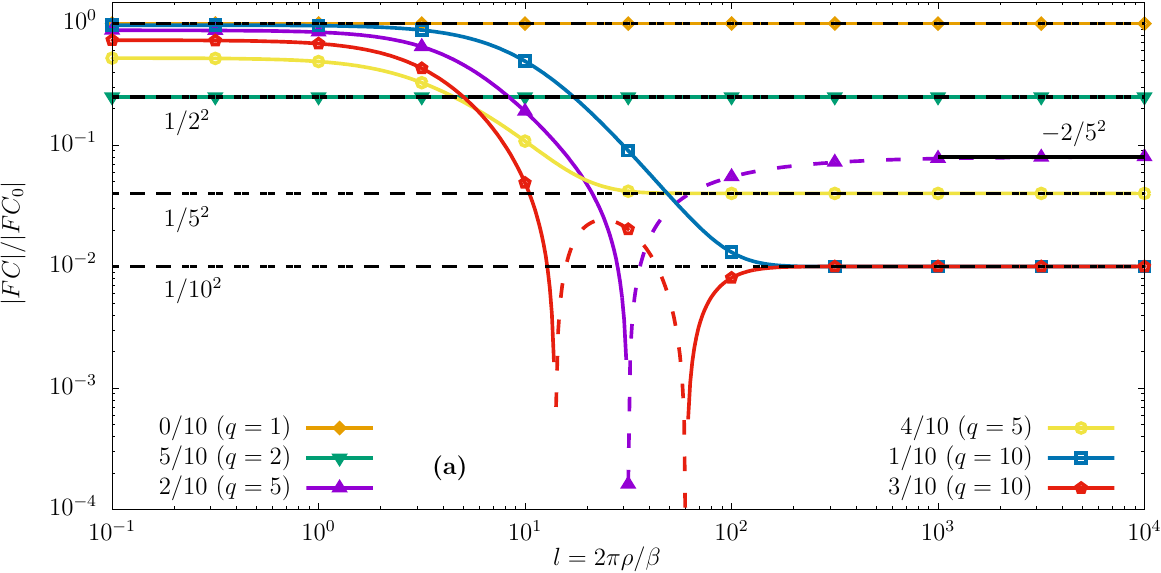}\\
\includegraphics[width=0.9\textwidth]{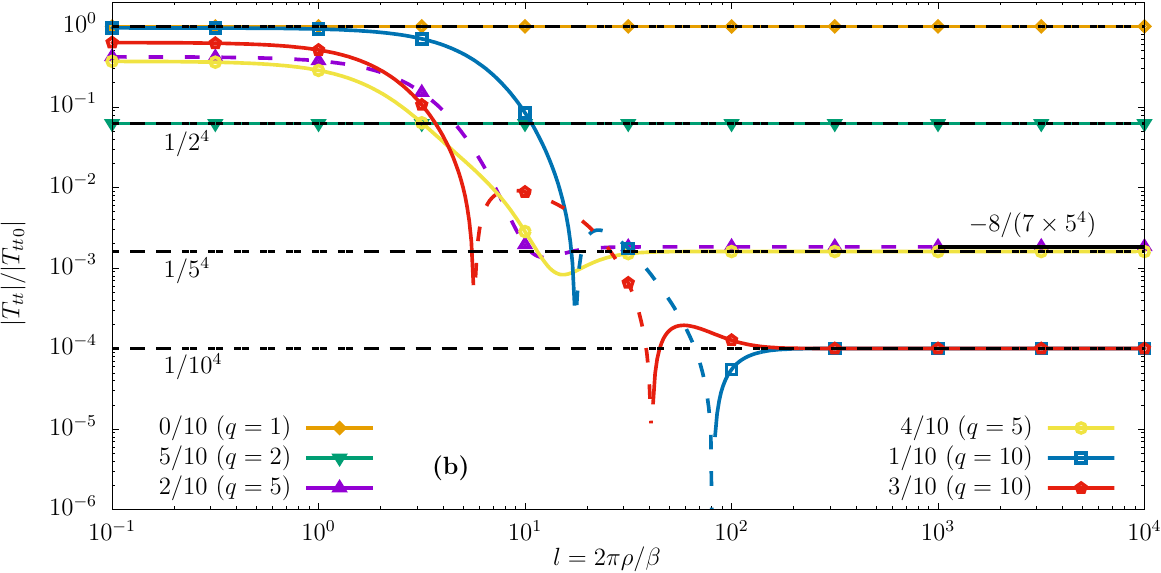}
\end{tabular}
\caption{The absolute values of \textbf{(a)} FC and \textbf{(b)} $T^{tt}$, evaluated using Eqs.~\eqref{eq:Aq_def}, \eqref{eq:fractal_q} and \eqref{eq:fractal_transient}, at vanishing chemical potential ($\mu = 0$), represented as functions of $l = 2\pi \rho/ \beta$, normalized with respect to their corresponding values at vanishing rotation. The legend indicates different rational values of the frequency $\nu = p' / 10 = p/q$, with $p/q$ being an irreducible fraction. The dashed segments of the result curves indicate negative expectation values. The horizontal dashed black lines indicate the large-$l$ asymptotes, given by the fractal terms shown in Eqs.~\eqref{eq:fractal_q} (the negative asymptotic values corresponding to even $k = p + q$ are shown with a horizontal solid black line).
\label{fig:fractal}
}
\end{figure*}

\subsection{Transient terms}\label{sec:fractal:dA}

When $1 \le r < q$, we employ the notations
\begin{subequations}\label{eq:sumQ}
\begin{align}
 \mathcal{Q}^r_k(x_r) &= \sum_{Q = 0}^\infty \frac{(-1)^{kQ}}{(Q + \frac{r}{q})^2 + x_r^2}, \label{eq:sumQ_Q}\\
 \mathcal{P}^r_k(x_r) &= \sum_{Q = 0}^\infty \frac{(-1)^{kQ} (Q + \frac{r}{q})}{(Q + \frac{r}{q})^2 + x_r^2}. \label{eq:sumQ_P}
\end{align}
\end{subequations}
For the remainder of this section, we will always employ the above functions with the argument $x_r$, i.e. $\mathcal{Q}^r_k \equiv \mathcal{Q}^r_k(x_r)$ and $\mathcal{P}^r_k \equiv \mathcal{P}^r_k(x_r)$. For brevity, we shall not display the argument explicitly in what follows.

The summations over $Q$ can be performed as explained in Appendix~\ref{app:psi}. Here we mention that the series with respect to $Q$ appearing in the expression for $\mathcal{P}^r_k$ is not absolutely convergent. Moreover, when $k = p+q$ is an even number, the series diverges logarithmically. The summation can be nevertheless performed as explained in Appendix~\ref{app:psi} and the results are
\begin{gather}
 \mathcal{Q}^r_{\rm odd} = -\frac{\Delta \psi_r^{\mathfrak{i}}}{2x_r}, \quad
 \mathcal{Q}^r_{\rm even} = \frac{1}{x_r} \psi^{\mathfrak{i}}_r, \quad
 \mathcal{P}^r_{\rm odd} = \frac{1}{2} \Delta \psi_r^{\mathfrak{r}},
 \nonumber\\
 \mathcal{P}^r_{\rm even} \to
 \frac{\pi}{2} \frac{\sin(2\pi r / q)}{\cosh(2\pi x_r) - \cos(2\pi r / q)},
 \label{eq:QP}
\end{gather}
 where the superscripts $\mathfrak{i}$, $\mathfrak{r}$ represent the imaginary and real parts of the functions $\psi_r$ and $\Delta \psi_r$, defined as:
\begin{align}
\psi_r &= \psi\left(\frac{r}{q} + i x_r\right),\nonumber\\
\Delta \psi_r &= \psi \left(\frac{\frac{r}{q}+i x_r+1}{2}\right)-\psi \left(\frac{\frac{r}{q}+i x_r}{2}\right),
\label{eq:delta_psi_r}
\end{align}
with $\psi(z) = \Gamma'(z) / \Gamma(z)$ being the polygamma function. The arrow $\to$ in the expression for $\mathcal{P}^r_{\rm even}$ indicates that the result is obtained by following the summation ordering prescription from Appendix~\ref{subsec:P_even}.

With respect to the above notation, we obtain
\begin{subequations}\label{eq:fractal_transient}
\begin{align}
 \delta J^\varphi_{V;r} &= -\frac{4i s_r}{q^4\beta^4 x_r} \frac{d}{dx_r} \left(\frac{s^{r\mu}}{x_r} \mathcal{P}^r_k\right), \\
 \delta J^z_{A;r} &= -\frac{i}{\rho q^2\beta^2} \frac{d(c^{r\mu} \mathcal{P}^r_k)}{dx_r},\label{eq:fractal_transient_JAz}\\
 \delta T^{t\varphi}_r &= \frac{i(1 + c_r^2)}{\rho q^4\beta^4} \frac{d}{dx_r} \left[\frac{1}{x_r} \frac{d(c^{r\mu} \mathcal{P}^r_k)}{dx_r} \right],\\
 \frac{\delta {\rm FC}_r}{M} &= \frac{2 c_r}{q^2 \beta^2} c^{r\mu} \mathcal{Q}^r_k, \\
 \delta J^t_{V;r} &= \frac{2c_r}{q^3 \beta^3 x_r} \frac{d}{dx_r}(x_r s^{r\mu} \mathcal{Q}^r_k),\\
 \label{eq: delta_Ttt_r}
 \delta T^{tt}_r &= -\frac{2c_r}{q^4 \beta^4 x_r} \frac{d^2}{dx_r^2}(x_r c^{r\mu} \mathcal{Q}^r_k),\\
 \delta T^{\rho\rho}_r &= -\frac{2 c_r}{q^4 \beta^4 x_r} \frac{d(c^{r\mu} \mathcal{Q}^r_k)}{dx_r}, \\
 \delta T^{\varphi\varphi}_r &= -\frac{2 c_r}{\rho^2 q^4 \beta^4} \frac{d^2 (c^{r\mu} \mathcal{Q}^r_k)}{dx_r^2}.
\end{align}
\end{subequations}
%########

Figure~\ref{fig:fractal} shows how the fermion condensate (panel a) and $T_{tt}$ (panel b) transition from their values on the rotation axis, which are consistent with an analytic continuation to the case of real rotation, towards their fractal asymptotes, given in Eqs.~\eqref{eq:fractal_q} and indicated with the horizontal black lines (the dashed and solid curves indicate positive and negative asymptotic values, respectively).

We now turn to the evaluation of the helicity current, $J^z_H$. Taking the $r = q$ limit in Eq.~\eqref{eq:DA_rQ_JHz}, it is not difficult to see that $J^z_{H;q,Q} = 0$. Since $J^z_{H;{\rm deg.}} = 0$, we have $J^z_{H;q} = 0$ and altogether,
\begin{align}
 J^z_H &= \sum_{r = 1}^{q-1} \frac{(-1)^{r+1}}{\pi^2} \delta J^z_{H;r}, \nonumber\\
 \delta J^z_{H;r} &=
 -\frac{i \sigma_\mu}{\rho q^2 \beta^2} \frac{d}{dx_r} \left[c^{r\mu} \mathcal{P}_k^r -
 \mathcal{B}_k^r(q\beta|\mu|;x_r)\right],
 \label{eq:fractal_transient_JHz}
\end{align}
where we introduced the notation
\begin{equation}
 \mathcal{B}_k^r(\zeta;x_r) = \sum_{Q = 0}^\infty
 (-1)^{kQ} \frac{e^{-\zeta (Q + \frac{r}{q})} (Q + \frac{r}{q})}{(Q + \frac{r}{q})^2 + x_r^2}.
 \label{eq:Bkr_def}
\end{equation}
At $\zeta > 0$, the above series is absolutely convergent. Using $a / (a^2 + b^2) = \frac{1}{2}[(a + ib)^{-1} + (a - ib)^{-1}]$, we can rewrite the above expression as
\begin{align}
 \mathcal{B}^r_k(\zeta;x_r) &= {\rm Re}\left[e^{-\zeta \frac{r}{q}} \Phi\left((-1)^k e^{-\zeta}, 1, \frac{r}{q} + i x_r\right)\right],
 \label{eq:Bkr_Phi}
\end{align}
with $\Phi(z,s,a) = \sum_{n = 0}^\infty z^n / (a + n)^s$ being Lerch's transcendent, defined for $|z| < 1$, ${\rm Re}(s) > 1$ and $a \neq 0, \pm 1, \pm 2, \dots$. The result can be expressed in terms of Gauss' hypergeometric function, ${}_2F_1(a,b;c;z) = \sum_{n = 0}^\infty z^n [(a)_n (b)_n / (c)_n]$, with $(a)_n = \Gamma(a + n) / \Gamma(a)$ being the Pochhammer symbol. For the particular case of $s = 1$, we have
\begin{equation}
 \Phi(z,1,a) = a^{-1} {}_2F_1(1, a; a+1; z).
\end{equation}

% TUDOR
\section{Fractal thermodynamics}\label{sec:thermo}

The grand canonical potential for a thermodynamic system of fermions under imaginary rotation is defined by
\[
  \Phi (\beta, \mu, \Omega_I) = - \beta^{-1} \log \mathcal{Z} (\beta, \mu, \Omega_I),
\]
where $\mathcal{Z} (\beta, \mu, \Omega_I) = \operatorname{tr} \{ \rho  (\beta, \mu, \Omega_I) \}$, with the trace running over the Fock space and the density operator being
\[
  \rho  (\beta, \mu, \Omega_I) = e^{- \beta (H - i \Omega_I J^z - \mu Q)}.
\]

In the macrocanonical ensemble, the energy $\mathcal{E} = \mathcal{Z}^{-1} {\rm tr}(\rho H)$ of a thermodynamic system of rotating fermions is given by the expectation value
\begin{equation}
\label{eq: average_energy}
\mathcal{E}
%= \frac{\operatorname{tr} \{ \rho H \}}{\operatorname{tr} \{ \rho \}}
= \frac{1}{\mathcal{Z}} \left(- \frac{\partial \mathcal{Z}}{\partial \beta}\right)_{\beta \Omega_I, \beta \mu} = \left[\frac{\partial \left(\beta \Phi\right)}{\partial \beta}\right]_{\beta \Omega_I, \beta \mu}.
\end{equation}
We construct the average energy $\om{E} = \mathcal{E} / V$ using the component $T^{tt}$ of the energy-momentum tensor, averaged over a fictitious cylinder of radius $R$ and length $L_z$ (volume $V = \pi R^2 L_z$):
\begin{equation}
\label{eq: spatial_average_energy}
\bar{\mathcal{E}} = \frac{1}{V} \int d V T^{tt} = \frac{2}{R^2} \int_0^R d \rho \rho T^{tt},
\end{equation}
allowing us to obtain the averaged thermodynamic potential,
\begin{equation}
 \overline{\Phi} = \frac{1}{\beta} \int d\beta (\om{E})_{\beta\Omega_I, \beta \mu},
\end{equation}
where the integration is performed while keeping $\beta \Omega_I$ and $\beta \mu$ as constants.

\subsection{Thermodynamic quantities}\label{sec:thermo:sumv}

Using the notation in Eqs.~\eqref{eq:decomposition_gen} and \eqref{eq:DA_v}, we find
\begin{align}
 \om{E} = \om{E}_{\rm deg.} + \Delta \om{E}, \quad
 \Delta \om{E} = \sum_{v = 1}^{\infty} \frac{(-1)^{v+1}}{\pi^2} \om{E}_v,
\end{align}
where $\om{E}_{\rm deg.} = \mu^4 / 4\pi^2$, by virtue of Eq.~\eqref{eq:deg}. In order to compute $\om{E}_v$, we insert Eq.~\eqref{eq: Ttt_v} into Eq.~\eqref{eq: spatial_average_energy} and change the integration variable to $\alpha_v = 2\rho s_v / v \beta$. Introducing the notation
\begin{equation}
 A_v = \alpha_v(R) = \frac{2R}{v \beta} s_v,
\end{equation}
we have $\frac{2}{R^2} d\rho \, \rho \to \frac{2}{A_v^2} d\alpha_v\, \alpha_v$ and the integral becomes trivially
\begin{equation}
 \om{E}_v = -\frac{4c_v}{v^4 \beta^4 A_v^2} \left[\frac{d}{dA_v} \left(\frac{A_v C^{v\mu}}{1 + A_v^2}\right) - 1\right],
\end{equation}
where $C^{v\mu} = \cos(v \beta \mu A_v)$ and we define, for later convenience, $S^{v\mu} = \sin(v \beta \mu A_v)$.

We are now in a position to compute $\overline{\Phi} = \overline{\Phi}_{\rm deg.} + \Delta \overline{\Phi}$. The former contribution is trivially
\begin{equation}
 \overline{\Phi}_{\rm deg.} = \frac{1}{\beta} \int d\beta (\om{E}_{\rm deg.})_{\beta \mu,\beta \Omega_I} = -\frac{\mu^4}{12\pi^2}.
\end{equation}
Taking into account that $\beta A_v$ depends on $\beta$ only through $\beta \mu$ and $\beta \Omega_I$, we can obtain $\overline{\Phi}_v$ as
\begin{align}
 \overline{\Phi}_v &= -\frac{4c_v}{v^4 \beta^3 A_v^2} \int \frac{d\beta}{\beta^2} \left[\frac{d}{dA_v}\left(\frac{A_v C^{v\mu}}{1 + A_v^2}\right) - 1 \right] \nonumber\\
 &= \frac{4c_v}{v^4 \beta^4 A_v^2} \left(\frac{C^{v\mu}}{1 + A_v^2} - 1\right),
\end{align}
where we employed the relation $dA_v = -(\beta A_v)d\beta / \beta^2$.

For a homogeneous system, the thermodynamic potential is equal to the negative of the system pressure. In the case of the rotating system, the transverse and parallel directions are not equivalent, such that the corresponding pressures have different values:
\begin{equation}
 \mathcal{P} dV \to 2\pi R L_z \mathcal{P}_\perp dR + \pi R^2 \mathcal{P}_z d L_z.
\end{equation}
The transverse pressure $\mathcal{P}_\perp = -\frac{1}{2R} \partial(R^2 \overline{\Phi}) / \partial R = \mathcal{P}_\perp^{\rm deg.} + \Delta \mathcal{P}_\perp$ has $\mathcal{P}_\perp^{\rm deg} = -\overline{\Phi}_{\rm deg.} = \mu^4 / 12\pi^2$, whereas
\begin{equation}
 \mathcal{P}_{\perp;v} = -\frac{2 c_v}{v^4 \beta^4 A_v} \frac{d}{dA_v} \left(\frac{C^{v\mu}}{1 + A_v^2}\right),
\end{equation}
while the vertical pressure is simply $\mathcal{P}_z = -\frac{1}{L_z} \partial (L_z \overline{\Phi}) / \partial L_z = -\overline{\Phi}$. It can be checked that the ultrarelativistic equation of state $\om{E} = 3 \mathcal{P}$ is satisfied, where $\mathcal{P} = \frac{2}{3} \mathcal{P}_\perp + \frac{1}{3} \mathcal{P}_z$ represents the average pressure.

The average charge $\om{Q} = - \partial \overline{\Phi} / \partial \mu = \om{Q}_{\rm deg.} + \Delta \om{Q}$ can be obtained as
\begin{equation}
 \om{Q}_{\rm deg.} = \frac{\mu^3}{3\pi^2}, \quad  \om{Q}_v = \frac{4 c_v S^{v\mu}}{v^3 \beta^3 A_v(1 + A_v^2)}.
\end{equation}
The average angular momentum $\om{M}_I = -\partial \overline{\Phi} / \partial \Omega_I$ has no degenerate contribution, $\om{M}_I^{\rm deg.} = 0$, while
\begin{multline}
 \om{M}_{I;v} = \frac{1}{R v^2 \beta^2 A_v} \left(\frac{C^{v\mu}}{1 + A_v^2} - 1\right) \\
 - \frac{4 c_v^2 R}{v^4 \beta^4} \frac{d}{dA_v} \left[\frac{1}{A_v^2} \left(\frac{C^{v\mu}}{1 + A_v^2} - 1\right)\right].
\end{multline}
It can be checked that the average entropy, $\om{S} = \beta^2 \partial \overline{\Phi} / \partial \beta$, satisfies the Euler relation, Eq.~\eqref{eq:Euler_Phi}.

\subsection{Fractalization properties}\label{sec:thermo:fractal}

\begin{figure*}[h]
\centering
\begin{tabular}{c}
\includegraphics[width=0.9\textwidth]{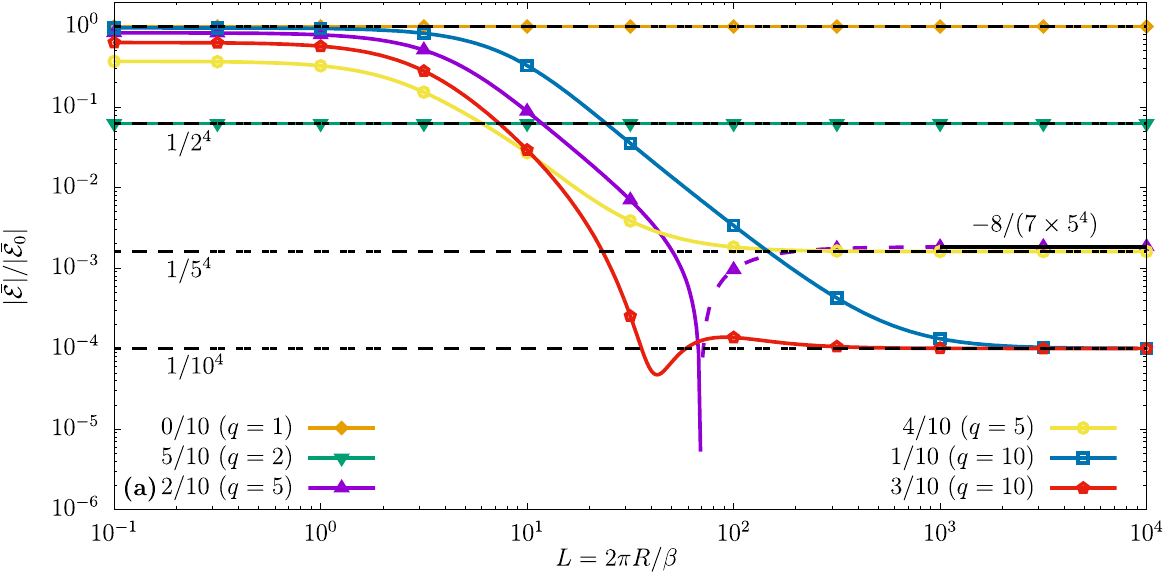}\\
\includegraphics[width=0.9\textwidth]{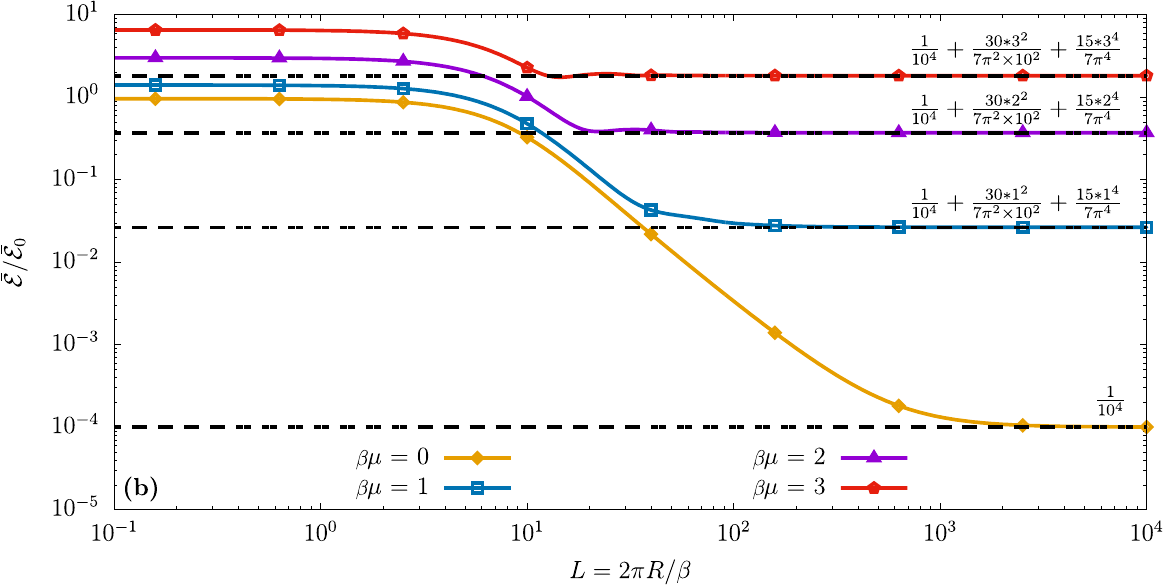}
\end{tabular}{c}
\caption{The average energy $\om{E}$ at \textbf{(a)} vanishing chemical potential $\mu = 0$ and various rotaion parameters $\nu = p'/10$, corresponding to irreducible fractions $p/q$ with $q \in \{1,2,5,10\}$; and \textbf{(b)} at $\nu = 1/10$ with various values of the chemical potential, as a function of $L = 2 \pi R / \beta$, where $R$ is the radius of the cylinder. The normalization is performed with respect to the energy density of non-rotating, neutral fermions: $\om{E}_0 = 7\pi^2 / 60\beta^4$. In panel (a), the dashed portions of the result curves correspond to negative values of $\om{E}$. The horizontal black dashed lines indicate the asymptotic values $\om{E}_q$ achieved at large $L$, computed using Eqs.~\eqref{eq:thermo_q_odd_E} and \eqref{eq:thermo_q_even_E}.
\label{fig:avg_E}
}
\end{figure*}

As discussed in Sec.~\ref{sec:fractal:split}, due to the periodicity of the harmonic functions $c_v$ and $s_v$, it is convenient to rewrite the summation parameter as $v = r Q + q$, where we consider $\nu = \beta \Omega_I / 2\pi = p / q$ as an irreducible fraction, while the summation over $v$ is replaced as $\sum_{v = 1}^\infty f_v \to \sum_{r = 1}^q \sum_{Q = 0}^\infty f_{v = rQ + q}$. Employing the relations in Eqs.~\eqref{eq:fractal_sv_cv}, \eqref{eq:shift_properties} and \eqref{eq:def_srmu}, we see that under the replacement $v \to r + pQ$, we have
\begin{align}
 A_v &\to \frac{(-1)^{pQ}}{Q +\frac{r}{q}} X_r, & X_r &\equiv x_r(R) = \frac{Ls_r}{\pi q},\nonumber\\
 S^{v\mu} &\to (-1)^{pQ} S^{r\mu}, &
 S^{r\mu} &\equiv s^{r\mu}(R) = \sin(q \beta \mu X_r), \nonumber\\
 C^{v\mu} &\to C^{r\mu}, &
 C^{r\mu} &\equiv c^{r\mu}(R) = \cos(q \beta \mu X_r),
 \label{eq:thermo_fractal_r}
\end{align}
where $s_r = \sin(\pi p r / q)$, as defined in Eq.~\eqref{eq:fractal_sv_cv}.

As discussed in Sec.~\ref{sec:fractal:q}, the terms corresponding to $r = q$ are ``asymptotic terms,'' as they are independent of system size. For an average thermodynamic quantity $\om{A} = \om{A}_{\rm deg.} + \sum_{v = 1}^\infty \frac{(-1)^{v+1}}{\pi^2} \om{A}_v$, we employ the separation in Eq.~\eqref{eq:fractal_notation} and write:
\begin{equation}
 \Delta \om{A} = \sum_{r = 1}^q  \frac{(-1)^{r+1}}{\pi^2} \sum_{Q = 0}^\infty (-1)^{kQ} \om{A}_{r,Q},
\end{equation}
where $\om{A}_{r,Q} = (-1)^{pQ} \om{A}_{v = r + qQ}$.
The $\om{A}_{r,Q}$ terms corresponding to the thermodynamic quantities considered in Subsec.~\ref{sec:thermo:sumv} read:
\begin{subequations}
\begin{align}
 \om{E}_{r,Q} &= -\frac{4c_r}{q^4 \beta^4 X_r^2} \left[\frac{d}{dX_r} \left(\frac{X_r C^{r\mu}}{(Q + \frac{r}{q})^2 + X_r^2}\right) - \frac{1}{(Q + \frac{r}{q})^2}\right], \\
 \overline{\Phi}_{r,Q} &= \frac{4c_r}{q^4 \beta^4 X_r^2} \left[\frac{C^{r\mu}}{(Q + \frac{r}{q})^2 + X_r^2} - \frac{1}{(Q + \frac{r}{q})^2}\right], \\
 \mathcal{P}^{\perp}_{r,Q} &= -\frac{2c_r}{q^4 \beta^4 X_r} \frac{d}{dX_r} \left[\frac{C^{r\mu}}{(Q + \frac{r}{q})^2 + X_r^2}\right],\\
 \om{Q}_{r,Q} &= \frac{4 c_r S^{r\mu}}{q^3 \beta^3 X_r} \frac{1}{(Q + \frac{r}{q})^2 + X_r^2},\\
 \om{M}^I_{r,Q} &= \frac{1}{R q^2 \beta^2 X_r} \left[\frac{C^{r\mu}(Q + \frac{r}{q})}{(Q + \frac{r}{q})^2 + X_r^2} - \frac{1}{Q + \frac{r}{q}}\right] \nonumber\\
 & - \frac{4 c_r^2 R}{q^4 \beta^4} \frac{d}{dX_r} \left[\frac{1}{X_r^2} \left(\frac{C^{r\mu}(Q + \frac{r}{q})}{(Q + \frac{r}{q})^2 + X_r^2} - \frac{1}{Q + \frac{r}{q}}\right)\right].
\end{align}
\end{subequations}

\subsection{Asymptotic terms} \label{eq:thermo:q}

We then employ Eq.~\eqref{eq:Aq_def} to group the degenerate contribution $\om{A}_{\rm deg.}$ and the asymptotic term into a system size-independent quantity $\om{A}_q$, such that
\begin{align}
 \om{A} &= \mathcal{A}_q + \sum_{r = 1}^{q - 1} \frac{(-1)^{r+1}}{\pi^2} \delta A_r, \nonumber\\
 \om{A}_q &= \om{A}_{\rm deg.} + \sum_{Q = 0}^\infty \frac{(-1)^{kQ + q + 1}}{\pi^2} \om{A}_{q,Q}, \nonumber\\
 \delta \om{A}_r &= \sum_{Q = 0}^\infty (-1)^{kQ} \om{A}_{r,Q}.
 \label{eq:thermo_fractal}
\end{align}
Noting that
\begin{gather}
 \om{E}_{q,Q} = \frac{12(-1)^p}{q^4 \beta^4 (Q+1)^4} + \frac{6\mu^2(-1)^p}{q^2 \beta^2 (Q+1)^2}, \nonumber\\
 \om{Q}_{q,Q} = \frac{4\mu(-1)^p}{q^2 \beta^2 (Q + 1)^2},
\end{gather}
while $\mathcal{P}^{\perp}_{q,Q} = \mathcal{P}^z_{q,Q} = -\overline{\Phi}_{q,Q} = \om{E}_{q,Q} / 3$, $\om{M}^I_{q,Q} = 0$ and $\om{S}_{q,Q} = \frac{4}{3} \beta \om{E}_{q,Q} - \beta \mu \om{Q}_{q,Q}$.
Inserting the above back into Eq.~\eqref{eq:thermo_fractal} and performing the summation over $Q$ using Eqs.~\eqref{eq:sumQ_q}, the system size-independent terms can be readily evaluated. Clearly, $\om{M}_{I;q} = 0$, while for the other quantities, we must distinguish between the cases when $k = p + q$ is odd or even, as discussed in Sec.~\ref{sec:fractal:q}. In the former case, when $k$ is odd, we have
\begin{subequations}\label{eq:thermo_q_odd}
\begin{align}
 \om{Q}_q &= \frac{\mu}{3q^2 \beta^2} + \frac{\mu^3}{3\pi^2}, \\
 \om{E}_q &= \frac{7\pi^2}{60q^4 \beta^4} + \frac{\mu^2}{2q^2 \beta^2} +
 \frac{\mu^4}{4\pi^2},\\
 \om{S}_q &= \frac{7\pi^2}{45 q^4 \beta^3} + \frac{\mu^2}{3q^2 \beta},
 \label{eq:thermo_q_odd_E}
\end{align}
\end{subequations}
while for even $k = p + q$, we get
\begin{subequations}\label{eq:thermo_q_even}
\begin{align}
 \om{Q}_q &= -\frac{2\mu}{3q^2\beta^2} + \frac{\mu^3}{3\pi^2}, \\
 \om{E}_q &= -\frac{2\pi^2}{15q^4 \beta^4} - \frac{\mu^2}{q^2 \beta^2} +
 \frac{\mu^4}{4\pi^2}, \\
 \om{S}_q &= -\frac{8\pi^2}{45q^4 \beta^3} - \frac{2\mu^2}{3q^2 \beta}.
\label{eq:thermo_q_even_E}
\end{align}
\end{subequations}
In both cases, the pressures and thermodynamic potential satisfy $\mathcal{P}_{\perp;q} = \mathcal{P}_{z;q} = -\overline{\Phi}_q = \om{E}_q / 3 \equiv \mathcal{P}_q$.

For odd $k = p + q$, the asymptotic average energy $\om{E}_q$, average charge $\om{Q}_q$, and pressure $\mathcal{P}_q = \mathcal{P}_{\perp;q} = \mathcal{P}_{z;q}$ are identical to those of a static fermion ensemble with effective temperature $T_q = T / q$, depending only on the denominator of the irreducible fraction $\nu = p / q$. As seen in the case of the asymptotic contributions to the local expectation values, computed in Sec.~\ref{sec:fractal:q}, the chemical potential $\mu$ does not fractalize, its contribution being independent of the rotation parameter.

In the case of even $k = p + q$, the thermal (fractal) contributions to $\om{E}_q$ and $\om{Q}_q$ are negative. At large $q \beta \mu$, when the degenerate chemical potential terms dominate, $\om{E}_q$ and $\om{Q}_q$ become positive, exceeding $0$ when $q \beta \mu > \pi \sqrt{2 + 2\sqrt{17/15}}$ and $q \beta \mu > \pi \sqrt{2}$, respectively.

We finally observe that the average entropy $\om{S}_q$ is a factor of $q$ smaller than the entropy corresponding to a static fermionic ensemble of temperature $T_q = T /q$ and chemical potential $\mu$, owing to its definition based on the temperature defining the grand canonical ensemble: $\om{S}_q = \beta^2 \partial \overline{\Phi}_q / \partial \beta$. One may introduce an effective entropy, $\om{S}_q^{\rm eff} = q \om{S}_q = \beta_q^2 \partial \overline{\Phi}_q / \partial \beta_q$, based on the effective temperature, which would indeed correspond to the entropy of the aforementioned static system. Since $\om{S}_q^{\rm eff} > \om{S}_q$, we can infer that the system under imaginary rotation does not represent a state of thermodynamic equilibrium, its entropy being lower than that of the equivalent equilibrium system having the same energy and charge density.

\begin{figure*}
\centering
\includegraphics[width=0.9\textwidth]{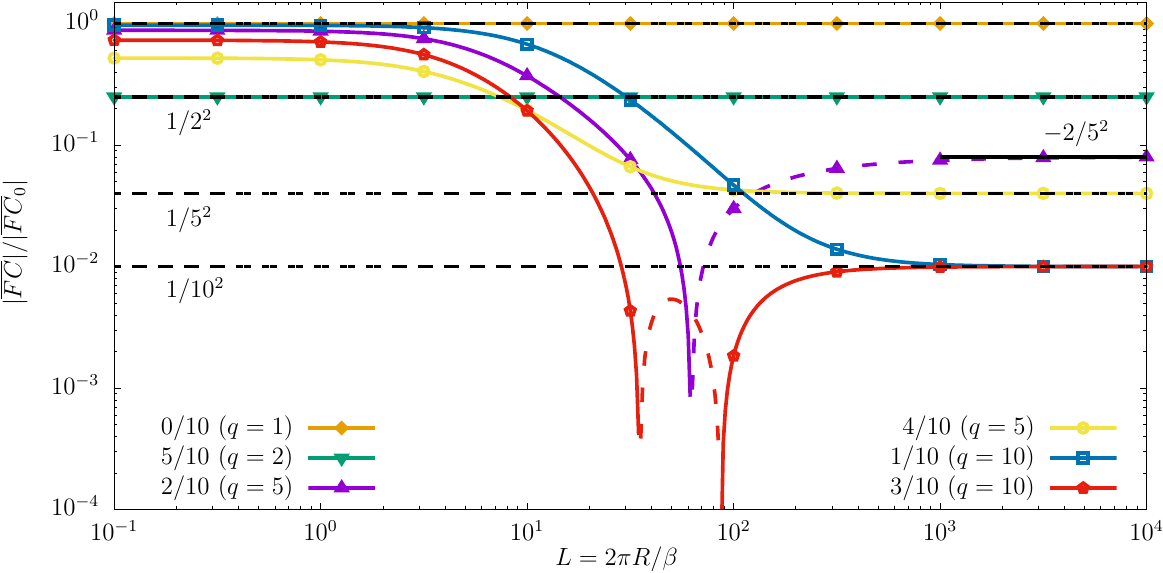}
\caption{
\label{fig:avg_FC}
The average fermion condensate $\overline{\rm FC}$ as a function of $L = 2\pi R / \beta$, computed for a neutral plasma ($\mu = 0$), normalized with respect to its non-rotating value, $\overline{\rm FC}_0 = M T^2 / 6$, for various rotation frequencies $\nu = p' / 10$, corresponding to irreducible fractions $p / q$ with $q \in \{1,2,5,10\}$. The dashed portions of the result curves indicate negative values of ${\rm FC}$. The horizontal black dashed lines indicate the asymptotic values ${\rm FC}_q$ taken at large $L$, corresponding to Eqs.~\eqref{eq:thermo_FC_q}.
}
\end{figure*}

\subsection{Transient terms}\label{sec:thermo:dA}

As discussed in Sec.~\ref{sec:fractal:dA}, the terms with $1 \le r < q$ make contributions that vanish as the distance to the rotation axis is increased. The same behavior will prevail for volume-averaged quantities. Performing the summation over $Q$ with respect to the functions $\mathcal{Q}^r_k$ and $\mathcal{P}^r_k$ introduced in Eqs.~\eqref{eq:sumQ},
we arrive at
\begin{subequations}
\begin{align}
 \delta \om{E}_r &= -\frac{4c_r}{q^4 \beta^4 X_r^2} \left[\frac{d(X_r C^{r\mu} \mathcal{Q}_k^r)}{dX_r} - \mathcal{Q}_k^r(0)\right], \\
 \delta \overline{\Phi}_r &= \frac{4c_r}{q^4 \beta^4 X_r^2} \left[C^{r\mu} \mathcal{Q}_k^r(X_r) - \mathcal{Q}_k^r(0)\right], \\
 \delta \mathcal{P}^{\perp}_r &= -\frac{2c_r}{q^4 \beta^4 X_r} \frac{d(C^{r\mu} \mathcal{Q}_k^r)}{dX_r},\\
 \delta \om{Q}_r &= \frac{4 c_r S^{r\mu}}{q^3 \beta^3 X_r} \mathcal{Q}_k^r,\\
 \delta \om{M}^I_r &= \frac{1}{R q^2 \beta^2 X_r} \left[C^{r\mu} \mathcal{P}_k^r(X_r) - \mathcal{P}_k^r(0)\right] \nonumber\\
 & - \frac{4 c_r^2 R}{q^4 \beta^4} \frac{d}{dX_r} \left[\frac{C^{r\mu} \mathcal{P}_k^r(X_r) - \mathcal{P}_k^r(0)}{X_r^2}\right].
\end{align}
\end{subequations}

To illustrate the behavior of the averaged thermodynamic quantities considered above, we show in Fig.~\ref{fig:avg_E} the average energy $\om{E}$ at (a) vanishing chemical potential and various rotation parameters; and (b) $\nu = 1/10$ and various chemical potentials. Starting from the value of $T^{tt}$ on the rotation axis, given in Eq.~\eqref{eq:axis_Ttt}, $\om{E}$ goes through a transient phase as the size $L$ of the averaging cylinder is increased, eventually tending to the value $\om{E}_q$ shown in Eqs.~\eqref{eq:thermo_q_odd_E} and \eqref{eq:thermo_q_even_E}. Panel (a) shows that for the neutral plasma with $\mu= 0$, the asymptotic value depends only on the denominator $q$ of the irreducible fraction $\nu = p /q$ representing the rotation parameter. This fractal-like behavior, reminiscent of the Thomae function, indicates the impossibility of analytically continuing results obtained in the thermodynamic limit of the system undergoing imaginary rotation to the case of real rotation. On the other hand, panel (b) demonstrates that the insensitivity of the asymptotic chemical potential terms to the value of $q$, as well as $\nu$, can lead to asymptotic values that are completely dominated by the degenerate term $\mu^4 / 4\pi^2$.

\subsection{Average fermion condensate}

In case of the fermion condensate, we start with the expression
\begin{equation}
 {\rm FC} = {\rm FC}_q + \sum_{r = 1}^{q - 1} \frac{(-1)^{r+1}}{\pi^2} \delta {\rm FC}_r,
\end{equation}
with ${\rm FC}_q / M$ and $\delta {\rm FC}_r / M$ given in Eqs.~\eqref{eq:fractal_q} and \eqref{eq:fractal_transient}, respectively. Performing the volume average, we obtain
\begin{equation}
 \overline{\rm FC} = \overline{\rm FC}_q + \sum_{r = 1}^{q - 1} \frac{(-1)^{r+1}}{\pi^2} \delta \overline{\rm FC}_r,
\end{equation}
where we have, for the cases of odd and even $k= p + q$, the following results:
\begin{subequations}\label{eq:thermo_FC_q}
\begin{align}
 \text{odd $k$:}& & \frac{\overline{\rm FC}_q}{M} &= \frac{T^2}{6q^2} + \frac{\mu^2}{2\pi^2}, \\
 \text{even $k$:}& & \frac{\overline{\rm FC}_q}{M} &= -\frac{T^2}{3q^2} + \frac{\mu^2}{2\pi^2},
\end{align}
\end{subequations}
while the transient term reads
\begin{align}
 \frac{\delta \overline{\rm FC}_r}{M} &= \frac{4 c_r}{\beta^2 q^2 X_r^2} \int_0^{X_r} dx_r \, x_r \cos \left(q \beta \mu x_r\right) \mathcal{Q}^r_k(x_r),
\end{align}
where $\mathcal{Q}^r_k(x_r)$ is given in Eqs.~\eqref{eq:QP}.
We illustrate the behavior of the average fermion condensate at vanishing chemical potential in Fig.~\ref{fig:avg_FC}. Starting from the value on the rotation axis, given in Eq.~\eqref{eq:axis_FC}, the average fermion condensate goes through a transient region where its value changes drastically as the dimensionless size $L$ of the averaging cylinder is increased, asymptoting to its expected value $\overline{\rm FC}_q$, given in Eqs.~\eqref{eq:thermo_FC_q}, as $L \rightarrow \infty$.

% DARIANA

\section{Polarization flux}\label{sec:flux}

In this section, we discuss the fate of the vortical effects that manifest as polarization currents along the rotation angular velocity vector. In this section, we will focus on computing the flux of the axial and helical currents through a disk of radius $R$, centered on the rotation axis:
\begin{align}
 \mathcal{F}_{A/H}(R) &= \int_{\rho < R} d \boldsymbol{\Sigma} \cdot \mathbf{J}_{A/H} = 2\pi \int_0^R d\rho \, \rho\, J^{z}_{A/H}.
 \label{eq:flux_def}
\end{align}
We proceed directly to considering the case of rational rotation parameter, represented as an irreducible fraction $\nu = p/q$, when
\begin{equation}
 J^z_{A/H} = \sum_{r = 1}^{q-1} \frac{(-1)^{r+1}}{\pi^2} \delta J^z_{A/H;r},
 \label{eq:JAHz_aux}
\end{equation}
where $\delta J^z_{A/H;r}$ can be found in Eqs.~\eqref{eq:fractal_transient_JAz} and \eqref{eq:fractal_transient_JHz}. Substituting Eq.~\eqref{eq:JAHz_aux} into Eq.~\eqref{eq:flux_def} gives
\begin{align}
 \mathcal{F}^{(p,q)}_{A/H}(R) &= \sum_{r = 1}^{q-1} \frac{(-1)^{r+1}}{\pi^2} \delta \mathcal{F}^{(p,q)}_{A/H;r},\nonumber\\
 \delta \mathcal{F}^{(p,q)}_{A/H;r} &= 2\pi \int_0^R d\rho\, \rho\, \delta J^z_{A/H;r},
 \label{eq:flux}
\end{align}
where we have added the $(p,q)$ superscript to indicate that the calculation is performed at rational rotation parameter, $\nu=\beta \Omega_I / 2\pi = p / q$.

\subsection{Axial flux}

\begin{figure*}[h!]
\centering
\begin{tabular}{c}
\includegraphics[width=0.9\textwidth]{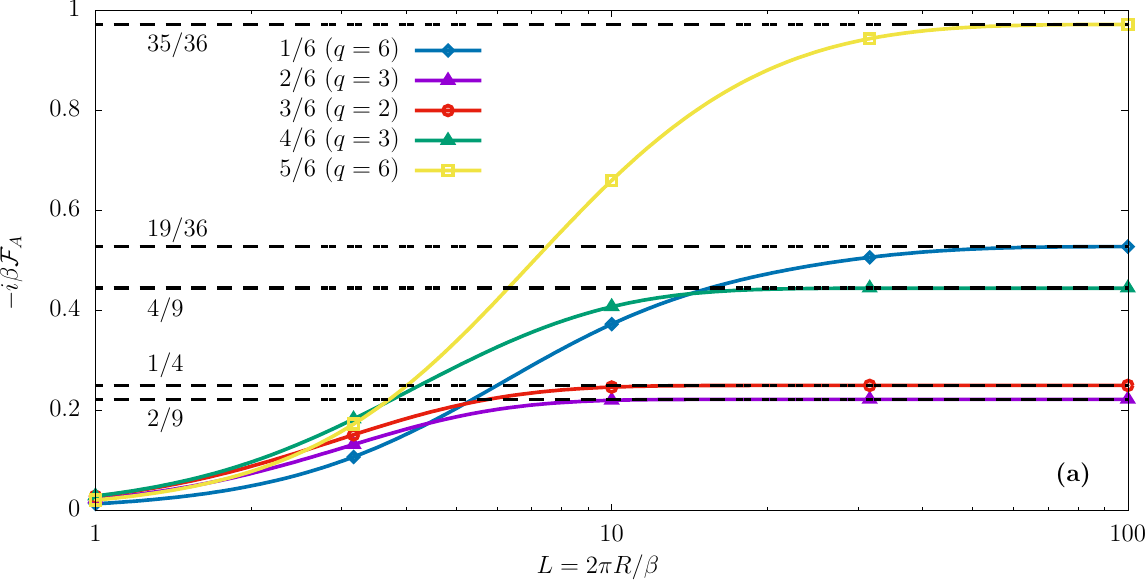} \\
\includegraphics[width=0.9\textwidth]{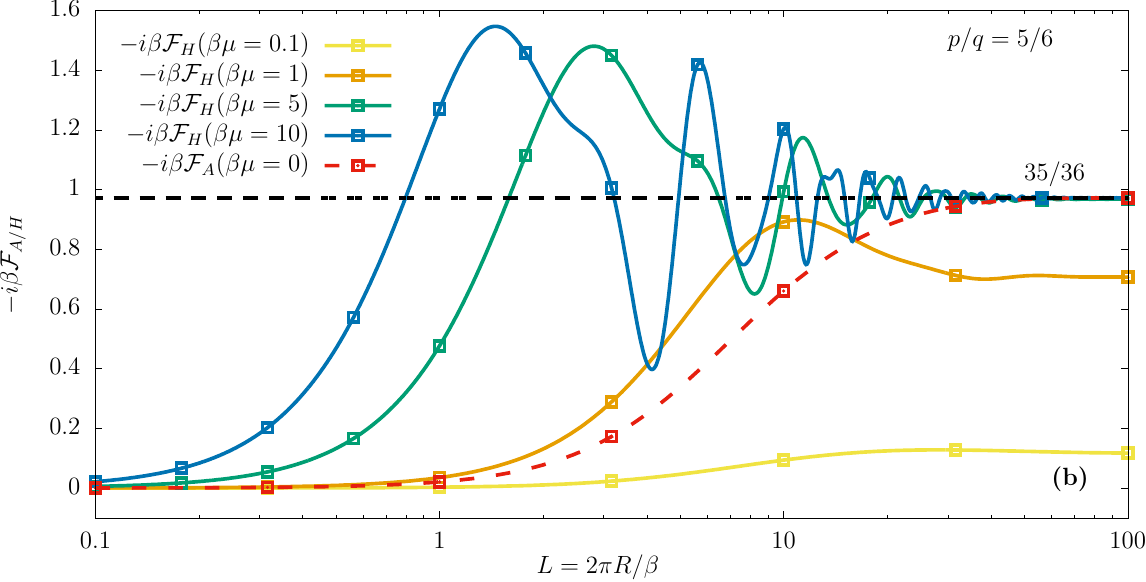}
\end{tabular}
\caption{Polarization fluxes through a disk of dimensionless radius $L = 2\pi R / \beta$. \textbf{(a)} Axial flux at vanishing chemical potential for various values of the rotation parameter, $\nu = p' / 6= p/q$, such that $q \in \{2,3,6\}$. \textbf{(b)} Helical flux at $\nu = p/q=5/6$, for various values of the dimensionless chemical potential, $\beta \mu \in \{0.1,1,5,10\}$. The dashed black lines denote the large-size asymptotes, given in Eqs.~\eqref{eq:FA_inf} and \eqref{eq:FH_inf}.
At large $\beta \mu$, the large-$L$ asymptote of the helical flux saturates, converging towards the asymptote of the axial flux, which is displayed for reference using the red dashed line in panel (b).
}
\label{fig:flux}
\end{figure*}

\begin{figure*}
\centering
\includegraphics[width=0.9\textwidth]{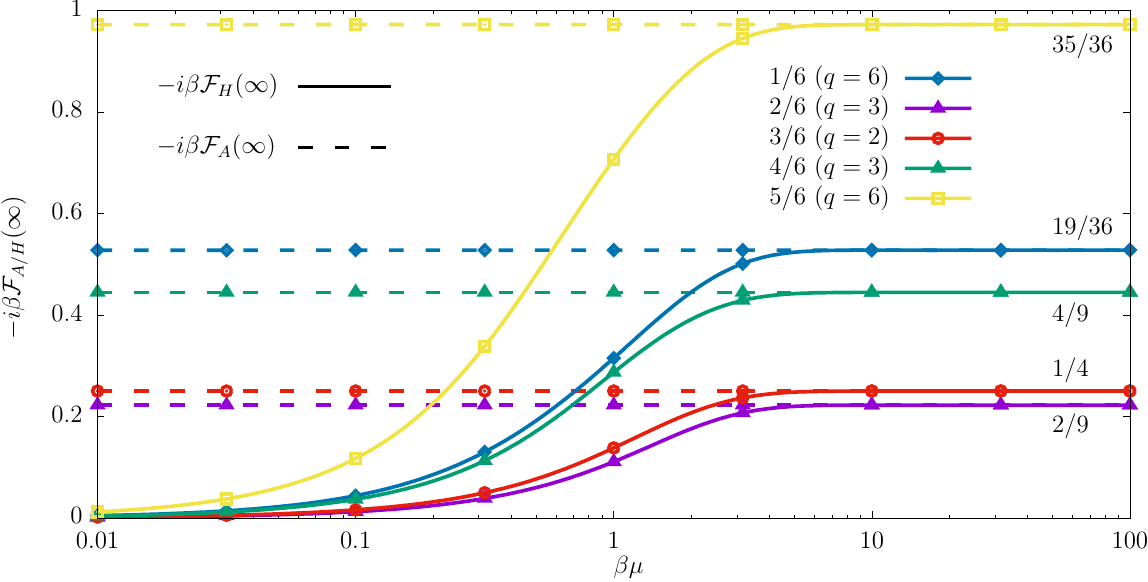}
\caption{The large-$L$ asymptotic values of the axial (horizontal dotted lines) and helical (solid lines) fluxes, computed using Eqs.~\eqref{eq:FA_inf} and \eqref{eq:FH_inf}, respectively, represented with respect to the dimensionless chemical potential, $\beta \mu$, for various values of the rotation parameter, $\nu = p'/6$, corresponding to irreducible fractions $p/q$ with $q \in \{2, 3, 6\}$.
}
\label{fig:flux_as}
\end{figure*}

Substituting Eq.~\eqref{eq:fractal_transient_JAz} into Eq.~\eqref{eq:flux}, we obtain:
\begin{align}
 \delta \mathcal{F}^{(p,q)}_{A;r} &= 2\pi \int_0^R d\rho\, \rho \left(-\frac{i}{\rho q^2 \beta^2}\right) \frac{d}{dx_r} \left(c^{r\mu} \mathcal{P}^r_k\right) \nonumber\\
 &= -\frac{\pi i}{q \beta s_r} \left[C^{r\mu} \mathcal{P}^r_k(X_r) - \mathcal{P}^r_k(0)\right],
 \label{eq:dFAr}
\end{align}
where we used the property $d\rho = (R / X_r) dx_r$, while $\mathcal{P}^r_k$ and $C^{r\mu} = \cos(q \beta \mu X_r)$ were introduced in Eqs.~\eqref{eq:sumQ} and \eqref{eq:thermo_fractal_r}, respectively.

In the case when $R \rightarrow \infty$, or equivalently $X_r \rightarrow \infty$,
%it can be seen that $\mathcal{P}^r_k(X_r) = \sum_{Q = 0}^\infty (-1)^{kQ} (Q + \frac{r}{q}) / [(Q + \frac{r}{q})^2 +X_r^2] \rightarrow 0$,
the first term in Eq.~\eqref{eq:dFAr} vanishes,
such that $\mathcal{F}^{(p,q)}_A$ asymptotes to
\begin{equation}
 \mathcal{F}^{(p,q)}_A(\infty) = \frac{i}{\pi q\beta} \sum_{r = 1}^{q-1} \frac{(-1)^{r + 1}}{s_r} \mathcal{P}^r_k(0).
\end{equation}
It is remarkable that the above expression is independent of the chemical potential $\mu$. The dependence on $\mu$ is fully contained in the $C^{r\mu} \mathcal{P}^r_k(X_r)$ term in Eq.~\eqref{eq:dFAr}, which has an oscillatory behavior due to the harmonic function $C^{r\mu} = \cos(q\beta\mu X_r) = \cos(2\mu R s_r)$.

In the case when $k = p+q$ is odd, we have
\begin{subequations}\label{eq:FA_inf}
\begin{equation}
 \mathcal{F}^{(p,q)}_A(\infty) = \frac{i}{2\pi q \beta} \sum_{r=1}^{q - 1} \frac{(-1)^{r+1}}{s_{r}} \left[ \psi \left(\frac{1+ \frac{r}{q}}{2}\right) - \psi\left(\frac{r}{2q}\right)\right],
\end{equation}
while for even $k=p + q$, $\mathcal{F}^{(p,q)}_A(\infty)$ becomes
\begin{equation}
 \mathcal{F}^{(p,q)}_A(\infty) = \frac{i}{2 q \beta} \sum_{r=1}^{q - 1} \frac{(-1)^{r+1}}{s_{r}} \cot\left(\frac{\pi r}{q}\right).
\end{equation}
\end{subequations}
It is instructive to evaluate $\mathcal{F}^{(p,q)}_A(\infty)$ for the special case $p = 1$. By direct computation, we find
\begin{equation}
 \mathcal{F}^{(1,q)}_A(\infty) =
% \frac{i}{\beta}
% \begin{cases}
     % \displaystyle \frac{q^2 + 2}{12q}, & q \text{ is even ($k$ is odd)}, \smallskip\\
     % \displaystyle \frac{q^2 - 1}{12q}, & q \text{ is odd ($k$ is even)}.
% \end{cases}
 \frac{i}{12 q\beta} \left(q^2 + \frac{1}{2} - \frac{3}{2} (-1)^k\right).
 \label{eq:FA_p1}
\end{equation}
Unexpectedly, the magnitude of the axial flux increases linearly with $q$ when $\nu = 1/q \to 0$. This behavior is seemingly in contradiction to our na\"ive expectation based on the vortical effects, that $J^z_A \sim \Omega_I$. One can trace this discrepancy to the non-commutativity of the $R \to \infty$ and $\nu \to 0$ limits of Eq.~\eqref{eq:dFAr}. It is noteworthy that the $\nu = 0$ limit implies $q = 1$ rather than $q \to \infty$, in which case Eq.~\eqref{eq:FA_p1} reproduces the correct limit, $\mathcal{F}^{(1,1)}_A(\infty) = 0$.

\subsection{Helical flux}

Substituting Eq.~\eqref{eq:fractal_transient_JHz} into Eq.~\eqref{eq:flux} leads to
\begin{multline}
 \delta \mathcal{F}^{(p,q)}_{H;r} = -\frac{\pi i \sigma_\mu}{q \beta s_r} [C^{r\mu} \mathcal{P}^r_k(X_r) - \mathcal{B}^r_k(q\beta|\mu|; X_r) \\
 - \mathcal{P}^r_k(0) + \mathcal{B}^r_k(q\beta|\mu|; 0)],
 \label{eq:dFHr}
\end{multline}
where $\mathcal{B}^r_k(\zeta; x_r)$ was introduced in Eqs.~\eqref{eq:Bkr_def} and \eqref{eq:Bkr_Phi}. We first note that, at vanishing chemical potential, $\mathcal{B}^r_k(0;x_r) \to \mathcal{P}^r_k(x_r)$, as can be seen by comparing Eqs.~\eqref{eq:sumQ_P} and \eqref{eq:Bkr_def}, such that $\mathcal{F}^{(p,q)}_{H}$ vanishes for a neutral plasma.

From the definition of $\mathcal{B}^r_k(\zeta;x_r)$ in Eq.~\eqref{eq:Bkr_def}, it can be seen that $\mathcal{B}^r_k(\zeta;X_r) \to 0$ when $X_r \to \infty$, such that the total helical flux through the transverse plane tends to
\begin{equation}
 \mathcal{F}^{(p,q)}_H(\infty) = \frac{i  \sigma_\mu}{\pi q\beta} \sum_{r = 1}^{q-1} \frac{(-1)^{r + 1}}{s_r} \left[\mathcal{P}^r_k(0) - \mathcal{B}^r_k(q\beta|\mu|; 0)\right].
\end{equation}

In the case when $k = p+q$ is odd, we have
\begin{subequations}\label{eq:FH_inf}
\begin{multline}
 \mathcal{F}^{(p,q)}_H(\infty) = \frac{i \sigma_\mu}{2 q \beta} \sum_{r=1}^{q - 1} \frac{(-1)^{r+1}}{\pi s_{r}} \left[ \psi \left(\frac{1+ \frac{r}{q}}{2}\right) - \psi\left(\frac{r}{2q}\right)\right. \\
 \left. - 2e^{-r \beta |\mu|} \Phi\left(-e^{-q \beta |\mu|}, 1, \frac{r}{q}\right)\right],
\end{multline}
whereas when $k=p + q$ is even, we have
\begin{multline}
 \mathcal{F}^{(p,q)}_H(\infty) = \frac{i \sigma_\mu}{2 q \beta} \sum_{r=1}^{q - 1} \frac{(-1)^{r+1}}{\pi s_{r}} \left[\pi \cot\left(\frac{\pi r}{q}\right) \right.\\
 \left. - 2 e^{-r \beta |\mu|} \Phi\left(e^{-q\beta|\mu|}, 1, \frac{r}{q}\right)\right].
\end{multline}
\end{subequations}

It can be seen that the helical flux \eqref{eq:dFHr} differs from the axial flux \eqref{eq:dFAr} by the term $\mathcal{B}^r_k(q\beta|\mu|;X_r) - \mathcal{B}^r_k(q\beta|\mu|;0)$. As can be seen from its definition in Eq.~\eqref{eq:Bkr_def}, this function is suppressed at large chemical potential by the factor $\sim e^{-\beta|\mu|}$. Therefore, $\mathcal{F}^{(p,q)}_H \to \sigma_\mu \mathcal{F}^{(p,q)}_A$ when $\beta |\mu|$ is very large. On the other hand, in the thermodynamic limit of large $R$, it can be seen that the asymptotic axial flux $\mathcal{F}^{(p,q)}_A(\infty)$ is independent of the chemical potential, providing an unambiguous limit for the helical flux when both the thermodynamic (large $R$) and dense (large $\mu$) limits are taken.

We illustrate the behavior of $-i\beta \mathcal{F}_A$ and $-i\beta \mathcal{F}_H$ with respect to increasing dimensionless system size $L = 2\pi R / \beta$ in Fig.~\ref{fig:flux}. Panel (a) shows the increase of the axial flux towards its asymptotic value $\mathcal{F}^{(p,q)}_A(\infty)$ at vanishing chemical potential, $\mu = 0$, for various rotation parameters $\nu = p' / 6$, corresponding to irreducible fractions $p/q$ with $q \in \{2,3, 6\}$ (when $q = 1$, $\mathcal{F}_A^{(p,1)} = 0$). The dependence on $L$ is smooth and monotonic, while the dependence on $\nu$ is non-monotonic. The red ($p/q = 1/2$), purple ($p/q = 1/3$) and blue ($p/q=1/6$) curves confirm Eq.~\eqref{eq:FA_p1}.

Panel (b) of Fig.~\ref{fig:flux} shows the effect of a finite chemical potential at the level of the helical flux, $\mathcal{F}^{(p,q)}_H$, for a fixed rotation parameter: $\nu = p/q = 5/6$. At $\beta \mu = 0$, $\mathcal{F}^{(p,q)}_H$ vanishes, as discussed at the beginning of the present subsection. As $\beta \mu$ is increased, the large-$L$ asymptote, $-i \beta \mathcal{F}^{(p,q)}_H(\infty)$, smoothly approaches $-i\beta \mathcal{F}^{(p,q)}_A(\infty)$. At finite $L$, an oscillatory behavior develops, the period of oscillations with respect to $L$ being roughly inversely-proportional to $\beta\mu$. While the figure focuses on the features of the helical flux, this transient behavior can be expected also for the axial flux, as in the large  $\beta \mu$ limit, the two observables coincide.

Finally, Fig.~\ref{fig:flux_as} shows how the large-$L$ asymptotic values of the helical flux converge towards those of the axial flux as $\beta \mu$ is increased.

\section{Conclusion} \label{sec:conc}

In this paper, we considered the properties of a finite-temperature ensemble of fermions, undergoing rigid rotation with an imaginary angular frequency. Our study is motivated by the fact that lattice simulations on the influence of rotation on the phase diagram of strongly-interacting matter are confined to the regime of imaginary rotation, due to the sign problem. The present paper complements a similar analysis done for the scalar field in Ref.~\cite{Ambrus:2023bid}.

The main result of our analysis is that a quantum system of fermions at finite temperature $\beta^{-1}$ and chemical potential $\mu$, undergoing rigid rotation with an imaginary angular frequency $\Omega_I$, exhibits fractal features in the thermodynamic limit of large volume. When the dimensionless rotation parameter $\nu = \beta \Omega_I / 2\pi$ is a rational number, represented as the irreducible fraction $\nu = p/q$, the expectation values tend at large distances from the rotation axis to those obtained in a stationary system at effective inverse temperature $\beta_q = q\beta$ and at the same chemical potential. The factor $q$ can vary wildly for small changes in $\nu$. For example, $\nu = 0.5 = 1/2$ has $q = 2$, while $\nu = 0.51 = 51/100$ gives $q = 100$.
Therefore, such states cannot be connected to states under real rotation by analytical continuation.
%For this reason, we can say that a system under imaginary rotation is\change{}{, in general,}{V} non-analytic, and the \change{extrapolation of the results obtained therein to the case of real rotation, e.g. via analytic continuation, is not reliable}{analytic continuation of the results obtained herein to the case of real rotation is not reliable and different procedures/method must be employed to describe a real system. One procedure suggested was to consider a system composed of two subsystems rotating counterclockwise, as to obtain real thermal expectation values. However, such a composite system is no longer in thermal equilibrium}{T}.

To better understand the context of the above result, we considered several limits. A classical (non-quantum) prediction can be obtained in the frame of relativistic kinetic theory (RKT). A system under rigid rotation is a global equilibrium state and therefore its thermodynamic state can be inferred by means of the Tolman-Ehrenfest law. With imaginary rotation, the local temperature and chemical potential scale as $T_\rho = \gamma_I / \beta$ and $\mu_\rho = \gamma_I \mu$, where the Lorentz factor $\gamma_I = 1/\sqrt{1 + \rho^2 \Omega_I^2}$ of an observer under imaginary rotation decreases as the distance $\rho$ to the rotation axis is increased.

In the quantum field theoretical (QFT) calculation, the quantization of the angular momentum leads to an important topological difference compared to the classical case: the system is periodic with respect to $\nu = \beta \Omega_I / 2\pi$ when $\nu \to \nu \pm 2$. In the limit of large temperature, the QFT calculation is in agreement with the RKT one (up to quantum corrections of order $\nu^2$ or higher), however the periodicity is not retained explicitly. Also in this limit, the QFT result can be analytically continued to real rotation, being in agreement with computations previously reported in the literature. Evaluating the QFT expressions on the rotation axis reveals the importance of these quantum corrections, which become dominant as $\nu$ is varied over its domain of periodicity.

Due to the nature of the system under imaginary rotation, the QFT calculation can be performed analytically to a large degree. One important feature is that, when $\nu = p/q$ is a rational number, the QFT result for an observable $A$ can be expressed as a ``fractal'', coordinate-independent term $A_q$, plus $q-1$ ``transient'' terms that interpolate as the dimensionless distance $l = 2\pi\rho/ \beta$ between the result on the rotation axis, $l = 0$, and the fractal term $A_q$, at large $l$. As already mentioned above, the fractal term corresponds to the expectation value of $A$ computed in a static system with inverse temperature $q \beta$, as also observed in the case of the scalar field~\cite{Ambrus:2023bid}, and at the same chemical potential. Thus, we can conclude that the chemical potential contribution breaks the fractalization.

The above discussion is valid when the quantity $k = p + q$, given in terms of the constituents of the irreducible fraction $\nu = p/q$, is odd. If $k$ is even, the fractal terms are vastly different: the thermal contributions (i.e., those that vanish when $\beta \to\infty$) are different, having the opposite sign compared to the case when $k$ is odd.

To further characterize fermionic systems under imaginary rotation, we constructed the thermodynamic functions under the grand canonical ensemble when considering the part of the system confined within a fictitious cylinder of finite radius $R$ and vertical extent $L_z$. No boundary conditions were imposed on this cylinder and therefore the thermodynamic system is not well-posed, however this issue dissapears when $R \to \infty$. We again found fractalization features that develop once the dimensionless radius of the cylinder, $L = 2\pi R / \beta$, becomes sufficiently large.

Finally, we analyzed the properties of the vortical effects for imaginary rotation. In particular, a charged (non-vanishing $\mu$), unpolarized (no chiral or helical chemical potentials) plasma develops a flow of chirality and helicity, manifested as non-vanishing expectation values for the charge current components $J^z_A$ and $J^z_H$, respectively. Both these observables have vanishing asymptotic value. Their fluxes through a transverse-plane disk of radius $R$, $\mathcal{F}_{A/H}(R)$, are non-vanishing. Our analysis showed that $\mathcal{F}_{A/H}(R)$ is finite, even when $R \to \infty$. The total axial flux through the transverse plane, $\mathcal{F}_{A}(\infty)$, is independent of the chemical potential. Its value is a function of both $p$ and $q$, obtained from the irreducible fraction representation of $\nu = p/q$. An analytic computation for the particular case when $p = 1$ revealed that the axial flux grows linearly with $q = \nu^{-1}$, in such a way that when $\nu = 1 /q \to 0$, $\mathcal{F}_A(\infty) \to \infty$. In the case of strictly-vanishing rotation, $q = 1$ and the axial flux vanishes identically, showing that the $\nu \to 0$ limit taken as $q \to \infty$ for $\nu = 1/q$ is disconnected from the particular point where $\nu = 0$.

\begin{acknowledgments}
We thank P. Aasha for fruitful discussions. This work was funded by the EU’s NextGenerationEU instrument through the National Recovery and Resilience Plan of Romania - Pillar III-C9-I8, managed by the Ministry of Research, Innovation and Digitization, within the project entitled ``Facets of Rotating Quark-Gluon Plasma'' (FORQ), contract no. 760079/23.05.2023 code CF 103/15.11.2022.
\end{acknowledgments}

\appendix

\section{Properties of polylogarithms} \label{app:Li}

In this section of the appendix, we derive some important properties of the polylogarithm function, defined as
\begin{equation}
 {\rm Li}_n(z) = \sum_{k = 1}^\infty \frac{z^k}{k^n}. \label{eq:Lin_def}
\end{equation}
In particular, we are looking for closed form expressions for the combinations ${\rm Li}_n(-e^{a}) + (-1)^n {\rm Li}_n(-e^{-a})$.

In the case $n = 1$, taking advantage of the relation
\begin{equation}
 {\rm Li}_1(z) = \sum_{k =1}^\infty \frac{z^k}{k} = -\ln(1 - z),
\end{equation}
we establish:
\begin{equation}
 {\rm Li}_1(-e^a) - {\rm Li}_1(-e^{-a}) = -\ln\left(\frac{1 + e^a}{1 + e^{-a}}\right) = -a.
 \label{eq:Li1_id_aux}
\end{equation}

For higher values of $n$, we first note that the polylogarithm function satisfies the recurrence relation:
\begin{equation}
 \frac{d{\rm Li}_n(z)}{dz} = \frac{1}{z} {\rm Li}_{n-1}(z),
\end{equation}
which leads to the following relation:
\begin{equation}
 \frac{d}{da} \left[{\rm Li}_n(-e^a) \pm {\rm Li}_n(-e^{-a})\right] = {\rm Li}_{n-1}(-e^a) \mp {\rm Li}_{n-1}(-e^{-a}).
\end{equation}

Integrating Eq.~\eqref{eq:Li1_id_aux} with respect to $a$, we arrive at
\begin{equation}
 {\rm Li}_2(-e^a) + {\rm Li}_2(-e^{-a}) = -\frac{a^2}{2} + C_2.
\end{equation}
The integration constant $C_2$ can be fixed by evaluating the LHS for $a = 0$. Using the definition of the alternating zeta function,
\begin{equation}
 {\rm Li}_n(-1) = \sum_{k = 1}^\infty \frac{(-1)^k}{k^n} = -\eta(n),
\end{equation}
as well as the relation $\eta(n) = (1 - 2^{1-n}) \zeta(n)$ between the alternating and usual zeta functions, we obtain $C_2 = -2\eta(2) = -\zeta(2)$, with $\zeta(2) = \pi^2 / 6$. We thus arrive at
\begin{equation}
 {\rm Li}_2(-e^a) + {\rm Li}_2(-e^{-a}) = -\frac{a^2}{2} - \frac{\pi^2}{6}.
 \label{eq:Li2_id_aux}
\end{equation}

Integrating now Eq.~\eqref{eq:Li2_id_aux} with respect to $a$ gives
\begin{equation}
 {\rm Li}_3(-e^a) - {\rm Li}_3(-e^{-a}) = -\frac{a^3}{6} - \frac{\pi^2 a}{6}.
 \label{eq:Li3_id_aux}
\end{equation}
There is no integration constant as the LHS vanishes when $a =0$. Integrating the above again with respect to $a$, and using $\zeta(4) = \pi^4 / 90$, as well as ${\rm Li}_4(-1) = -\eta(4) = -7\pi^4 / 720$, we get
\begin{equation}
 {\rm Li}_4(-e^a) + {\rm Li}_4(-e^{-a}) = -\frac{a^4}{24} - \frac{\pi^2 a^2}{12} - \frac{7\pi^4}{360}.
 \label{eq:Li4_id_aux}
\end{equation}
Finally, in the case when $n = 5$, we find
\begin{equation}
 {\rm Li}_5(-e^a) - {\rm Li}_5(-e^{-a}) = -\frac{a^5}{120} - \frac{\pi^2 a^3}{36} - \frac{7\pi^4 a}{360}.
 \label{eq:Li5_id_aux}
\end{equation}

Inserting now $a = i \pi \nu$, we employ the notation $\operatorname{Li}^{\nu, \mathfrak{r}(\mathfrak{i})}_s$ introduced in Eqs.~\eqref{eq:Li_r_i_def} and reexpress Eqs.~\eqref{eq:Li1_id_aux} and \eqref{eq:Li2_id_aux}--\eqref{eq:Li5_id_aux} as
\begin{subequations}
\label{eq:polylogs}
\begin{align}
 \operatorname{Li}^{\nu, \mathfrak{i}}_1 & = - \frac{\pi \nu}{2}, \label{eq:Li1_id} \\
 \operatorname{Li}^{\nu, \mathfrak{r}}_2 &= - \frac{\pi^2}{12} + \frac{\pi^2 \nu^2}{4}, \label{eq:Li2_id} \\
 \operatorname{Li}^{\nu, \mathfrak{i}}_3 &= - \frac{\pi^3 \nu}{12} + \frac{\pi^3 \nu^3}{12}, \label{eq:Li3_id} \\
 \operatorname{Li}^{\nu, \mathfrak{r}}_{4} &= -\frac{7\pi^4}{720} + \frac{\pi^4 \nu^2}{24} - \frac{\pi^4 \nu^4}{48},\label{eq:Li4_id} \\
 \operatorname{Li}^{\nu, \mathfrak{i}}_{5} &= -\frac{7\pi^5 \nu}{720} + \frac{\pi^5 \nu^3}{72} - \frac{\pi^5 \nu^5}{240}.\label{eq:Li5_id}
\end{align}
\end{subequations}

\section{Polygamma summations}\label{app:psi}

In this section, we evaluate the terms $\mathcal{Q}^r_k$ and $\mathcal{P}^r_k$ defined in Eq. \eqref{eq:sumQ} and show that they lead to Eq.~\eqref{eq:QP}. The results will be expressed with the help of the digamma function, $\psi(z) = d\ln \Gamma(z) / dz = \Gamma'(z) / \Gamma(z)$.

\subsection{The \texorpdfstring{$\mathcal{Q}^r_k$}{Qrk} calculation}

\subsubsection{Odd \texorpdfstring{$k = p + q$}{k=p+q}}

In the case when $k = p+q$ is odd, Eq.~\eqref{eq:sumQ_Q} becomes
\begin{equation}
 \mathcal{Q}^r_{\rm odd}(x_r) = \sum_{Q = 0}^\infty \frac{(-1)^Q}{(Q + \frac{r}{q})^2 + x_r^2}.
 \label{eq:sumQ_Qodd}
\end{equation}
To compute the above series, we employ Eq.~(5.7.7) of Ref.~\cite{DLMF}, given below:
\begin{equation}
 \psi\left(\frac{z + 1}{2}\right) - \psi\left(\frac{z}{2}\right) = 2\sum_{k = 0}^\infty \frac{(-1)^k}{k + z}.
 \label{eq:Dpsi_ser}
\end{equation}
To arrive at Eq.~\eqref{eq:sumQ_Qodd}, we substitute $z = \frac{r}{q} + i x_r$ and $z = \frac{r}{q} - i x_r$ in the above relation and take the difference between the two expressions. Employing the notation $\Delta \psi_r = \psi \left(\frac{\frac{r}{q}+i x_r+1}{2}\right)-\psi \left(\frac{\frac{r}{q}+i x_r}{2}\right)$ introduced in Eq.~\eqref{eq:delta_psi_r}, we find
\begin{equation}
 \Delta \psi_r - \Delta \psi_r^* = -4i x_r \mathcal{Q}^r_{\rm odd}.
\end{equation}
Upon division by $-4ix_r$, we arrive at the first relation in Eq.~\eqref{eq:QP}.

\subsubsection{Even \texorpdfstring{$k =p +q$}{k=p+q}}

In the case when $k = p+q$ is even, Eq.~\eqref{eq:sumQ_Q} becomes
\begin{equation}
 \mathcal{Q}^r_{\rm even}(x_r) = \sum_{Q = 0}^\infty \frac{1}{(Q + \frac{r}{q})^2 + x_r^2}.
\end{equation}
Using the series representation of $\psi(z)$, given in Eq.~(5.7.6) of Ref.~\cite{DLMF} and reproduced below,
\begin{equation}
 \psi(z) = -\gamma + \sum_{k =0}^\infty \left(\frac{1}{k+1} - \frac{1}{k + z}\right),
 \label{eq:digamma_ser}
\end{equation}
where $\gamma \simeq 0.577$ is the Euler-Mascheroni constant, it is not difficult to see that
\begin{equation}
 \psi\left(\frac{r}{q} + i x_r\right) - \psi\left(\frac{r}{q} - ix_r\right) = -2ix_r \mathcal{Q}^r_{\rm even}(x_r),
\end{equation}
thereby establishing the second relation in Eq.~\eqref{eq:QP}.

\subsection{The \texorpdfstring{$\mathcal{P}^r_k$}{Prk} calculation}

\label{subsec:P}

\subsubsection{Odd \texorpdfstring{$k = p + q$}{k=p+q}}

In the case when $k$ is odd, Eq.~\eqref{eq:QP} reduces to
\begin{equation}
 \mathcal{P}_{\rm odd}^r = \sum_{Q = 0}^\infty \frac{(-1)^Q(Q + \frac{r}{q})}{(Q + \frac{r}{q})^2 + x_r^2}.
\end{equation}
Starting from Eq.~\eqref{eq:Dpsi_ser}, setting $z = \frac{r}{q} + ix_r$  and $z = \frac{r}{q} - ix_r$ and adding the results, we arrive at
\begin{equation}
 \Delta \psi_r + \Delta \psi_r^* = 4 \mathcal{P}^r_{\rm odd}.
\end{equation}
Dividing by $4$ establishes the third relation in Eq.~\eqref{eq:QP}.\\

\subsubsection{Even \texorpdfstring{$k = p + q$}{k=p+q}}
\label{subsec:P_even}

In the case when $k$ is even, Eq.~\eqref{eq:QP} becomes
\begin{align}
 \mathcal{P}_{\rm even}^r &= \sum_{Q = 0}^\infty \frac{Q + \frac{r}{q}}{(Q + \frac{r}{q})^2 + x_r^2} = {\rm Re} \left(\sum_{Q = 0}^\infty \frac{1}{Q + \frac{r}{q} + i x_r}\right).
\end{align}
As explained also in the main text, the above series is divergent for fixed $r/q$. However, it can be regularized by making use of the fact that it is under a summation over $r$ from $1$ to $q - 1$, which can be rewritten as follows:
\begin{equation}
\label{eq:trick}
\sum_{r = 1}^{q - 1} f_r = \frac{1}{2} \sum_{r = 1}^{q - 1} (f_r + f_{q - r}).
\end{equation}
The functions $f_r$ containing $\mathcal{P}_{\rm even}^r$ as a factor can be read from Eqs.~\eqref{eq:fractal_transient}, being of the form $\frac{1}{\pi^2} (-1)^{r+1} \delta A_r$, with $A \in \{J^\varphi_V, J^z_A, T^{t\varphi}_r\}$. Employing the reflection properties
\begin{subequations}
\begin{align}
    s_{q - r} &= (-1)^{p + 1} s_r, \\
    x_{q - r} &= (-1)^{p + 1} x_r, \\
    s^{r\mu}  &= (-1)^{p + 1} s^{r\mu},
\end{align}
\end{subequations}
and writing $f_r = \widetilde{f}_r \mathcal{P}^r_{\rm even}$, it can be seen that for all quantities considered above, we have
\begin{equation}
 \widetilde{f}_{q-r} = -\widetilde{f}_r,
\end{equation}
keeping in mind that $k = p + q$ is even. We can thus replace the original function $\mathcal{P}^r_{\rm even}$ by
\begin{align}
 \mathcal{P}^r_{\rm even} &\to \widetilde{\mathcal{P}}^r_{\rm even} =
 \frac{1}{2} \left(\mathcal{P}^r_{\rm even} - \mathcal{P}^{q-r}_{\rm even}\right) \nonumber\\
 &= \frac{1}{2} {\rm Re} \left[\sum_{Q = 0}^\infty \left(\frac{1}{Q + \frac{r}{q} + i x_r} - \frac{1}{Q + 1 - \frac{r}{q} - ix_r} \right)\right].
\end{align}
The expression within the square brackets can be expressed in terms of Eq.~\eqref{eq:digamma_ser},
\begin{equation}
 \widetilde{\mathcal{P}}^r_{\rm even} = \frac{1}{2} {\rm Re}\left[\psi\left(\frac{r}{q} + ix_r\right) - \psi\left(1 - \frac{r}{q} - ix_r\right)\right].
\end{equation}
Using now the reflection formula $\psi(z) - \psi(1 - z) = -\pi / \tan(\pi z)$, given as Eq.~(5.5.4) in Ref.~\cite{DLMF}, we arrive at
\begin{equation}
 \widetilde{\mathcal{P}}^r_{\rm even} = \frac{\pi}{2} {\rm Re} \left\{\cot \left[\pi \left(\frac{r}{q} + i x_r\right)\right]\right\}.
\end{equation}
Evaluating the real part in the above expression establishes the fourth relation in Eq.~\eqref{eq:QP}.
\\
\bibliography{plasma}

\end{document}